\documentclass[aps,prb,twocolumn,superscriptaddress,showpacs,showkeys,floatfix]{revtex4}

\usepackage{graphicx,color}
\graphicspath{{figs/}}
\newcommand{\jwj}[1]{\textcolor{red}{#1}}

\begin{document}
\title{A Review on Flexural Mode of Graphene: Lattice Dynamics, Thermal Conduction, Thermal Expansion, Elasticity, and Nanomechanical Resonance}
\author{Jin-Wu Jiang}
    \altaffiliation{Corresponding author: jiangjinwu@shu.edu.cn; jwjiang5918@hotmail.com}
    \affiliation{Shanghai Institute of Applied Mathematics and Mechanics, Shanghai Key Laboratory of Mechanics in Energy Engineering, Shanghai University, Shanghai 200072, People's Republic of China}
\author{Bing-Shen Wang}
    \affiliation{State Key Laboratory of Semiconductor Superlattice and Microstructure and Institute of Semiconductor, Chinese Academy of Sciences, Beijing 100083, China}
\author{Jian-Sheng Wang}
    \affiliation{Department of Physics and Centre for Computational Science and Engineering, National University of Singapore, Singapore 117551, Republic of Singapore}
\author{Harold S. Park}
    \affiliation{Department of Mechanical Engineering, Boston University, Boston, Massachusetts 02215, USA}
\date{\today}
\begin{abstract}
Single-layer graphene is so flexible that its flexural mode (also called the ZA mode, bending mode, or out-of-plane transverse acoustic mode) is important for its thermal and mechanical properties. Accordingly, this review focuses on exploring the relationship between the flexural mode and thermal and mechanical properties of graphene. We first survey the lattice dynamic properties of the flexural mode, where the rigid translational and rotational invariances play a crucial role. After that, we outline contributions from the flexural mode in four different physical properties or phenomena of graphene -- its thermal conductivity, thermal expansion, Young's modulus, and nanomechanical resonance.  \jwj{We explain how graphene's superior thermal conductivity is mainly due to its three acoustic phonon modes at room temperature, including the flexural mode.} Its coefficient of thermal expansion is negative in a wide temperature range resulting from the particular vibration morphology of the flexural mode.  We then describe how the Young's modulus of graphene can be extracted from its thermal fluctuations, which are dominated by the flexural mode. Finally, we discuss the effects of the flexural mode on graphene nanomechanical resonators, while also discussing how the essential properties of the resonators, including mass sensitivity and quality factor, can be enhanced. 

\end{abstract}

\maketitle
\tableofcontents

\section{Introduction}
This review focuses on the connection between the fundamental lattice dynamics and the thermal and mechanical properties of the unique two-dimensional material graphene.  The lattice dynamical properties, which comprise the phonon spectrum or phonon modes, give fundamental information regarding the atomic interaction within the material. The framework for lattice dynamics was established by Born in the 1920s, and further developed by Debye, Einstein, Mott and others in the following decades. An interesting point is that while lattice dynamics is an atomic scale theory, direct connections can be made to macroscopic phenomena and properties. Readers are referred to the book by Born and Huang for a comprehensive description and discussion regarding lattice dynamic theory.\cite{BornM}

We focus on the flexural mode, which is a characteristic vibration mode in solid plates or rods.\cite{LandauLD} Elastic waves in solid plates are guided by the two outer surfaces, at which the component of stress is zero in the perpendicular direction. Kichhoff formulated the exact equation and boundary conditions for the flexural vibration of a plate in 1850.\cite{KirchhoffG1850} A practical theoretical model of the flexural motion was first developed by Rayleigh in 1885.\cite{RayleighL1885} In 1917, Lamb completely solved the surface vibration problem.\cite{LambH1917prsl} It was found that the free surface boundary condition restricts the elastic waves in the plate into two infinite sets of Lamb waves. The motion of one set of the Lamb wave is symmetrical about the midplane of the plate, while the other is anti-symmetric about the midplane. The zero-order (lowest-frequency) wave from the antisymmetric Lamb set is the flexural mode, whose defining characteristic is its parabolic dispersion curve. In this sense, flexural mode is one particular form of the Lamb wave in solid plates. Because the flexural mode has the lowest frequency among all Lamb waves, it is the easiest to be excited and carries most of the vibrational energy.

In microscopic lattice dynamics theory, the surface phonon mode is well known, and the long wave limits of the flexure phonon mode is the acoustic flexural wave in elastic mechanics. For three-dimensional bulk solid with volume $L^3$, the ratio of surface phonon mode number to total phonon mode number is proportional to $1/L$. Hence, in large piece of bulk materials, surface phonon modes (including the flexural mode) are not so important, except for some specific surface-related topics such as surface reconstruction, surface adsorption, and surface chemistry, etc. However, nanomaterials have large surface to volume ratio, so surface modes play an important role. As an extreme case, all atoms in graphene are exposed on the surface, so all phonon modes are essentially ``surface modes", and the flexural mode occupies one sixth of all phonon modes. That is the origin for the importance of the flexural mode in graphene.

The understanding and study of flexural modes in other novel materials have intensified significantly in recent years due to the recent isolation of the thinnest possible two-dimensional, one-atom-thick ``plate", graphene. Graphene is the best-known one-atom-thick two-dimensional material, which earned a Nobel prize in physics for Novoselov and Geim in 2010.\cite{GeimAK2007nm} Accordingly, substantial effort has been expended in achieving a basic understanding of the lattice dynamical properties of single-layer graphene due to its status as the thinnest possible two-dimensional material. A key outcome of its low-dimensional structure is the existence of a flexural mode (also named ZA mode, bending mode, or out-of-plane transverse acoustic mode) in graphene. As the long wave flexural mode has the lowest frequency among all phonon modes, it is the easiest to be excited in graphene.

A complete understanding of the effect that the flexural mode has on the thermal and mechanical properties of graphene will be essential for many of the key applications graphene has been envisioned for. For example, graphene has been touted as the next-generation replacement for silicon in integrated circuits, though its bandgap is zero unless mechanical strain, doping, or finite width nanoribbons are created.\cite{NovoselovKS2005nat} For these electronic applications, it is important to efficiently remove heat from the transistor during its high speed operation. Graphene possesses a superior thermal conductivity, which is crucial to prevent graphene based transistors from suffering thermally-induced malfunctioning during operation.\cite{BalandinAA2008,NikaDL2009prb,NikaDL2009apl,Balandin2011nm,JiangJW2009direction} It is now clear that the high thermal conductivity of graphene is contributed by its three acoustic modes at room temperature, including flexural mode.

The flexural mode also controls the behavior and properties of graphene nanoelectromechanical systems (NEMS) and nanoresonators, which have been proposed for various sensing applications, in particular ultrasensitive mass sensing and detection. However, the performance of these NEMS, and in particular its quality factor and energy dissipation, is strongly controlled by the flexural mode. Therefore, understanding how the limits imposed by the flexural mode as well as methods to circumvent these flexural mode-related loss mechanisms will be essential to enabling graphene NEMS devices.  

From the above, it is clear that the flexural mode plays a critical role in governing the thermal and mechanical properties of graphene that will be critical for the success of graphene as a commercially-viable material. Due to the ongoing interest in graphene, it is an appropriate time to review the topic of the flexural mode in graphene, and its impact on graphene's physical properties.

\section{Lattice Dynamics}

\subsection{Introduction}
The phonon dispersion of graphene is calculated through the diagonalization of its dynamical matrix, which is a $6\times 6$ complex matrix, due to two inequivalent carbon atoms in its primitive unit cell. The dynamical matrix is constructed based on the interatomic interaction and the space group symmetry of the system.\cite{BornM} The eigenvalue solution of the dynamical matrix gives the frequency and the eigen vector (vibration morphology) for all phonon modes in the system. The vibration morphology of the phonon mode actually represents a particular symmetric mechanical vibration of the graphene. Phonon modes can be used to decompose a general movement of the graphene. In this sense, phonon modes are usually called normal modes.

The lattice dynamics provides fundamental information for mechanical and thermal properties in graphene. For instance, the flexural mode is characteristic by its parabolic spectrum, which is related to its superior thermal conductivity.\cite{BalandinAA2008} The special vibration morphology of the flexural mode is the origin for the strong thermal contraction effect in graphene in a wide temperature range.\cite{BaoW2009nn} The flexural mode's long life time results in a high quality (Q) factor of the graphene nanomechanical resonator (GNMR).\cite{JensenK} These few examples demonstrate the importance of the flexural mode in graphene. Hence, our first task in present review is to recall the computation of the phonon dispersion for graphene.

\subsection{Lattice Dynamics for Flexural Mode}

\subsubsection{Bloch's Theorem}
In this subsection, we take the honeycomb lattice structure of graphene as an example to explain the application of the Bloch's theorem. Graphene has a honeycomb lattice structure as shown in Fig.~\ref{fig_cfg_vffm}. Its symmetry is described by the $D_{6h}$ point symmetry group.\cite{TangH2011book} According to its space group, the whole honeycomb lattice can be obtained by repeating a rhombus primitive unit cell, with two bases $\vec{a}_{1}$ and $\vec{a}_{2}$. The lattice constant\cite{SaitoR} $|\vec{a}_{i}|=2.46$~{\AA}. Each cell is denoted by a pair of integers ($l_{1}$, $l_{2}$). Carbon atoms are indexed by ($l_{1}$, $l_{2}$, $s$), where $s$=A or B are the two inequivalent carbon atoms in each unit cell. The locations of atoms A and B in the unit cell (0, 0) are $\vec{\tau}_{A}=(\vec{a}_{1}+\vec{a}_{2})/3$ and $\vec{\tau}_{B}=2(\vec{a}_{1}+\vec{a}_{2})/3$. The position of an arbitrary carbon atom is determined by $\vec{r}(l_{1},l_{2},s)=\vec{R}_{l_{1}l_{2}} + \vec{\tau}_{s}$. The lattice vector is $\vec{R}_{l_{1}l_{2}}=l_{1}\vec{a}_{1}+l_{2}\vec{a}_{2}$.

The structure of an infinite graphene sheet is unchanged after it is shifted for a lattice vector $\vec{R}_{l_{1}l_{2}}$. That is graphene has lots of translation symmetries. All of these translation operations construct an Abelian translation group\cite{RotmanJJ1995book} $\textbf{T}$. The group elements are the translation operations $\hat{T}_{l_{1}l_{2}}$. For convenience, in practical calculations, the infinite system is usually replaced by a finite system with periodic boundary conditions in the in-plane directions. The dimension of the finite graphene is $N_{1}\vec{a}_{1}\times N_{2}\vec{a}_{2}$, so integers ($l_{1}$, $l_{2}$) have finite possible values, i.e., $l_{1}=0, 1, ..., N_{1}-1$ and $l_{2}=0, 1, ..., N_{2}-1$.

There are $N_{1}N_{2}$ elements in the translation group $\textbf{T}$, with $\hat{T}_{1,0}$ and $\hat{T}_{0,1}$ as the two generators for this cyclic group. Using the character table of the group, it can be found that there are $N_{1}N_{2}$ one-dimensional irreducible representations for the group $\textbf{T}$. The irreducible representation can be labeled by the wave vector $\vec{k}$. The representation form of the translation operation is $\hat{T}_{l_{1}l_{2}}=e^{i\vec{k}\cdot \vec{R}_{l_{1}l_{2}}}$ in the irreducible representation $\vec{k}$. For convenience, the reciprocal space is usually introduced via the definition of the reciprocal basis vectors $(\vec{b}_{1}, \vec{b}_{2})$ as,
\begin{eqnarray}
\vec{a}_{i}\cdot \vec{b}_{j}=2\pi \delta_{ij},
\end{eqnarray}
where $i$ and $j$ are 1 or 2. $\delta_{ij}$ is the Kronecker delta. The wave vector $\vec{k}$ can be written in terms of the reciprocal bases as $\vec{k}=k_{1}\vec{b}_{1}+k_{2}\vec{b}_{2}$. Using the cyclic property of the translation group, we have $\hat{T}_{l_{1}0}^{N_{1}}=1$ and $\hat{T}_{0l_{2}}^{N_{2}}=1$, so we get $k_{1}=j_{1}/N_{1}$ and $k_{2}=j_{2}/N_{2}$ with $j_{1}=$ 0, 1, 2, ..., $N_{1}-1$ and $j_{2}=$ 0, 1, 2, ..., $N_{2}-1$.

The Bloch's theorem says that, in the irreducible representation $\vec{k}$, the displacements of atoms in the unit cell ($l_{1}$, $l_{2}$) are related to atoms the (0, 0) unit cell by a phase factor $e^{i\vec{k}\cdot \vec{R}_{l_{1}l_{2}}}$; i.e., $u(l_1l_2s)=u(00s)e^{i\vec{k}\cdot \vec{R}_{l_{1}l_{2}}}$. A general displacement can be expanded in follow terms,
\begin{eqnarray}
\vec{u}(l_{1}l_{2}s) & = & \frac{1}{\sqrt{m_{s}}}\frac{1}{\sqrt{N_{1}N_{2}}}\sum_{\vec{k}}\sum_{\tau=1}^{6}\hat{Q}_{\vec{k}}^{(\tau)}e^{i\vec{k}\cdot\vec{R}_{l_{1}l_{2}}}\vec{\xi}^{(\tau)}(\vec{k}|00s).\nonumber\\
\end{eqnarray}
This formula works for the lattice using the translational part of the space group. For nanotubes with screw symmetries (line group), the Bloch's theorem should be generalized to include screw operations.\cite{MiloevicI1993prb,PopovVN1999prb,DobardzicE2003prb,JiangJW2006,DakicB2009jpamt} An explicit introduction on the lattice dynamics of nanotubes can be found in the book chapter by Tang, Wang, and Su.\cite{TangH2011book} In the above expansion, $\vec{\xi}^{(\tau)}(\vec{k}|00s)$ is actually the vibrational displacement for atom $(00s)$ in the $\vec{k}$ mode (irreducible representation). These eigen vectors are orthogonal to each other. The wave vector $\vec{k}$ is used to denote the phonon modes; $\tau$ runs over the six branches. There are two relations for the eigen vector $\vec{\xi}^{(\tau)}(\vec{k}|00s)$ and the quantum operator $\hat{Q}_{-\vec{k}}^{(\tau)}$:
\begin{eqnarray*}
 &  & \vec{\xi}^{(\tau)}(-\vec{k}|00s)=\vec{\xi}^{(\tau)}(\vec{k}|00s)^{*}\\
 &  & \hat{Q}_{-\vec{k}}^{(\tau)}=\hat{Q}_{\vec{k}}^{(\tau)\dagger}.
\end{eqnarray*}

\subsubsection{Dynamical Matrix}
The general potential energy of graphene is determined by positions of all atoms $(l_{1}, l_{2}, s)$, i.e., $V_{0}=V(\vec{r}_{00s}, ...\vec{r}_{l_{1}l_{2}s},... \vec{r}_{N_{1}-1 N_{2}-1 s})$. There will be some variance in the total potential energy, if there is a small displacement of atom $(l_{1}, l_{2}, s)$, i.e., $\vec{r}_{l_{1} l_{2} s} \rightarrow \vec{r}_{l_{1} l_{2} s}+\vec{u}_{l_{1} l_{2} s}$. Here $\vec{u}_{l_{1} l_{2} s}$ is a small displacement for atom $(l_{1}, l_{2}, s)$. The total potential energy can be expanded in a Taylor series of displacements $\vec{u}_{l_{1} l_{2} s}$,
\begin{eqnarray}
&&V(\vec{r}_{l_{1} l_{2} s}+\vec{u}_{l_{1} l_{2} s}) \nonumber\\
&=& V_{0} + \sum_{l_1=1}^{N_1}\sum_{l_2=1}^{N_2}\sum_{s=A,B}\sum_{\alpha=x,y,z}\frac{\partial V}{\partial u_{l_{1} l_{2} s}^{\alpha}}u_{l_{1} l_{2} s}^{\alpha}\nonumber\\
 &+& \frac{1}{2}\sum_{l_1,l'_1=1}^{N_1}\sum_{l_2,l_2'=1}^{N_2}\sum_{s,s'=A,B}\sum_{\alpha,\beta=x,y,z}\frac{\partial^{2} V}{\partial u_{l_{1} l_{2} s}^{\alpha} \partial u_{l_{1}' l_{2}' s'}^{\beta}}\nonumber\\
&\cdot&u_{l_{1} l_{2} s}^{\alpha}u_{l_{1}' l_{2}' s'}^{\beta} + ...,
\end{eqnarray}
where higher order nonlinear terms have been omitted. The first term $V_{0}$ is the minimum potential energy of the system at the optimized configuration. The second term vanishes due to equilibrium condition for the optimized structure. The third term describes a harmonic energy induced by the displacement (vibration). Coefficients in front of the third term are normally called the force constant matrix $K_{l_{1} l_{2} s\alpha;l_{1}' l_{2}' s'\beta}=\frac{\partial^{2} V}{\partial u_{l_{1} l_{2} s}^{\alpha} \partial u_{l_{1}' l_{2}' s'}^{\beta}}$. Applying the Bloch's theorem to all displacement vectors, we get the potential variance up to the second order,
\begin{eqnarray}
\delta V &=& \frac{1}{2}\sum_{\tau \tau'}\sum_{\vec{k}}\hat{Q}_{\vec{k}}^{(\tau)\dagger}\hat{Q}_{\vec{k}}^{(\tau')}\sum_{ss'=A,B}\sum_{\alpha,\beta=x,y,z}\nonumber\\
&\cdot&D_{s\alpha;s'\beta}\left(\vec{k}\right)\xi_{\alpha}^{(\tau)}(\vec{k}|00s)^{\star}\xi_{\beta}^{(\tau')}(\vec{k}|00s').
\end{eqnarray}
We have introduced the dynamical matrix,
\begin{eqnarray}
D_{s\alpha;s'\beta}\left(\vec{k}\right) & = & \frac{1}{\sqrt{m_{s}m_{s'}}}\sum_{l_{1}=1}^{N_{1}}\sum_{l_{2}=1}^{N_{2}}K_{00s\alpha;l_{1}l_{2}s'\beta}e^{i\vec{k}\cdot\vec{R}_{l_{1}l_{2}}},\nonumber\\
\end{eqnarray}
where the summation over $(l_1, l_2)$ can be truncated to the summation over neighboring atoms in case of short-range interactions. The eigenvalue of the dynamical matrix gives the eigen frequency,
\begin{eqnarray}
\sum_{s'\beta}D_{s\alpha;s'\beta}\left(\vec{k}\right)\xi_{\beta}^{(\tau')}(\vec{k}|00s') & = & \omega^{(\tau)2}(\vec{k})\xi_{\alpha}^{(\tau')}(\vec{k}|00s).
\end{eqnarray}

The dynamical matrix is Hermitian, since the force constant matrix $K$ is symmetric. As a result, all eigen values $\omega^{2}$ are real. However, there is no guarantee for the positive definiteness of the dynamical matrix, i.e., it is possible to encounter $\omega^{2}<0$. This positive definite property can actually be used to analyze the structure stability. The relation of $\omega^{2}<0$ leads to an imaginary frequency of the phonon mode, i.e., $\omega=i\gamma$ with real number $\gamma$. If atoms in the system are displaced according to the morphology of this imaginary mode, then the oscillation amplitude is proportional to $e^{-i\omega t}=e^{\gamma t}\longrightarrow +\infty$ in the limit of infinite time. An infinite vibration amplitude indicates the instability of the configuration. As an example, we note that this technique has been applied to predict the instability of nanowires,\cite{PeelaersH} the tension induced instability of graphene,\cite{LiuF2007prb} or the compression-induced buckling of the single-layer molybdenum disulphide.\cite{JiangJW2014mos2bandgap}

From the above, it is clear that the frequency from the dynamical matrix is a linear property, because it is extracted from the harmonic term. Phonon modes have infinite life time in the linear regime. The phonon life time can be limited by various scattering mechanisms. For instance, the phonon-phonon scattering becomes more important at high temperature, leading to a frequency shift and a finite value for the phonon life time. This nonlinear information can be accounted through mode coupling theory\cite{KuboR,WangJS2004pre}, effective phonon conception,\cite{BoniniN,MarianiE} or Boltzmann equation description.\cite{ZimanJM}

\subsection{Origin of Flexural Mode}

\subsubsection{Valence Force Field Model}
\begin{figure}[tb]
    \begin{center}
        \scalebox{0.8}[0.8]{\includegraphics[width=8cm]{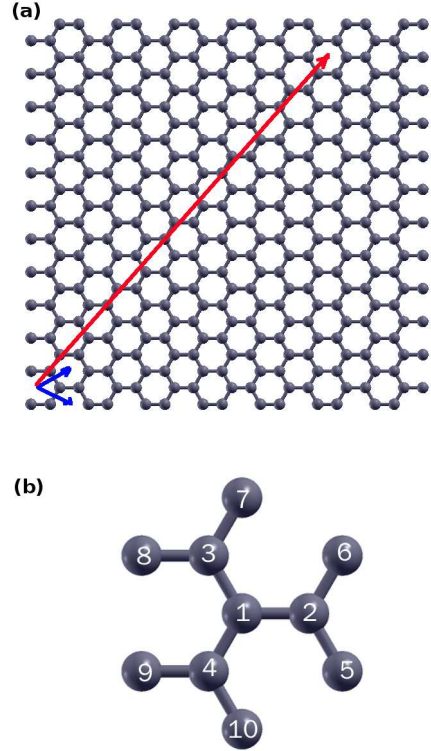}}
    \end{center}
    \caption{Graphene structure. (a) Honeycomb lattice of graphene. Bases $\vec{a}_{1}$ and $\vec{a}_{2}$ are displayed by two short blue arrows. The long red arrow illustrates a lattice vector $\vec{R}_{-6,15}=-6\vec{a}_{1}+15\vec{a}_{1}$. (b) Sketch of the local environment of atom 1; i.e., three first-nearest and six second-nearest neighboring carbon atoms.}
    \label{fig_cfg_vffm}
\end{figure}
There are two major ingredients in the dynamical matrix. The first one is a phase factor, which is contributed by the space group symmetry of the system. As we known, all unit cells can be repeated by the (0,0) unit cell via a corresponding symmetric operation from the space group. The phase factor, $e^{i\vec{k}\cdot \vec{R}_{l_1l_2}}$, carries the relationship between the vibration displacement of the (0,0) unit cell and the $(l_1, l_2)$ unit cell.

The second ingredient in dynamical matrix is the force constant matrix $K_{00s\alpha;l_{1}l_{2}s'\beta}$. The force constant matrix can be obtained mainly through three approaches; i.e., first-principles calculations, empirical potential, or the force constant model. For ionic materials, the shell model\cite{CochranW} or the bond charge model\cite{MartinRM} can be useful for the description of the charge interaction.

In the first two methods, the force constant matrix is calculated by $K_{l_{1} l_{2} s\alpha;l_{1}' l_{2}' s'\beta}=\frac{\partial^{2} V}{\partial u_{l_{1} l_{2} s}^{\alpha} \partial u_{l_{1}' l_{2}' s'}^{\beta}}$, where $V$ is the total potential energy from an empirical potential or the Columb interaction in the first-principles calculations. $u_{l_{1} l_{2} s}^{\alpha}$ is the displacement of the degree of freedom $(l_{1} l_{2} s \alpha)$. This formula is realized numerically by calculating the energy change after displacing a small value for the degrees of freedom $(l_{1} l_{2} s \alpha)$ and $(l'_{1} l'_{2} s' \beta)$. There are some existing packages for such numerical calculation. For instance, some common empirical potentials have been implemented in the lattice dynamic properties package GULP.\cite{gulp} It gives the force constant matrix or the phonon dispersion directly. The first-principles package SIESTA\cite{siesta} also gains some success in the calculation of the force constant matrix or phonon dispersion.

For the third method, there are two popular force constant models for the force constant matrix in graphene, i.e., mass-spring model\cite{JishiRA1993cpl} and valence force field model (VFFM).\cite{AizawaT} In the mass-spring model, each atom is denoted by a mass, that is connected to other atoms via springs. The force constant of the spring governs the force constant matrix. This model includes two-body interaction. It has been shown that the fourth-nearest neighbors should be included in the calculation of the phonon dispersion of graphene.\cite{SaitoR}

The VFFM aims to capture contribution from the valence electrons on the vibration frequency. Energy variations corresponding to both bond length and bond angle are included in this model. As pointed out by Yu in his book,\cite{YuPY} a big advantage of the VFFM is its transferability; i.e., force constant parameters in the model are almost the same for the same bonds within different materials. We will illustrate the explicit form of a VFFM for graphene in the following. The VFFM was successful in diamond\cite{MusgraveMJP} and C$_{d}$S.\cite{NusimoviciMA} A simplified version has been proposed by Keating,\cite{KeatingPN} which has gained success in many covalent semiconductors.

There are five VFFM terms corresponding to five typical vibration motions for the graphene sheet.\cite{AizawaT} The equilibrium position for atom $i$ is $\vec{r}_{i}$. The vector pointing from atoms $i$ to $j$ is $\vec{r}_{ij}=\vec{r}_j-\vec{r}_i$. The distance between atoms $i$ and $j$ is the modulus $r_{ij}$. We will write out interaction for one bond or one angle explicitly. Fig.~\ref{fig_cfg_vffm}~(b) illustrates the three first-nearest-neighboring atoms (2-4) and six second-nearest-neighboring atoms (5-10) for atom 1. The interaction for other bonds or other angles can be obtained analogously. The general expressions can be found in Refs.~\onlinecite{AizawaT,JiangJW2006,JiangJW2008}.

(1) The bond stretching interaction between atoms 1-2,
\begin{eqnarray}
    V_{l}=\frac{k_{l}}{2}[(\vec{u}_{2}-\vec{u}_{1})\cdot \vec{e}_{12}^{~l}]^{2}.
    \label{Potential1}
\end{eqnarray}
$k_{l}$ is the force constant parameter. $ \vec{e}_{12}^{~l}=\vec{r}_{12}/|\vec{r}_{12}|$ is a unit vector from atom 1 to atom 2. This is the bond stretching interaction between two first-nearest-neighboring atoms. There are similar interactions for other first-nearest-neighboring carbon-carbon bonds, i.e., bond 1-3, 1-4, 2-5, and 2-6.

(2) The bond stretching interaction between atoms 1-5,
\begin{eqnarray}
    V_{sl}=\frac{k_{sl}}{2}[(\vec{u}_{5}-\vec{u}_{1})\cdot\vec{e}_{15}^{~l}]^{2}
    \label{Potential2}
\end{eqnarray}
with $k_{sl}$  the corresponding force constant parameter. This term describes the bond stretching interaction between two second-nearest-neighboring atoms. There are similar interactions for other second-nearest-neighboring carbon-carbon bonds.

(3) The angle bending interaction for $\angle 213$ is $V_{BB}$,
\begin{eqnarray}
    V_{BB}=\frac{k_{BB}}{2}(\cos\theta'_{213}-\cos\theta_{213})^{2};.
    \label{Potential3}
\end{eqnarray}
where $\theta_{213}$ is the equilibrium angle and $\theta'_{213}$ is the angle in vibration. This interaction term describes the bending of angles, which are formed by two first-nearest-neighboring C-C bonds. There are similar interactions for the other angles: $\angle 213$, $\angle 214$, $\angle 314$, $\angle 125$, $\angle 126$, and $\angle 526$.

(4) The out-of-plane bond bending is a four-body interaction. It describes the interaction between atom 1 and its neighboring atoms 2-4. If atom 1 moves out of the plane, then its neighboring atoms 2-4 will try to drag it back to the plane. This potential is, $V_{rc}$,
\begin{equation}
    V_{rc}=\frac{k_{rc}}{2}[(3\vec{u}_{1}-(\vec{u}_{2}+\vec{u}_{3}+\vec{u}_{4}))\cdot\vec{e}_{z}]^{2}.
    \label{Potential4}
\end{equation}
$\vec{e}_{z}$ is the unit vector in the out-of-plane direction. Similar interaction is also applied to atom 2.

(5) If the carbon-carbon bond 1-2 is twisted, then the following twist potential will try to react the twisting motion,
\begin{eqnarray}
    V_{tw}=\frac{k_{tw}}{2}[(\vec{u}_{3}-\vec{u}_{4}-(\vec{u}_{6}-\vec{u}_{5}))\cdot\vec{e}_{z}]^{2}.
    \label{Potential5}
\end{eqnarray}
There are similar twisting interactions for the other first-nearest-neighboring carbon-carbon bonds: 1-3, 1-4, 2-5, and 2-6.

\subsubsection{Rigid Translational and Rotational Invariance}

\begin{figure}[tb]
  \begin{center}
    \scalebox{0.8}[0.8]{\includegraphics[width=8cm]{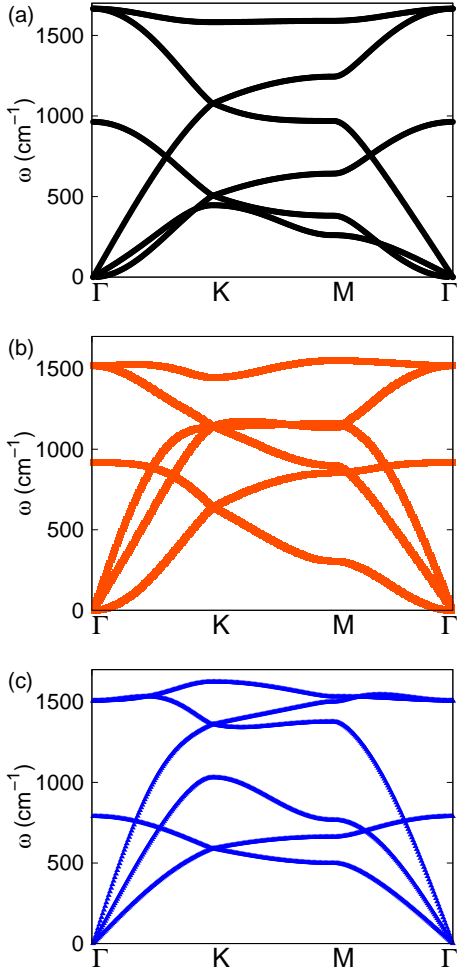}}
  \end{center}
  \caption{(Color online) The phonon dispersion in graphene along high symmetry lines in the Brillouin zone from (a) Brenner potential, (b) VFFM, and (c) mass spring model.}
  \label{fig_dispersion_graphene}
\end{figure}

\begin{figure}[tb]
    \begin{center}
        \scalebox{1}[1]{\includegraphics[width=8cm]{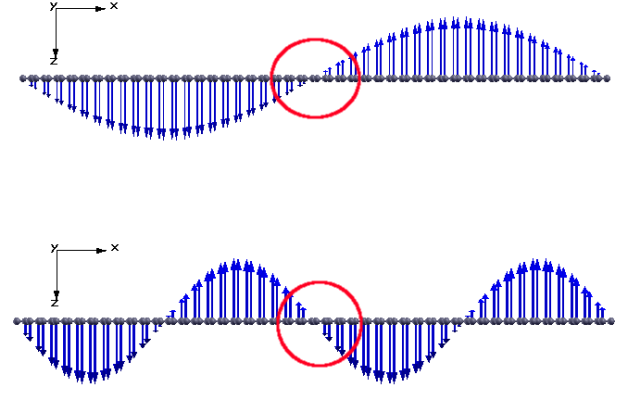}}
    \end{center}
    \caption{Vibration displacement for two flexural modes in graphene. Arrow on top of each atom represents the vibration component of the atom in this vibration mode. Red circles enclose small pieces of graphene, which are effectively rotated around $y$-axis.}
    \label{fig_u_flexuremode}
\end{figure}
The crystal is rigid in the sense that its total potential energy should not vary if the system is rigidly translated or rotated.\cite{BornM} According to this requirement, the empirical potential energy should satisfy two conditions.

\begin{itemize}
\item The rigid translational invariance. It says that, if $\vec{u}_{i}=\vec{u}_{0}$ is a constant vector for all atoms, then we should have $\delta V=0$.

\item The rigid rotational invariance. It says that, if the system is rotated by $\vec{u}_{i}=\delta\vec{\omega}\times\vec{r}_{i}$, then we should also have $\delta V=0$. Here, the rotation angle is $|\delta\vec{\omega}|$ and the rotation direction is $\frac{\delta\vec{\omega}}{|\delta\vec{\omega}|}$.
\end{itemize}

We can check that the five terms in the above VFFM satisfy both translational and rotational invariance.

For the translational invariance, the following relationship can be easily found,
\begin{eqnarray}
\vec{u}_{i}&=&\vec{u}_{j}\\
\vec{u}_{i}-\vec{u}_{j}&=&0.
\end{eqnarray}
As a result, the translational invariance is satisfied, i.e.,
\begin{eqnarray}
V_{l}=V_{sl}=V_{BB}=V_{rc}=V_{tw}=0.
\end{eqnarray}

We will now illustrate the rigid rotational invariance for the above five VFFM potential terms.\cite{JiangJW2006} During a rigid rotation motion, the displacement for atom $i$ is
\begin{eqnarray}
\vec{u}_{i}=\delta\vec{\omega}\times\vec{r}_{i}.
\end{eqnarray}
As a result, we get following relationship,
\begin{eqnarray}
 \vec{u}_{i}-\vec{u}_{j}=\delta\vec{\omega}\times(\vec{r}_{i}-\vec{r}_{j})=\delta\vec{\omega}\times\vec{r}_{ji}.
\label{rotational}
\end{eqnarray}
Using Eq.~(\ref{rotational}), we can get
\begin{eqnarray}
(\vec{u}_{j}-\vec{u}_{i})\cdot\vec{e}_{ij}^{~l} =r_{ij}(\delta\vec{\omega}\times\vec{e}_{ij}^{~l})\cdot\vec{e}_{ij}^{~l} =0.
\end{eqnarray}
As a result, we find that (\ref{Potential1}) and (\ref{Potential2}) are zero under a rigid rotational motion, i.e., 
\begin{eqnarray}
 V_{l}=V_{sl}=0.
\end{eqnarray}
For the the other three potential terms (\ref{Potential3}), (\ref{Potential4}) and (\ref{Potential5}), they become summation over the following expressions, if the system is rotated rigidly,
\begin{eqnarray}
&V_{BB}&\sim\frac{k_{BB}}{4}[\delta\vec{\omega}\cdot(\vec{e}_{12}^{~l}\times\vec{e}_{13}^{~l}+\vec{e}_{13}^{~l}\times\vec{e}_{12}^{~l})]^{2}=0;\\
&V_{rc}&\sim\frac{k_{rc}}{2}[\delta\vec{\omega}\times(\vec{r}_{12}+\vec{r}_{13}+\vec{r}_{14})\cdot\vec{e}_{1}^{~z}]^{2}=0;\\
&V_{tw}&\sim\frac{k_{tw}}{2}[\delta\vec{\omega}\times(\vec{r}_{43}-\vec{r}_{56})\cdot\vec{e}_{12}^{~z}]^{2}=0.
\end{eqnarray}

Fig.~\ref{fig_dispersion_graphene} shows the phonon dispersion in graphene along high symmetric Brillouin line $\Gamma$KM$\Gamma$ using three models. The dispersion in panel (a) is calculated from the Brenner potential.\cite{brennerJPCM2002} In particular, the lowest branch around $\Gamma$ point is the flexural branch with a parabolic spectrum. Panel (b) shows that the spectrum of flexural mode is also parabolic using the VFFM. Parameters in the VFFM are $k_{l}=5.8337$~{eV\AA$^{-2}$}, $k_{sl}=5.2936$~{eV\AA$^{-2}$}, $k_{BB}=10.2245$~{eV}, $k_{rc}=14.8$~N m$^{-1}$, and $k_{tw}=6.24$~N m$^{-1}$. However, panel (c) shows that phonon spectrum of the flexural mode from the mass spring model is linear instead of parabolic. It is because the rigid rotation symmetry is violated in the mass spring model. The longitudinal and transverse parameters are considered up to the fourth nearest-neighboring atoms.\cite{JishiRA1993cpl} These parameters for the mass spring model are $(k_{l}, k_{\perp})=$ (27.7521, 4.4753)~{eV\AA$^{-2}$}, (6.8350, 0.1728)~{eV\AA$^{-2}$}, (0.5054, 0.1021)~{eV\AA$^{-2}$}, and (0.2665, 0.0012)~{eV\AA$^{-2}$}. The comparison in Fig.~\ref{fig_dispersion_graphene} demonstrates that the parabolic spectrum for the flexural mode is closely related to the rigid rotational invariance. 

The vibration displacement of the flexural mode is actually directly related to the rigid rotational motion style. Fig.~\ref{fig_u_flexuremode} shows the vibration displacement of two flexural modes. The arrow on top of each atom represents the vibration displacement of each atom in this mode. We can divide the system into lots of small pieces along the $x$-axis. It can be shown that each piece is effectively rotated around the y-axis in the flexural mode. Red circles in the figure illustrate two graphene pieces, which are effectively rotated around the y-axis. The rigid rotational invariance leads to zero recovery force for this vibration to first order. That is the micro-origin for the parabolic dispersion of the flexural mode.

For nanotubes with cylindrical hollow structure, the rigid rotational invariance guarantees both the zero frequency of twisting mode and the existence of flexural mode.\cite{MahanGD2004prb} The flexural mode (parabolic dispersion) turns into acoustic mode (linear dispersion) gradually with the increase of the thickness of the thin plate. We have used few-layer graphene as an example to show this dimensional crossover phenomenon.\cite{JiangJWcorssover}

\jwj{All three of the phonon spectra in Fig.~\ref{fig_dispersion_graphene} are computed based on short-range empirical interaction potentials, which are known to be less accurate than first-principles calculations.\cite{DubayO2003prb,MounetN,GillenR2009prb}  This may cause some errors because the long-range interactions may impact the flexural mode. For instance, the long-range dipolar interaction force was considered in a model for the flexural mode in graphene.\cite{MetlovKL2010prb}  In particular, the crossover in the two in-plane high frequency optical dispersion deviate from the first-principles results.}

\jwj{Although the phonon spectra from empirical potentials less accurate than first principles methods, they have been widely used in practice for many physical phenomena.\cite{JishiR1982prb,MohrM2007prb} For instance, classical molecular dynamics simulations are commonly used for the study of the thermal transport in graphene. In principle, the interatomic force can be calculated from first-principles calculations, but the associated computational cost renders such approaches infeasible for systems larger than a thousand or so atoms.  Hence, the interatomic force is usually computed from an efficient empirical potentials like the Brenner potential\cite{brennerJPCM2002} or the Stillinger-Weber potential.\cite{StillingerFH} The thermal conductivity obtained from the molecular dynamics simulations can be explained by the phonon spectra of graphene, which should also be calculated based on the same empirical potentials for consistency.  In this sense, the phonon spectra that are calculated from empirical potentials are certainly useful, though they are not as accurate as from first-principles calculations.}

\jwj{The inaccuracy in the optical branches in the phonon spectra from empirical potentials will impact some of the computed properties of graphene. For instance, the thermal conductivity of pure graphene is mainly limited by phonon-phonon scattering at room temperature. A typical phonon-phonon scattering process requires the involvement of the optical phonon, so the inaccuracy in the optical branches will lead to some influence on such scattering processes. However, these empirical potentials can give accurate acoustic branches in the phonon spectra, as can be seen from Fig.~\ref{fig_dispersion_graphene}. Acoustic phonon branches are important for many physical phenomena, such as thermal transport. The thermal conductivity in graphene is mainly contributed by its three acoustic phonon branches. As a result, the inaccuracy in the optical branches in the phonon spectra has a much smaller effect on the computed values for the thermal conductivity.}

\section{Thermal Conduction}

\subsection{Introduction}
Thermal transport occurs in the presence of temperature gradient. In metals, both electrons and phonons are important thermal energy carriers to deliver thermal energy. In insulators or semiconductors, phonons carry most of the thermal energy, while electrons only make limited contribution. The thermal conductivity contributed by phonons is called lattice thermal conductivity. Graphene is a well-known semiconductor with zero electronic band gap. Experiments show that the electronic thermal conductivity is around\cite{YigenS} 10~{W/m$\cdot$K}, which is less than 1\% of the overall thermal conductivity in graphene.\cite{SaitoK} As a result, the electronic contribution can be safely ignored in the study of the thermal conductivity in graphene. In the following, we focus on the lattice thermal conductivity in graphene.

The thermal conductivity ($\kappa$) and thermal conductance ($\sigma$) are two related concepts that are useful in different thermal transport conditions. They are related to each other from their definitions: $\kappa/L=\sigma/s$, where $L$ and $s$ are the length and the cross-sectional area, respectively. It should be noted that for quasi-two-dimensional materials like graphene, the concept of cross-sectional area is not a well defined concept, because it is only one atom thick. It is crucial to use the same thickness in the comparison of thermal conductivity from different measurements or calculations. The thermal conductance is useful in the ballistic thermal transport regime, which is typically the primary transport mechanism in nanoscale structures.  The ballistic transport also happens at very low temperature, where the phonon density is too weak for phonon-phonon scattering. During ballistic transport, each phonon mode delivers a quanta of thermal energy $\hbar \omega$ across the system without scattering. As a result, the thermal conductance is quantized in the ballistic regime.\cite{SchwabK} The ballistic thermal conductance does not depend on the length of the system, because of the infinite phonon mean free path.

The diffusive thermal transport happens in large systems and/or at high temperatures. The thermal transport ability is mainly limited by phonon related scattering mechanisms in the diffusive regime. In this regime,  phonon modes have finite life time and finite mean free path. In other words, the thermal energy carried by a phonon mode get dissipated with increasing distance. In the diffusive regime, the thermal conductivity is a constant with respect to the length of the structure.

\jwj{The thermal transport in graphene has attracted significant interest after the experimental observation of superior thermal conductivity by Balandin et al. in 2008.\cite{BalandinAA2008,GhoshS2008apl} The quasi-ballistic thermal transport was reported in suspended single-layer graphene below room temperature.\cite{XuXF2010} The measured temperature dependence of the thermal conductance scales as $~T^{1.5}$, which is consistent with the contribution from the flexural mode to the thermal conductance.\cite{MingoN2005prl,SaitoK,JiangJW2009direction} The ballistic thermal conductance in graphene was found to be anisotropic,\cite{JiangJW2009direction,XuY2009apl}  while the thermal transport is size-dependent in graphene\cite{MunozE,WangJ2012jpcm,NikaDL2012nl,XuX2014nc} following a theoretical two-dimensional disk model.\cite{XiongD} The isotopic doping effect on the thermal conductivity of graphene was investigated theoretically\cite{JiangJW2010isotopic} and verified experimentally,\cite{ChenSS} and the thermal rectification phenomenon was observed in graphene with asymmetric structures and nonlinear scattering.\cite{YangN2009apl,HuJ,JiangJW2010epl,JiangJW2010topology,HuJ2011apl,ChehJ2012} }

Experiments showed that the thermal conductivity in few-layer graphene decreases exponentially with increasing layer number and eventually crosses over to that exhibited by bulk, three-dimensional graphite value.\cite{GhoshS} This dimensional crossover phenomenon has received intensive theoretical effort as intrigued by the experimental work by Ghosh et al..\cite{SinghD,LindsayL,KongBD,ZhangG2011nns,ZhongWR20111,ZhongWR20112,RajabpourA,CaoHY,SunT,JiangJW2014flgthermal} For the single-layer graphene, all of these theoretical works have shown that the single-layer graphene has the highest thermal conductivity among all few-layer graphene systems. For few-layer graphene with layer number above two, most theoretical calculations demonstrated a monotonic decrease of the thermal conductivity with increasing layer number; while the thermal conductivity was found to be independent of the layer number in a recent molecular dynamics simulations.\cite{JiangJW2014flgthermal}

It is important for nano devices to spread the thermal energy generated during its operation. Graphene has a superior thermal conductivity, which can be very helpful in delivering heat from these devices. For instance, experiments have shown that the graphene based composites have a much higher thermal conductivity owning to graphene's superior thermal transport ability.\cite{YangZ,GoyalV,ShahilKMF}

Lots of approaches have been proposed to manipulate the thermal conductivity of graphene, including the application of axial or bending strain,\cite{LiX,JiangJW2011negf,WeiN,GunawardanaKGSH,ZhangJ} edge reconstruction,\cite{JiangJW2009edge,LanJH2009,JiangJW2010edgemode,SavinAV,TanZW,HuJ2010apl} interfaces,\cite{JiangJW2011minimum,JiangJW2011bngra,ChehJ} points defects,\cite{JiangJW2011defect,HaoF,ZhangH,AdamyanV,SerovAY} wrinkles,\cite{ChenSS2012nano} substrate coupling,\cite{CaiWW2010nl,ChenSS2011acsn,LeeJU,GuoZX2011,OngZY2011,HuangSY,QiuB,GuoZX} and asymmetric interactions,\cite{XiongD,ChenS2012arxiv,ChenS2013arxiv} amongst others.

In the following, we focus on the role of the flexural modes on the thermal conductivity in graphene. There have been a bunch of reviews focusing on different aspects of the thermal conductivity of graphene. Wang et al. discuss the non-equilibrium Green's function (NEGF) approach in the calculation of the thermal conductivity for nanomaterials.\cite{WangJSnegf,WangJS2013fp} The comparison between the thermal conductivity in graphene and other carbon materials was summarized in Ref.~\onlinecite{Balandin2011nm}. Different theoretical approaches to calculating the thermal conductivity in graphene is outlined in Ref.~\onlinecite{NikaDL2012jpcm}. Several reviews have been devoted to the discussion of the anomalous thermal transport in low dimensional nanoscale systems including graphene.\cite{CahillDG,DharA2008,LiuS,LiNB2012rmp,YangN2012aipa,LuoTF,MarconnetAM} Some basic issues on the heat transport in microscopic level are surveyed in Ref.~\onlinecite{DubiY} by Dubi and Ventra. Zhang and Li discuss the isotopic doping effect on thermal properties on nanomaterials, including the isotopic doping effect on the thermal conductivity in graphene.\cite{ZhangG2010nns} Heat dissipation in nanoscale electronic device is outlined by Pop in Ref.~\onlinecite{PopE2010}.

\subsection{Simulation of Thermal Transport}
There are several available approaches for the calculation of thermal conductivity. The ballistic thermal conductance can be predicted rigorously by the NEGF approach.\cite{OzpineciA,MingoN2003prb,YamamotoT,WangJSnegf} For diffusive transport, several approaches are useful, such as the NEGF method,\cite{MingoN2006prb,WangJSnegf} the mode coupling theory,\cite{WangJS2004pre} the Boltzmann transport equation method,\cite{SpohnH,LindsayL} the classical molecular dynamic (MD) simulation,\cite{LepriS} and the quantum non-equilibrium MD simulation.\cite{WangJS2007prl,WangJS2009prb,CeriottiM,CeriottiM2,DammakH} The thermal conductivity value can be obtained directly from some open source simulation packages, such as LAMMPS.\cite{lammps} In this section, we illustrate the simulation details for thermal transport using direct MD simulation. The thermal conductivity is determined by the Fourier law. 

\subsubsection{Fourier's Law}
Fourier's law is a linear empirical law based on observation. It states that the heat flows from high-temperature region to low-temperature region, and that the thermal current density is proportional to the temperature gradient. We consider the in-plane heat transport along $x$ direction in Fig.~\ref{fig_cfg_md}. In this situation, the Fourier law says,
\begin{eqnarray}
J = -\kappa \frac{dT}{dx},
\label{eq_fourier_law}
\end{eqnarray}
where $J$ is the thermal current divided by the cross-sectional area and $\kappa$ is the thermal conductivity. The minus sign on the right-hand side indicates that the heat flows from the high-temperature region to the low-temperature region.

It should be noted that the Fourier law is valid only when the local thermal equilibrium is achieved. This linear law requires the thermal conductivity to be a constant with respect to the structure dimension. However, it has been found that the Fourier law is violated in nanomaterials. More explicitly, the thermal conductivity calculated from the Fourier law is size-dependent.\cite{ChangCW2008prl,DubiY2009pre,YangN2010nt,XiongDX2010pre,ChenSD2013pre}

According to the Fourier law, the thermal conductivity can be extracted based on the knowledge of the thermal current density and the temperature gradient. In the following, we show how to calculate these two quantities from direct MD simulations.

\begin{figure}[tb]
  \begin{center}
    \scalebox{0.8}[0.8]{\includegraphics[width=8cm]{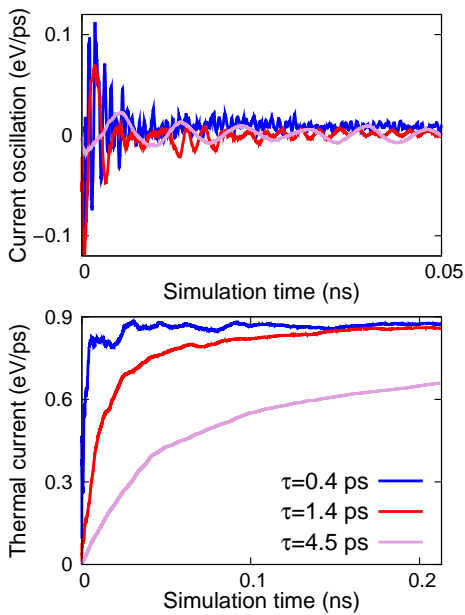}}
  \end{center}
  \caption{(Color online) The thermal current for different relaxation time ($\tau$) Top: thermal fluctuation. Bottom: the averaged thermal current.}
  \label{fig_current}
\end{figure}
\begin{figure}[tb]
  \begin{center}
    \scalebox{0.8}[0.8]{\includegraphics[width=8cm]{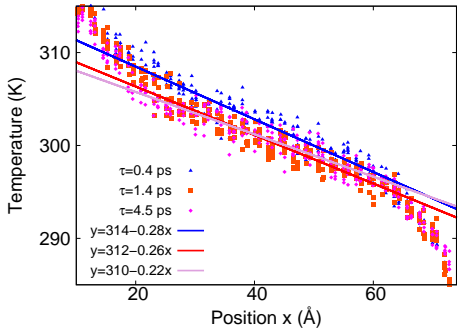}}
  \end{center}
  \caption{(Color online) Temperature profile. Lines are least squares fits.}
  \label{fig_dTdx}
\end{figure}

\subsubsection{Equations of Motion}
\begin{figure}[tb]
  \begin{center}
    \scalebox{1.0}[1.0]{\includegraphics[width=8cm]{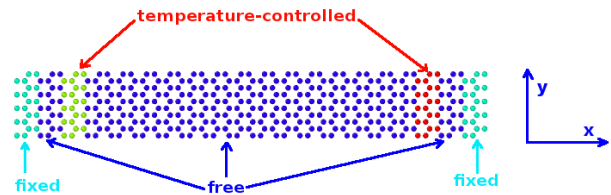}}
  \end{center}
  \caption{(Color online) The graphene is divided into three different regions, including the two fixed ends, the two temperature-controlled regions, and the free region.}
  \label{fig_cfg_md}
\end{figure}

In the classical direct MD simulation, typically, structure is divided into three regions, i.e., the high temperature-controlled region, low temperature-controlled region, and the free central region. The graphene nanoribbon shown in Fig.~\ref{fig_cfg_md} has armchair edges. The dimension is $185\times 12.3$~{\AA}. The thickness of the graphene is taken to be 3.35~{\AA}. This is the inter-layer distance in the three-dimensional graphite.\cite{SaitoR} The fixed boundary is applied in the x-direction, i.e., both left and right ends are fixed during the simulation. Periodic boundary condition is applied in the y-direction. Free boundary is applied in the out-of-plane direction.

In the central region, the degrees of freedom for atom $i$ ($\vec{r}_{i}$, $\vec{v}_{i}$) are controlled by the following equation of motion,
\begin{eqnarray}
\frac{d\vec{r}_{i}}{dt} & = & \vec{v}_{i},\\
\frac{d\vec{v}_{i}}{dt} & = & -\frac{1}{m_{i}}\frac{\partial V}{\partial \vec{r}_{i}}.
\label{eq_dynamics}
\end{eqnarray}
The Brenner potential is used to describe the interatomic interaction in this calculation.\cite{brennerJPCM2002} The time evolution for each atom can be obtained by solving Eq.~(\ref{eq_dynamics}) numerically.

In the temperature-controlled regions, the motion of the atom is influenced by the heat bath besides the inter-atomic force. The heat bath helps to keep a constant temperature for these regions. The heat bath is described by the thermostat degrees of freedom. There are various thermostat algorithms for temperature controlling, such as N\'ose-Hoover heat bath,\cite{Nose,Hoover} the classical or quantum Langevin heat bath,\cite{WangJS2007prl,WangJS2009prb,CeriottiM,CeriottiM2,DammakH} amongst others. We take the N\'ose-Hoover heat bath as an example in the following demonstration of the simulation for the thermal transport. The movements of these atoms in the temperature-controlled regions are governed by the following coupled dynamic equations,
\begin{eqnarray}
\frac{d\vec{r}_{i}}{dt} & = & \vec{v}_{i},\\
\frac{d\vec{v}_{i}}{dt} & = & -\frac{1}{m_{i}}\frac{\partial V}{\partial \vec{r}_{i}}-\eta_{i} \vec{v}_{i},\\
\frac{d\eta_{i}}{dt} & = & \left(\sum_{j}m_{j}v_{j}^{2}-gk_{B}T\right)/Q.\\
Q & = & gk_{B}T\tau^{2},
\label{eq_nh}
\end{eqnarray}
$g$ is the total degrees of freedom in the temperature-controlled region. $\tau$ represents the interaction strength between the heat bath and the system, so it is a kind of thermal relaxation time for the N\'ose-Hoover heat bath. This thermal relaxation time has some direct effect on the thermal current across the system as shown in Fig.~\ref{fig_current}. The top panel is for the thermal fluctuation and the bottom panel is the average thermal current. For $\tau=4.5$~{ps}, the response from the heat bath is very slow, so it corresponds to a weak coupling between the heat bath and the system. In this case, a longer thermalization time is needed to realize a stable temperature distribution across the system, although the influence from the heat bath is smaller. For $\tau=0.4$~{ps}, the heat bath can respond very fast, so the heat bath couples with the system strongly. As a result, shorter thermalization time is required, but the heat bath will induce more influence to the system. Similar thermal current are obtained for $\tau=0.4$~{ps} and 1.4~{ps}. Simulations with different relaxation time result in almost the same thermal conductivity.

\subsubsection{Thermal Current and Temperature Gradient}

The fundamental effect of the heat baths is to inject thermal energy into the system through the high-temperature region, and pump out the same amount of thermal energy in the low-temperature region. Due to energy conservation, the thermal current across the system should equal to the energy exchange between the heat bath and system in the temperature-controlled regions, as long as there is no energy accumulation in the system.\cite{PoetzschRHH,JiangJW2010isotopic}

From Eq.~(\ref{eq_nh}), it can be found that,
\begin{eqnarray}
m_{i}v_{i}\frac{dv_{i}}{dt} & = & -v_{i}\frac{\partial V}{\partial r_{i}}-\eta_{i} m_{i} v_{i}^{2},
\end{eqnarray}
so we have the following equation,
\begin{eqnarray}
\frac{d}{dt}\left(\sum_{i}\frac{1}{2}m_{i}v_{i}^{2} \right) + \frac{dV}{dt} & = & -\sum_{i}\eta_{i} m_{i} v_{i}^{2}.
\end{eqnarray}
The left-hand side is the rate of change in the total energy. As a result, we get the energy flowing from the heat bath to the system,
\begin{eqnarray}
E_{ex} & = & -\int_{t_{0}}^{t_{0}+\Delta t} \sum_{i}\eta_{i} m_{i} v_{i}^{2}dt.
\label{eq_ex_energy}
\end{eqnarray}
$\Delta t$ is the total simulation time. The summation index $i$ runs over all atoms in the temperature-controlled region. We can thus compute the thermal current in a more symmetric manner,
\begin{eqnarray}
J=\frac{1}{s}\frac{E_{ex}^{high}-E_{ex}^{low}}{2\Delta t},
\label{eq_current}
\end{eqnarray}
with $s$ as the cross-sectional area. $E_{ex}^{high}$ and $E_{ex}^{low}$ are energy from heat bath on the left and right sides. The thermal current is shown in Fig.~\ref{fig_current}. The thermal fluctuation is $dJ=\frac{E_{ex}^{high}-E_{ex}^{low}}{2\Delta t}$. The cross-sectional area is not included for these data shown in the figure. For $\tau=$ 0.4, 1.4, and 4.5~{ps}, the thermal currents are 0.87, 0.86, and 0.66~{eV/ps}, respectively.

Eqs.~(\ref{eq_dynamics}) and (\ref{eq_nh}) are solved iteratively, by discretizing the time with a small time step, where the time step in MD simulations for graphene is usually on the order of femtoseconds. The highest-frequency phonon mode in graphene is the in-plane optical mode, with frequency around 300~{THz}. For a time step of 1.0~{fs}, there are about 20 simulation steps within one oscillation cycle of the optical mode. The trajectory of each atom and the thermostat parameter from the iterative solution are used to compute the thermal current.

\begin{figure*}
  \begin{center}
    \scalebox{0.8}[0.8]{\includegraphics[width=\textwidth]{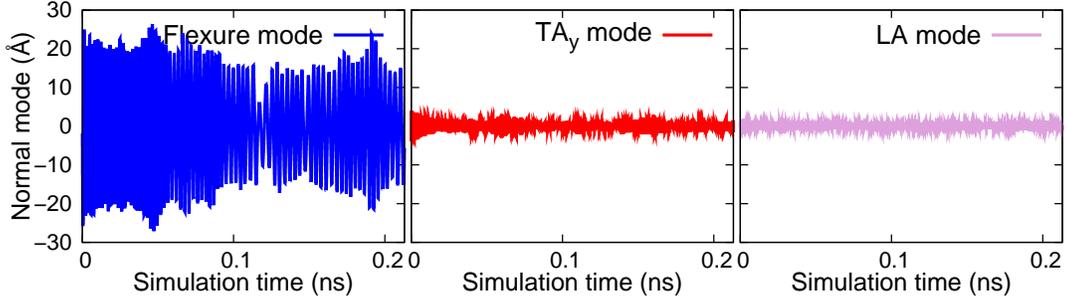}}
  \end{center}
  \caption{(Color online) Normal mode coordinate extracted from MD simulation at room temperature for the first flexural mode (left), the first in-plane transverse acoustic mode (center), and the first longitudinal acoustic mode (right).}
  \label{fig_normal_mode}
\end{figure*}

The average kinetic energy for each atom gives the temperature for the atom according to the equipartition theorem $\frac{1}{2}(v_{x}^{2} + v_{y}^{2} + v_{z}^{2})=\frac{3}{2}k_{B}T$, where $k_{B}$ is the Boltzmann constant. Hence, the temperature profile $T(\vec{r}_{i})$ is obtained simultaneously from the MD simulation. Fig.~\ref{fig_dTdx} shows a typical temperature profile in graphene. The temperature profile within the central region $x\in[0.25L, 0.75L]$ is linearly fitted to give the temperature gradient within the system. The resulted temperature gradients are -0.28, -0.26, and -0.22~{K/\AA}, for the three different relaxation times $\tau=$ 0.4, 1.4, and 4.5~{ps}. The thermal conductivity can then be extracted through the Fourier law in Eq.~(\ref{eq_fourier_law}). The obtained thermal conductivity values are 120.9, 128.7, and 116.7~{W/m$\cdot$K}.

\subsection{Contribution from Flexural Mode}

\subsubsection{Long Lifetime for Flexural Mode}
It was shown in several recent works that the Fourier law is not valid in nanoscale structures,\cite{CahillDG,DharA2008,LiuS,XuX2014nc}  where the thermal conductivity becomes size-dependent and increases with increasing length. For quasi-one-dimensional nanostructures, the thermal conductivity can be written as a power function of the length, i.e., $\kappa \propto L^{\beta}$, where the exponent $\beta=0$ for purely diffusive transport and $\beta=1$ for purely ballistic transport. The exponent deviates from 0, and becomes size-dependent in nanomaterials; i.e., the Fourier law is violated.

The contribution from each phonon mode to the thermal conductivity can be collected as follows,\cite{HoneJ,GuY,JiangJW2011bntube}
\begin{eqnarray}
\kappa_{\rm ph} = \frac{1}{V} \sum_{\vec{k}} \tau_{\vec{k}}^{\sigma}C_{\rm ph}(\omega)v_{\vec{k}}^{2}.
\label{eq_conductivity}
\end{eqnarray}
$V$ is the volume of the system. $C_{\rm ph}=k_{B}x^{2}e^{x}/(e^{x}-1)^{2}$ is the heat capacity. $x=\hbar\omega/(k_{B}T)$. $v_{\vec{k}}$ is phonon group velocity in the thermal flow direction. The lifetime for each phonon in this formula can be obtained using the single mode relaxation time approximation. Following formula gives the the lifetime for phonon mode $\vec{k}$ due to phonon-phonon scattering,
\begin{eqnarray}
\frac{1}{\tau_{\rm ps}} & = & \left(\frac{4}{3\rho_{L}}\right)\left(\frac{\hbar\omega\gamma^{2}}{v_{z}^{2}}\right)\sum_{n'\sigma'}^{\prime}\frac{1}{v_{g}}\omega'\omega''N\left(\omega',\omega''\right).
\label{eq_pslifetime}
\end{eqnarray}
In this formula, $\rho_{L}$ is the mass per length. $v_{z}$ is the phonon velocity along the thermal current direction. $v_{g}=|v'-v''|$ is the group velocity. The energy and momentum conservation is implicated by the prime over the summation.

The phonon-phonon scattering is weak at low temperature, so the boundary scattering becomes more important, especially for systems with small size. Hence, it is also important to consider the boundary scattering process,\cite{ZimanJM}
\begin{eqnarray}
\frac{1}{\tau_{\rm bs}} & = & \frac{v_{\vec{k}}^{\sigma}}{L}\times\frac{1-p}{1+p},
\label{eq_bslifetime}
\end{eqnarray}
where $p$ is the spectacular parameter. The overall phonon lifetime can be obtained as,
\begin{eqnarray}
\frac{1}{\tau_{\rm tot}} & = & \frac{1}{\tau_{\rm ps}}+\frac{1}{\tau_{\rm bs}}.
\label{eq_totlifetime}
\end{eqnarray}

The above formula gives the phonon lifetime and thermal conductivity due to boundary scattering and phonon-phonon scattering. It was found that, in the frequency range [50, 80]~{cm$^{-1}$}, the lifetime for the flexural mode is dominated by the boundary scattering, as the phonon-phonon scattering is weak.\cite{JiangJW2011bntube} These phonons with long lifetime transport across the system almost ballistically, while the other phonons with short lifetime behavior diffusively. As a result, the overall behavior for the thermal conductivity is sandwiched between the ballistic and diffusive transport regimes, i.e., the power factor $\beta$ sits in $[0, 1]$.

\jwj{The thermal conductivity in graphene was found to increase with increasing size even in the $\mu$m range.\cite{NikaDL2009prb,NikaDL2009apl} When the width of the graphene is enlarged by a factor of 3, the thermal conductivity increases by about a factor of 1.8. There is still no conventionally accepted argument for such violation of the Fourier law in graphene.}

\jwj{It is clear that the three acoustic phonon branches make the largest contribution to the 
the thermal conductivity for graphene. However, there is still no universally accepted fact on the relative contribution from the flexural mode to the thermal conductivity in graphene. Typically, the contribution from the flexural mode depends on the temperature or defect density or the substrate for the graphene sample. For perfect graphene, using Boltzmann transport theory, Lindsay et al. found that the flexural mode dominates the thermal conductivity of the graphene.\cite{LindsayL} The flexural mode contributes about 70\% of the thermal conductivity in graphene, while each of the other two acoustic phonon modes contribute 10\%. The large contribution from the flexural mode is attributed to the large density of states of the flexural mode, and the strict symmetry selection rule imposed on the flexural mode, which leads to very long lifetime of the flexural mode. We note that the symmetry selection rules are demonstrated in the first Brillouin zone, which was combined with the phonon-phonon scattering formula Eq.~(\ref{eq_pslifetime}) to give the extremely long lifetime for the flexural mode in graphene.\cite{NikaDL2009prb}}

\jwj{In other works, the flexural mode has been found to make a smaller contribution than the other two acoustic (LA and in-plane TA) modes, especially at higher temperature.\cite{NikaDL2009prb,AksamijaZ,NikaDL2012jpcm,ChenL2012jap} Aksamijaa and Knezevicb found that the flexural mode has about a 50\% contribution at temperatures bellow 130~K for graphene with rough edges.\cite{AksamijaZ} However, the flexural mode contribution decreases quickly with increasing temperature, and the contribution from the flexural mode becomes less than 20\% at 400~K. Chen and Kumar found that the LA mode dominates the thermal conductivity of both isolated and supported graphene on Cu substrate.\cite{ChenL2012jap}}

\jwj{Furthermore, in practice, the symmetry selection rule will be relaxed in experiments, where the high symmetry of the perfect graphene is broken. For instance, the symmetry will be lowered in suspended graphene samples, which is inevitably bent during measurement. In this situation, the lifetime of the flexural mode should be considerably reduced. For deformed graphene samples, it is difficult to use the Boltzmann transport theory to compute the thermal conductivity, because the symmetry selection rule no longer valid.  Therefore, classical MD simulations can be used to study the thermal transport in the deformed graphene or graphene with defects.}

\jwj{Several theoretical works have shown that the temperature dependence will scale as $~T^{1.5}$ at low temperature for the thermal conductance contributed by the flexural mode in the ballistic regime.\cite{MingoN2005prl,SaitoK,JiangJW2009direction} In a recent experiment, Xu et al. measured the thermal conductivity of the graphene in a quasi-ballistic regime and found that the temperature dependence of the thermal conductivity scales as\cite{XuXF2010} $~T^{1.5}$, which is the same as the theoretical predictions. This consistency gives one piece of evidence that the flexural mode makes an important contribution to the thermal transport in graphene in the quasi-ballistic regime. However, it should be noted that, in practice, the sample quality plays an important role on the temperature-dependence of the thermal conductivity. In particular, the temperature-dependence for the thermal conductivity in graphene is sensitive to the defect densities or the grain size of the graphene sample.\cite{AdamyanV,SerovAY}}

\subsubsection{Projection Operator for Flexural Mode}
We provide a normal mode projection operator for analyzing the contribution from the flexural mode to the thermal conductivity in MD simulations. From the lattice dynamic properties of the flexural mode discussed in Sec.II, we can determine the eigenvector for the flexural mode. The position of atom $i$ is determined by the vector $\vec{r}_{i}$. The eigenvectors can be used to define the normal mode projection operator, $P^{k}$,
\begin{eqnarray}
P^{k} = (\vec{\xi}_{1}, \vec{\xi}_{2}, \vec{\xi}_{3}, ..., \vec{\xi}_{N})
\label{eq_projector}
\end{eqnarray}
where $N$ is the total number of atoms. From MD simulations, we have the time history of the vibration displacement of each atom,
\begin{eqnarray}
\vec{u}_{i}(t) = \vec{r}_{i}(t)-\vec{r}_{i}^{0}.
\label{eq_u}
\end{eqnarray}
where $\vec{r}_{i}^{0}$ is the initial position for atom $i$. Applying the normal mode projection operator, $P^{k}$, onto the vibration displacement vector will give us a scalar normal mode coordinate, $Q^{k}(t)$,
\begin{eqnarray}
Q^{k}(t) = \sum_{i=1}^{N} \vec{\xi}_{i}^{*}\cdot \vec{u}_{i}(t).
\label{eq_Q}
\end{eqnarray}
The normal mode projection technique can selectively disclose the contribution to the thermal conductivity from each phonon mode. Fig.~\ref{fig_normal_mode} compares the normal mode coordinate of the first flexural mode, in-plane transverse acoustic mode, and longitudinal acoustic mode from the MD simulation at 300~K. It is clear that the flexural mode has much larger normal mode amplitude than the other two modes, indicating that the flexural mode is the most important vibration morphology in the graphene during the MD simulation. It shows explicitly that the major contribution is from the flexural mode to the thermal conductivity.

\subsubsection{Tuning the Thermal Conductivity Via Flexural Mode}
\begin{figure}[tb]
  \begin{center}
    \scalebox{0.6}[0.6]{\includegraphics[width=8cm]{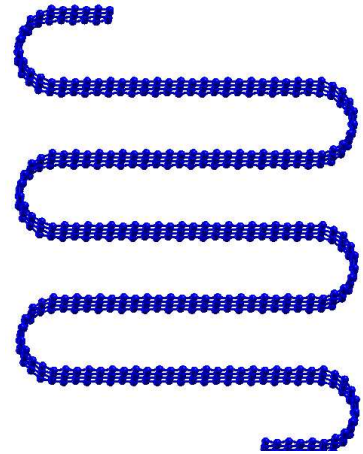}}
  \end{center}
  \caption{(Color online) Graphene is folded, introducing the inter-layer coupling for the flexural mode.}
  \label{fig_fold}
\end{figure}

The graphene is very flexurable in the out-of-plane direction, so it is easier to modify the flexural mode. Hence, it will be an efficient way to manipulate the thermal conductivity through the flexural motion.

The first method is to introduce inter-layer coupling for the flexural mode; e.g., coupling different graphene monolayers to form a few-layer graphene. Indeed, it was observed in experiment that the thermal conductivity is considerably reduced by increasing layer numbers in few-layer graphene.\cite{GhoshS} The thermal conductivity in bilayer graphene is about 30\% lower than that of the single-layer graphene. The reduction of the thermal conductivity is attributed to the enhanced phonon-phonon scattering for the flexural mode in thicker few-layer graphene, where more scattering channels are available.

The second direct method to manipulate the flexural mode is to fold the graphene as shown in Fig.~\ref{fig_fold}. A fold will introduce inter-layer coupling to the flexural mode.\cite{OuyangT2011apl,YangNfold} This coupling can be further increased by compression the inter-layer space in the folds. For a flat graphene, flexural mode is difficult to scatter with phonon modes from the high-frequency optical branches, because the band gap between the flexural mode and the optical branches is very large due to the parabolic spectrum of the flexural mode. However, the phonon-phonon scattering channels are considerably increased when the inter-layer space is compressed in the folds. The compression of the inter-layer space leads to considerable shifting up of the flexural phonon spectrum, thus narrowing the band gap between the flexural mode and the optical branches. As a result, the phonon-phonon scattering channels increase significantly. This results in a reduced thermal conductivity in folded graphene.

The third method to manipulate the flexural mode is to put the graphene on a substrate.\cite{CaiWW2010nl,ChenSS2011acsn,LeeJU,GuoZX2011,OngZY2011,HuangSY,QiuB,GuoZX} The thermal conductivity can be reduced by an order of magnitude due to damping of the flexural mode by the substrate.\cite{OngZY2011} Furthermore, the increase in the coupling strength between graphene and substrate will enhance the thermal conductivity. It is because of the coupling of the flexural mode to the substrate Rayleigh waves, which results in a hybridized mode with linear spectrum and higher group velocity than the original flexural mode in graphene. Quite recently, Amorim and Guinea examined the flexural mode of graphene on different substrates, with the consideration of the dynamics of the substrate.\cite{AmorimB2013prb}

\section{Thermal Expansion}

\subsection{Introduction}
Negative coefficient of thermal expansion (CTE) occurs in many materials, typically at low temperature. For example, bulk Si and Ge are two well-known semiconductors with negative CTE at temperatures below 100~K, which is attributed to the inter-play between the bond-stretching and bond-bending forces.\cite{XuCH1991prb,WeiS} Ref.~\onlinecite{BarreraGD} summarizes the negative CTE in different materials and various possible mechanisms underlying this phenomenon.

The CTE for graphene was also found to be negative. On the experimental side, in 2009, the CTE is found to be about $-7\times 10^{-6}$~{K$^{-1}$} at room temperature as measured by Bao et al.\cite{BaoW2009nn} The room temperature CTE is about $-8\times 10^{-6}$~{K$^{-1}$} in the experiment by Yoon et al..\cite{YoonD}  The measured CTE is negative in a wide temperature range. Singh et al. found that graphene has a negative CTE for temperatures bellow 300~K,\cite{SinghV} while Yoon et al. obtained a negative CTE for graphene for temperatures bellow 400~K.\cite{YoonD}

\jwj{On the theoretical side, there are several standard approaches to compute the CTE. The lattice constant is temperature dependent, which can be used to extract the CTE value.\cite{MounetN,ZakharchenkoKV,PozzoM} These calculations show a minimum lattice constant at a transition temperature. Below this transition temperature, the CTE decreases with increasing temperature. Above this transition temperature, the CTE increases with increasing temperature. This implies that CTE is negative below the transition temperature and becomes positive above the transition temperature. In our recent work, the NEGF approach was implemented to compute the CTE of graphene.\cite{JiangJW2009expansion} As an advantage of the NEGF approach, this method is able to account for the quantum zero-point vibration effect in a quite natural manner. Another advantage of the NEGF approach is that it is straightforward to decompose the contribution of each phonon mode to the total CTE value. We will mainly discuss this NEGF approach in the following.}

\subsection{Green's Function Approach for Thermal Expansion}
The thermal expansion is usually studied by MD simulations or the standard Gr\"uneisen method.\cite{GruneisenE1926} In our recent work, the NEGF approach is used to examine the thermal expansion phenomenon in the single-walled carbon nanotube and graphene, where the calculated CTE agrees quite well with experiments.\cite{JiangJW2009expansion} In this section, we review some key steps in this NEGF approach for the thermal expansion.

\subsubsection{Green's Function Approach}

The potential energy associated with the phonon modes are assumed to be,
\begin{eqnarray}
V &=& \sum_{ij}\frac{K_{ij}}{2}u_{i}u_{j} + H_{n}.
\label{eq_hamiltonian}
\end{eqnarray}
It includes linear and nonlinear interactions. Both third and fourth orders nonlinear interactions are considered,
\begin{eqnarray}
H_{n} = \sum_{lmn}\frac{k_{lmn}}{3}u_{l}u_{m}u_{n}+\sum_{opqr}\frac{k_{opqr}}{4}u_{o}u_{p}u_{q}u_{r}.
\label{eq_nonlinear}
\end{eqnarray}
$K_{lmn}$ and $K_{opqr}$ are force constant matrix elements. We have extracted the nonlinear coefficients $k_{lmn}$ and $k_{opqr}$ from the Brenner potential.\cite{brennerJPCM2002}

Details for the Green's function approach in thermal properties can be found in Refs.~\onlinecite{WangJSnegf,WangJS2013fp}. We utilize the following two GFs for the study of the thermal expansion,
\begin{eqnarray}
G_{j}(\tau) & = & -\frac{i}{\hbar}\langle T_{\tau}u_{j}^{H}(\tau)\rangle,
\label{eq_gone}\\
G_{jk}(\tau,\tau') & = & -\frac{i}{\hbar}\langle T_{\tau}u_{j}^{H}(\tau)u_{k}^{H}(\tau')\rangle.
\end{eqnarray}
$u_{j}^{H}(\tau)$ is the vibrational displacement. For convenience, $u_{j}^{H}(\tau)$ also includes the square root of the atom's mass. It is very nice that $G_{j}$ can be obtained analytically,
\begin{eqnarray}
G_{j}  = \sum_{lmn}k_{lmn}G_{lm}^{>}(0)\tilde{G}_{nj}^{r}[0].
\label{eq_gj}
\end{eqnarray}
$G^{>}(0)$ is the greater GF in time domain. $\tilde{G}^{r}[0]$ is the retarded GF in frequency domain. These two GFs can be computed without any integration. Using Eq.~(\ref{eq_gone}), we get the average vibrational displacement for atom $j$, i.e., $\langle u_{j} \rangle$. CTE can be computed from the derivative of the displacement with respect to the temperature.

It is more convenience to work in the normal mode space for the derivative of the one-point GF. After the Fourier transformation, we get,
\begin{eqnarray}
\frac{dG_{j}}{dT} & = & (-i)\sum_{lmn}k_{lmn}\left(S\left(\begin{array}{ccc}
\ddots\nonumber\\
 & \frac{1}{\omega_{\mu}}(\frac{df}{dT})\nonumber\\
 &  & \ddots\end{array}\right)S^{\dagger}\right)_{lm}\nonumber\\
&&\times\left(-K^{-1}\right)_{nj},
\label{eq_gj_mode_space}
\end{eqnarray}
where $f$ is the Bose distribution function. $K$ is the force constant matrix. $S$ stores the eigen vector of $K$, i.e., $S^{\dagger}KS$ is diagonal with diagonal elements $\omega^{2}_{\mu}$.

A particular boundary condition is used for the NEGF treatment of the thermal expansion phenomenon. Specifically, the left end is fixed, while the right end is free. Periodic boundary condition is applied in the in-plane lateral direction. It was shown that graphene has a negative CTE for temperatures bellow 600~K.\cite{JiangJW2009expansion} The minimum CTE value is achieved around 200~K. These results are in good agreement with the experimental findings. The NEGF approach is able to examine the substrate coupling effect, which has been found to be important for the  CTE value of graphene.\cite{PozzoM}

\subsubsection{Green's Function Approach and Gr\"uneisen Method}
We will show that the NEGF approach is equivalent to the traditional Gr\"uneisen method in the weak nonlinear limit, if the system is isotropically and uniformly deformed in the thermal expansion phenomenon.\cite{WangJS2014thermalexpansion}

From above, the CTE from NEGF for atom $N$ is,
\begin{eqnarray}
\alpha_{N} & = & \frac{1}{L}\sum_{lmn\mu}k_{lmn}S_{l\mu}S^{*}_{m\mu}\frac{c_{\mu}}{\omega_{\mu}^{2}}\left(-K_{nN}^{-1}\right),
\end{eqnarray}
where $c_{\mu}=\hbar\omega_{\mu}df/dT=C_{rm ph}(\omega_{\mu})$ is the heat capacity for phonon mode $\mu$ as introduced in Eq.~(\ref{eq_conductivity}). The Gr\"uneisen parameter for mode $\mu$ is,
\begin{eqnarray}
\gamma_{\mu} & = & -\frac{\partial \ln\omega_{\mu}}{\partial \ln V}\nonumber\\
\nonumber\\
 & = & -\frac{L}{3\omega_{\mu}^{2}}\sum_{lmn}S_{l\mu}S^{*}_{m\mu}k_{lmn}\epsilon_{n},
\end{eqnarray}
where $\epsilon_{n}=\delta R_{n}/L$ is the change of position for atom $n$ with respective to total length $L$. The CTE from Gr\"uneisen method is then,
\begin{eqnarray}
\alpha & = & \frac{1}{3B}\left(\frac{\partial P}{\partial T}\right)_{V}\nonumber\\
\nonumber\\
 & = & \frac{1}{3BV}\sum_{\mu}\gamma_{\mu}c_{\mu}\nonumber\\
\nonumber\\
 & = & -\frac{1}{B}\sum_{lmn\mu}\epsilon_{n}k_{lmn}S_{l\mu}S^{*}_{m\mu}\frac{c_{\mu}}{\omega_{\mu}^{2}},
\end{eqnarray}
where $P$ is pressure, $B$ is bulk modulus, $V$ is volume, and an isotropic and uniform deformation has been assumed for the system during the thermal expansion phenomenon.

To show the equivalence between these two methods, let's assume the system undergoes a uniform deformation by external force $F$. In this situation, the displacement of atom $n$ can be obtained from the force constant matrix, or equivalently from the bulk modulus,
\begin{eqnarray}
u_{n} & = & K_{nN}^{-1}F=\epsilon_{n}\frac{FL}{B},
\end{eqnarray}
so,
\begin{eqnarray}
\frac{1}{B}\epsilon_{n} & = & \frac{K_{nN}^{-1}}{L}.
\end{eqnarray}
As a result, we find that, 
\begin{eqnarray*}
\alpha & = & \alpha_{N}.
\end{eqnarray*}

From the comparison, we show that the NEGF approach has three advanced properties over the Gr\"uneisen method. First, the NEGF approach can be applied to study structures which have a nonuniform deformation, while the Gr\"uneisen method can only be applied for structures with uniform deformation in the thermal expansion phenomenon. Second, the NEGF approach is applicable for nanostructures without periodicity, while the Gr\"uneisen method requires periodicity with a well-defined bulk modulus. Third, the NEGF approach is suitable for systems with anisotropic CTE, while Gr\"uneisen method only works for systems with isotropic CTE.

In the meantime, we point out two disadvantages for the NEGF approach as compared with the Gr\"uneisen method. First, the calculation of $K^{-1}$ is computationally expensive in the NEGF approach, especially for large systems. Hence, in terms of computation cost, the Gr\"uneisen method is better than the NEGF approach for materials with isotropic CTE, where both methods are applicable. Second, the NEGF approach is based on a perturbation theorem and high-order nonlinear terms have been omitted; while Gr\"uneisen method includes an overall effect from all high-order nonlinear interaction terms.

\subsection{Contribution from Flexural Mode}

\begin{figure}[tb]
  \begin{center}
    \scalebox{1.0}[1.0]{\includegraphics[width=8cm]{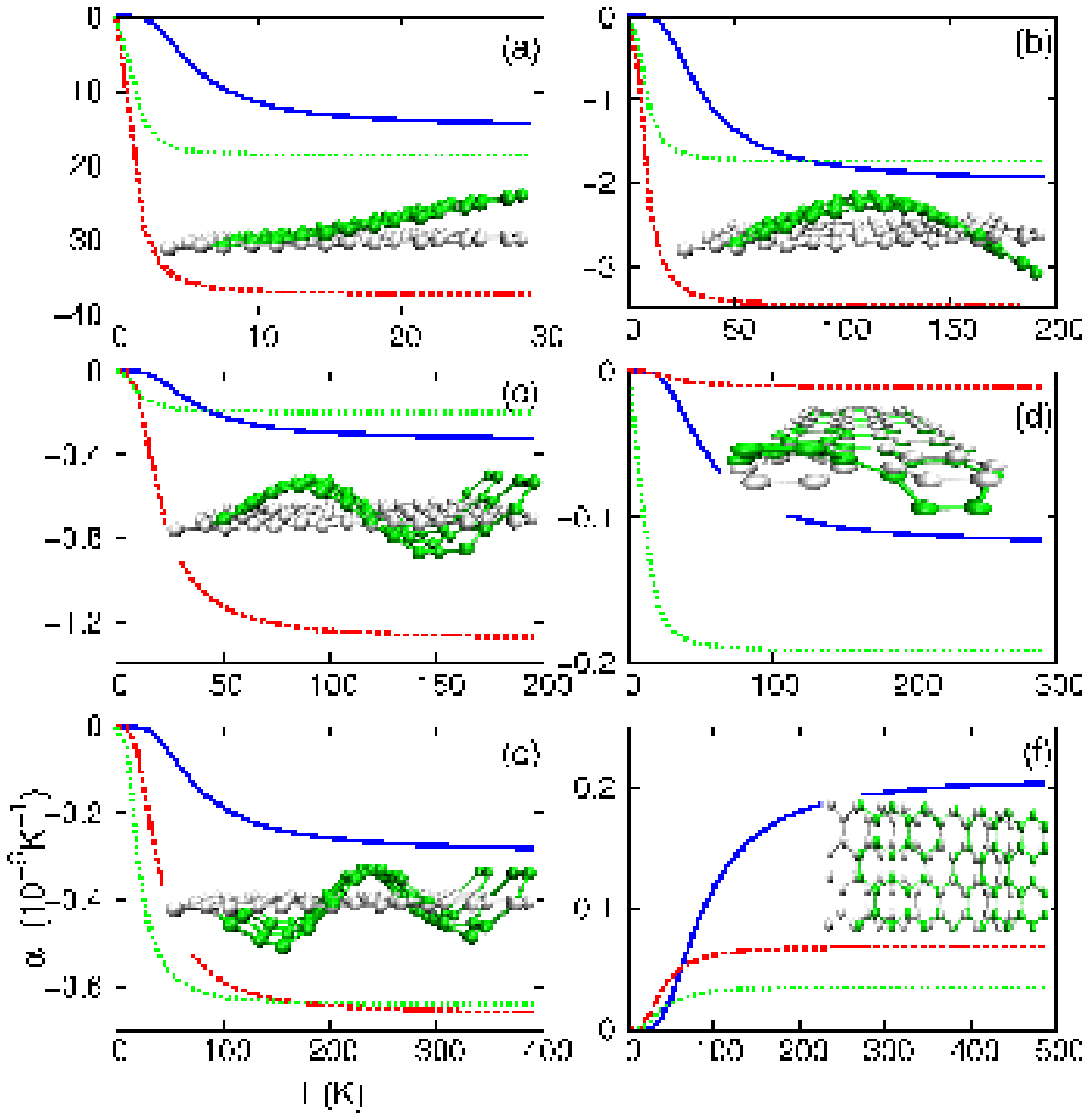}}
  \end{center}
  \caption{Contribution to CTE from six lowest-frequency phonon modes in graphene sheet without substrate interaction. Blue solid line is for graphene with (length, width)=(10, 8.5)~{\AA}, green dashed line is for (20, 17)~{\AA} and red dotted line is for (20, 8.5)~{\AA}. Insets are the vibrational morphology for the corresponding mode. (a), (b), (c) and (e) are the first four bending modes. (d) is a tearing mode. (f) is the longitudinal phonon mode.}
  \label{fig_from_each_mode}
\end{figure}
\begin{figure}[tb]
  \begin{center}
    \scalebox{1.0}[1.0]{\includegraphics[width=8cm]{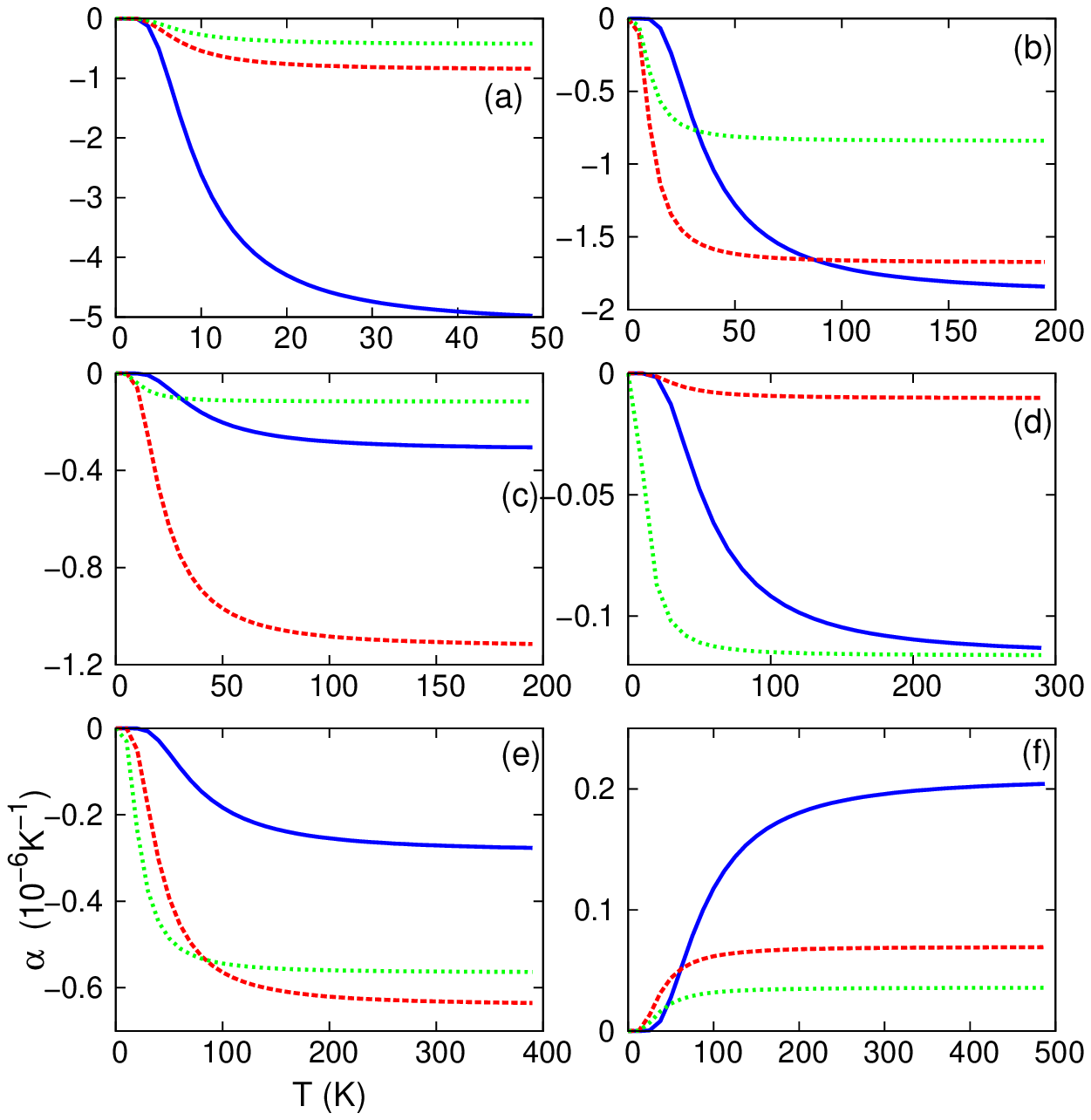}}
  \end{center}
  \caption{Contribution to CTE from six lowest-frequency phonon modes in graphene sheet with substrate interaction $\gamma=0.001$.}
  \label{fig_from_each_mode_gamma_0.001}
\end{figure}

We will now demonstrate that the large negative CTE in graphene is due to its flexural modes. Eq.~(\ref{eq_gj_mode_space}) includes the contribution from all phonon modes to the CTE through the diagonal matrix,
\begin{eqnarray}
\left(
\begin{array}{ccc}
\ddots\nonumber\\
 & \frac{1}{\omega_{\mu}}(\frac{df}{dT})\nonumber\\
 &  & \ddots
\end{array}
\right).
\end{eqnarray}
This matrix is diagonal, which indicates that each phonon mode makes separate contributes for the negative CTE. If this matrix has a single nonzero element, $\frac{1}{\omega_{\mu}}(\frac{df}{dT})$, then Eq.~(\ref{eq_gj_mode_space}) yields the sole contribution from the mode $\mu$. Hence, we can distinguish the independent contribution from each phonon mode to the CTE.

Fig.~\ref{fig_from_each_mode} shows the contribution from the six lowest-frequency phonon modes. There is no substrate interaction in this figure, i.e., $\gamma=0$. The inset in each panel is the corresponding vibration morphology of the phonon mode. (a), (b), (c) and (e) are the first four bending modes and (d) is an interesting tearing mode. Among all of these six phonon modes, the first bending mode shown in (a) has $90\%$ contribution to CTE. Due to its bending morphology, this mode induces a contraction effect in the graphene sheet, and thus the negative CTE.

If the substrate interaction is nonzero as shown in Fig.~\ref{fig_from_each_mode_gamma_0.001}, the CTE is clearly enhanced and the size effect becomes weaker. The contribution from the second bending mode is also very important. The substrate interaction is more important for larger piece of graphene, because the bending movement in larger graphene is more serious than smaller piece of graphene when it is fixed on one edge. In case of strong substrate interaction, graphene is so difficult to be bent that all bending modes do not contribute. In this situation, the CTE is dominated by the sixth mode, which leads to a positive CTE. As a result, the CTE is positive in whole temperature range, and the size effect on the CTE becomes smaller.

\section{Young's Modulus}

\subsection{Introduction}
Graphene has many remarkable mechanical properties. Readers are referred to Refs.~\onlinecite{LiH2009ieee,TerronesM2010nt,YangZ2012nml,LauCN2012mt,TuZC2008jctn} for comprehensive reviews on various mechanical properties of graphene. For instance, the Young's modulus for graphene is on the order of TPa. There are several different approaches for the investigation of the Young's modulus. The atomic force microscope can measure the force-displacement relationship for graphene, from which the Young's modulus can be extracted. This method is used in the experiment in 2008, which found the Young's modulus for graphene to be $1.0\pm 0.1$~TPa. This method has also been adopted in some theoretical works to study the Young's modulus in graphene.\cite{luPRL1997,LierGV2000cpl, KudinKN2001prb,KonstantinovaE2006prb,ReddyCD2006nano, huangPRB2006,LiuF2007prb,KhareR2007prb,WeiXD2009prb,JiangJW2010young,YiLJ2012scpma,ZhouL2013jpcm}

\jwj{The classical theory of elasticity has been widely used to predict the mechanical properties for graphene. Finite element simulations, which are based on a numerical discretization of the equations of elasticity, were used to explain the edge stress induced warping phenomenon in the graphene nanoribbon.\cite{ShenoyVB} The finite element method has also been coupled with atomistic calculations to simulate the crack propagation in graphene in an efficient way.\cite{KhareR2007prb} The elasticity approach has also been combined with atomic potentials to study elastic properties for finite size graphene.\cite{ReddyCD2006nano,DuanWH2009nano}}

\jwj{As an ultra-thin plate, graphene's flexural mode is directly related to its Young's modulus.\cite{LandauLD}  The flexural mode has the lowest frequency, so this mode is the easiest to be excited by thermal vibration. As a result, the thermal vibration of graphene is closely related to its Young's modulus. Using this relationship, it is possible to extract the Young's modulus from the thermal vibrations.  In 1996, this idea was implemented by Treacy et al. to measure the Young's modulus of carbon nanotubes.\cite{TreacyMMJ1996nat,KrishnanA} In our recent work, we applied this method to compute the Young's modulus from graphene's thermal vibration.\cite{JiangJW2009young} In this section, we will review this approach and emphasize the relationship between the flexural mode and the Young's modulus of graphene.}

\subsection{Flexural Mode and Young's Modulus}
In the elasticity theory, the flexural mode is directly related to the Young's modulus of a plate. Here we review some key derivation steps for the Young's modulus of graphene.\cite{JiangJW2009young} The flexure mode for the elastic plate is governed by the following equation,\cite{LandauLD}
\begin{eqnarray}
\rho\frac{\partial^{2}z}{\partial t^{2}}+\frac{D}{h}\Delta^{2}z  =  0,
\label{eq_fm}
\end{eqnarray}
where $\rho$ is the mass density. $D=\frac{1}{12}Yh^{3}/(1-\mu^{2})$ is the bending modulus. $Y$ is the Young's moduls. $\mu$ is the Poisson ratio. $h$ is the thickness of the plate. This partial differential equation can be solved after the following boundary conditions are applied,
\begin{eqnarray}
z(t,x=0,y) & = & 0,\nonumber\\
z(t,x=L,y) & = & 0,\\
z(t,x,y+L) & = & z(t,x,y).\nonumber
\label{eq_boundary}
\end{eqnarray}
According to this boundary condition, we can get following solution,\cite{PolyaninAD}
\begin{eqnarray}
\omega_{n} & = & k_{n}^{2}\sqrt{\frac{Yh^{2}}{12\rho(1-\mu^{2})}},\nonumber\\
z_{n}(t,x,y) & = & u_{n}\sin (k_{1}x)\cdot\cos (k_{2}y)\cdot\cos(\omega_{n}t),\\
\vec{k} & = & k_{1}\vec{e}_{x}+k_{2}\vec{e}_{y}.\nonumber
\label{eq_eigen}
\end{eqnarray}
The two wave vector components are $k_{1}=\pi n_{1}/L$ and $k_{2}=2\pi n_{2}/L$. This flexural mode is characteristic for its parabolic phonon dispersion in the long wave limit. This solution shows explicitly the relationship between the frequency of the flexural mode and the Young's modulus.

At finite temperature, the thermal vibration of the plate is mainly controlled by the lowest-frequency flexural mode. The Young's modulus can be related to the thermal mean square vibration amplitude $\langle\sigma^{2}\rangle$ as follows,
\begin{eqnarray}
Y & = & 0.3\times\frac{S}{h^{3}}\times\frac{k_{B}T}{\langle\sigma^{2}\rangle}.
\label{eq_Y}
\end{eqnarray}
The thickness is chosen as $h=3.35$~{\AA}. The Poisson ratio for graphene is $\mu=0.17$.\cite{BlaksleeOL1970jap, PortalD1999prb} $S$ is the area. Eq.~(\ref{eq_Y}) can be used to extract the value for the Young's modulus of graphene. MD simulations are performed to obtain the thermal vibration quantity $\langle\sigma^{2}\rangle$.

Eq.~(\ref{eq_Y}) is derived based on the elasticity theory shown in Eq.~(\ref{eq_fm}). In principle, the elastic theory works only for very large system. However, it has been shown that the elastic continuum theory is still valid in a very small piece of graphene.\cite{KudinKN2001prb} Such elasticity theory has also been successfully applied to compute the elastic ripple-like deformation at free edges in finite graphene nanoribbons and other mechanics properties.\cite{ReddyCD2006nano,ShenoyVB,ReddyCD,WeiXD2009prb,CadelanoE2009prl}

From the above, we are aware that the flexural mode, especially the first flexural mode, is closely related to the Young's modulus of graphene. More specifically, the frequency of the flexural mode is proportional to the square root of the Young's modulus. The flexural mode describes the out-of-plane bending movement of graphene. However, it is quite interesting that this out-of-plane property is governed by the in-plane mechanical property, Young's modulus. This is due to the special bending vibration of the flexural mode.

\subsection{Manipulation for Young's Modulus}
The Young's modulus for the bulk material is a constant value with respect to the system size. However, for a small piece of graphene, the dimension of the system has important effect on the value of Young's modulus.\cite{JiangJW2009young} The Young's modulus increases with increasing size. Similar size effect was also found by Zhao et al..\cite{ZhaoH2009nl} The experimental value for the Young's modulus is around $1 \pm 0.1$~{TPa}.\cite{LeeC2008sci}

\jwj{In the above approach, the Young's modulus is calculated based on the elasticity equations, so the atomic orientation dependence cannot be predicted by this approach. The Young's modulus in graphene nanoribbons with free edges was found to be orientation dependent. Armchair and zigzag are the two common orientation directions in graphene. Zhao et al. used the molecular mechanics method to compute the Young's modulus for graphene nanoribbon, using the Brenner potential.\cite{ZhaoH2009nl} They found that the Young's modulus is larger in the armchair direction than that in the zigzag direction. Based on the Tersoff potential, Zhao and Xue also found a larger Young's modulus in the armchair direction.\cite{ZhaoS2013jpdap} For graphene nanoribbons with periodic boundary conditions, the Young's modulus is insensitive to the orientation.\cite{ZhouL2013jpcm}}

\jwj{Due to grapheme's exceptional mechanical properties, it is often used to synthesize hybrid structures. These composites usually have good mechanical performance. It was shown that the mechanical properties for the graphene/h-BN heterostructure is superior to pure h-BN.\cite{ZhaoS2013jpdap} More recently, we have found that graphene/MoS$_{2}$ heterostructures also  have enhanced mechanical properties, with a larger Young's modulus than pure MoS$_{2}$.\cite{JiangJW2014gmgyoung}}

\section{Nanomechanical Resonator}

\subsection{Introduction}
The one-atom thick graphene has large Young's modulus.\cite{LeeC2008sci,JiangJW2009young,JiangJW2010young} Lots of experimental and theoretical works have demonstrated the application of graphene in the NMR field. The GNMR has several advantages compared to their micron-sized counterparts, which are typically made of silicon. For mass sensor application, GNMR has a large surface to vomlune ratio for the adsorption of more atoms. Furthermore, GNMR has very low mass density, so it has a very high mass sensitivity.

The GNMR can also serves as a good platform for the study of quantum mechanics problems\cite{ConnellADO}, quantum information storage,\cite{PalomakiTA} electron pumping,\cite{LowT} gas sensing,\cite{Sakhaee-PourA,RumyantsevS,AvdoshenkoSM} or as a test for the classical Fermi-Pasta-Ulam problem.\cite{MidtvedtD2014prl}

The resonant frequency and the quality (Q) factor are two important factors for the description of the GNMR samples. High Q-factor is essential for practical applications of the GNMR. Experiments have achieved considerable success in the preparation of the GNMR samples. Bunch et al. demonstrated the electromechanical resonant oscillation of the graphene sheets in 2007.\cite{BunchJS2007sci} The GNMR samples are prepared in the experiment in a large dimension.\cite{RobinsonJT,ZandeAMVD} Currently, the resonant motion of the GNMR can be detected using various techniques,\cite{Reserbat-PlanteyA,Ruiz-VargasCS,SanchezDG,GiritCO} and the strain within the GNMR can also be measured by the Raman spectroscopy.\cite{Reserbat-PlanteyA} It is found that the Q-factor increases with decreasing temperature.\cite{RobinsonJT,ChenC2009nn,ZandeAMVD} A substantial increases in Q is obtained by increasing the size of the GNMR.\cite{BartonRA}

\jwj{It is important to understand the energy dissipation mechanism for the GNMR, so that the Q-factor can be enhanced. The resonant oscillation of the GNMR is actually the vibration of the flexural mode. Hence, the scattering between the flexural mode and in-plane phonon modes becomes an important intrinsic nonlinear energy dissipation in GNMR.\cite{CroyA,AtalayaJ,AtalayaJ2010epl} The grain boundary and imposed mechanical strain also strongly impact the Q factor of the GNMR.\cite{KimSYnanotechnology,QiZ2012nns,KwonaOK} The inter-layer van der Waals interaction can be useful in the modulation of the Q factor in GNMRs.\cite{HeXQ2005nano,KimSY2009apl} The energy dissipation was found to be dominated by ohmic losses in the GNMR with large electronic current.\cite{SeoanezC} The free edges can lead to extremely strong energy dissipation in the GNMR, due to the instability of imaginary edge modes.\cite{KimSY2009nl,JiangJW2012jap} A temperature scaling law can be induced by the adsorbate diffusion in the Q factor for GNMRs.\cite{JiangJW2013diffusion}}

In this review, we concentrate on the connection between the GNMR and the flexural mode. We focus on the explanation of the actuation of the resonant oscillation with the usage of the bending-like vibration of the flexural mode. For comprehensive reviews on NEMS resonators, readers are referred to Ref.~\onlinecite{EkinciKL,VargheseSH,EomK,ArlettJL,ImbodenM2013pr}.

\subsection{Flexural Mode and Nanomechanical Resonance}
The mechanical oscillation of the GNMR is actuated following the vibration morphology of the first flexural mode in graphene. In following, we discuss the actuation that follows the vibration morphology of the first flexural mode. We note that some studies actuate the resonant oscillation using high order flexural modes.\cite{PauloAS2008jpcs,UnterreithmeierQP2010prl}

\begin{figure*}
  \begin{center}
    \scalebox{1.0}[1.0]{\includegraphics[width=\textwidth]{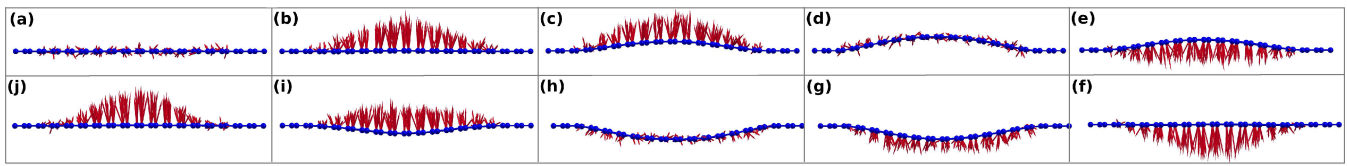}}
  \end{center}
  \caption{(Color online) The actuation of GNMR. Arrows (red online) are velocities. (a) The system is thermalized at a constant temperature 4.2~K within NVT ensemble. (b) The GNMR is actuated by adding a velocity distribution on the system. (d)-(j) The mechanical oscillation of the GNMR within NVE ensemble.}
  \label{fig_gnmr}
\end{figure*}

\subsubsection{Resonant Frequency and Quality Factor}
There are two characteristic quantities for a nanomechanical resonator, i.e., its resonant frequency ($f$) and the Q factor. During the mechanical oscillation, the potential energy and kinetic energy exchange between each other at a frequency of $2f$. From the lattice dynamic analysis in Sec.II, we have computed the frozen frequency for the flexural mode, i.e., the frequency at zero temperature. This is an elastic property for the graphene. The temperature dependence of the flexural mode can be obtained after the consideration of the phonon-phonon scattering. The resonant frequency of a GNMR is the frequency of the flexural mode at finite temperature.

The amplitude of the mechanical oscillation decays gradually. According to this decay, the mechanical oscillation energy transforms into the random thermal vibration energy of the GNMR. The temperature increases in the system as a result of the decay of the resonant oscillation. The Q factor is directly related to the decay rate of this energy oscillation amplitude. There are several equivalent definitions for the Q factor. In MD simulations, the Q factor is usually defined with respect to the ratio of the initial mechanical oscillation energy to the dissipated energy. Its explicit formula is,\cite{JiangH2004prl}
\begin{eqnarray}
Q=2\pi\frac{E_{\rm MR}^{0}}{\Delta E_{\rm MR}}\nonumber,
\end{eqnarray}
where $E_{\rm MR}^{0}$ is the initial oscillation energy. The remaining oscillation energy after $n$ cycles becomes,
\begin{eqnarray}
E_{\rm MR}^{n}=E_{\rm MR}^{0}\left(1-\frac{2\pi}{Q} \right)^{n}.
\label{eq_Q_energy}
\end{eqnarray}
This formula is usually applied to extract the Q factor from MD simulations.

In another alternative computation of Q factor, the resonant mechanical oscillation of the GNMR is the vibration of the flexural mode in graphene. The flexural mode has angular frequency ($\omega=2\pi f$) and lifetime ($\tau$) at finite temperature. The lifetime can be interpreted as the critical time, after which the vibration of the flexural mode decays significantly. The Q factor is essentially the total oscillation cycle number before the decay of the flexural mode,
\begin{eqnarray}
Q=\omega\tau.
\label{eq_Q_lifetime}
\end{eqnarray}
This method was used to calculate the Q factor of graphene torsional resonators.\cite{JiangJWtorsion}

\subsubsection{GNMR Actuation}
MD simulations and continuum elastic modeling are both useful approaches for the investigation of GNMRs. We herein demonstrate the MD simulation procedure for GNMRs. The Q factor is calculated following Eq.~(\ref{eq_Q_energy}). The actuation procedure is illustrated in Fig.~\ref{fig_gnmr}. Both left and right ends are fixed during the whole simulation. Periodic boundary condition is applied in the lateral in-plane direction. There are normally following three steps for the actuation of the GNMR.
\begin{itemize}
\item Firstly, Fig.~\ref{fig_gnmr}~(a) shows the thermalization of the system to a constant pressure and temperature within the NPT (i.e. the particles number $N$, the pressure $P$, and the temperature $T$ of the system are constant) ensemble. The N\'ose-Hoover\cite{Nose,Hoover} heat bath can be used to control both temperature and pressure. 

\item Secondly, in Fig.~\ref{fig_gnmr}~(b), the mechanical oscillation is actuated by adding a velocity distribution to the system. The overall shape of the velocity distribution is actually the same as the morphology of the first flexural mode in graphene.

\item Finally, Fig.~\ref{fig_gnmr}~(d)-(j) display a free oscillation of the system within the NVE (i.e. the particles number $N$, the volume $V$, and the energy $E$ of the system are constant) ensemble.
\end{itemize}
Simulation data from the final NVE stage will be used in the analysis of the mechanical oscillation of the GNMR. In particular, both resonant frequency and Q-factor can be extracted from the time history of the kinetic energy.

\subsection{Energy Dissipation Mechanisms}
A high Q factor is crucial for the practical application of GNMRs. Hence, it is meaningful to understand energy dissipation mechanisms for the resonant oscillation. In this section, we review some energy dissipation mechanisms in the GNMR.

\textit{Phonon-phonon scattering} -- This dissipation mechanism is a nonlinear effect. It is the result of the phonon-phonon scattering phenomenon. In a pure and perfect GNMR without free edge or adsorbates, the phonon-phonon scattering is the only intrinsic energy dissipation mechanism. All phonon modes are in thermal equilibrium state prior to the actuation of the mechanical oscillation. After actuation, the first flexural mode is driven into a highly non-equilibrium state. The energy of the first flexural mode will flow into the other phonon modes with the assistance of the phonon-phonon scattering. It has been shown that the flexural mode is seriously scattered by the in-plane phonon modes.\cite{CroyA,MathenyMH} As a result, the mechanical oscillation energy of the GNMR decays and leads to the temperature increase in the system. As a result of the phonon-phonon scattering, the Q factor in the GNMR is typically inversely proportional to the temperature.\cite{KimSY2009nl,JiangJW2012jap}

\textit{Edge effect} -- Kim and Park pointed out the importance of the free edge on the Q factor of GNMRs.\cite{KimSY2009nl} The free edge is able to reduce the Q factor of the GNMR by two orders. This effect was explained in detailed by Jiang and Wang via the lattice dynamic analysis.\cite{JiangJW2012jap} The imaginary edge modes are found to be responsible for such a large reduction in the Q-factor. In these imaginary edge modes, the edge atoms have large vibration amplitude, while the other inner atoms has very weak vibration amplitude. Owning to its localization property, these imaginary modes will localize the thermal energy. As a result, the edge atoms will oscillate at larger amplitude than the inner atoms. It means that the edge atoms break the resonant oscillation of the whole GNMR. This contradiction leads to a fast decay of the resonant oscillation.

\begin{figure}[tb]
  \begin{center}
    \scalebox{1.0}[1.0]{\includegraphics[width=8cm]{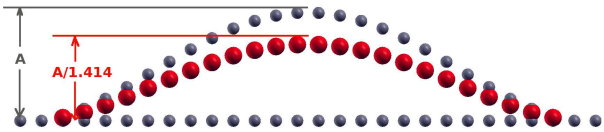}}
  \end{center}
  \caption{(Color online) Geometry of the GNMR. $A$ is the actuation amplitude. The effective amplitude is $A/\sqrt{2}$. The effective strain is determined by the difference between the length of the effective shape and the initial shape. From reference ~\onlinecite{JiangJW2012nanotechnology}.}
  \label{fig_cfg_gnmr}
\end{figure}

\textit{Effective strain} -- In a recent work, we found that the mass sensitivity of the GNMR-based mass sensor can be enhanced by driving the resonant mechanical oscillation with large actuation energy.\cite{JiangJW2012nanotechnology} As illustrated in Fig.~\ref{fig_cfg_gnmr}, the oscillating GNMR shape is equivalent to a stationary shape, which is longer than the initial GNMR. The difference between the effective shape and the initial shape yields the effective strain during the resonant oscillation of the GNMR,
\begin{eqnarray}
\epsilon_{\alpha}= \frac{3}{4}\pi^{2}\alpha\frac{E_{k}^{0}}{m\omega^{2}L^{2}}.
\label{eq_effective_strain}
\end{eqnarray}
It was shown that this effective strain has the same effect as the mechanical strain, i.e., the frequency of the GNMR can be enhanced by the effective strain.

\textit{Adsorbate migration} -- The temperature dependence for the Q-factor of GNMRs has been measured by several experiments. There is an interesting scaling phenomenon. The Q-factor increases exponentially with decreasing temperature, and the exponent value will change at a transition temperature $T_{z}$.\cite{ZandeAMVD,ChenC2009nn} In a recent work, we have attributed this temperature scaling phenomenon to the adsorb migration effect on the surface of the GNMR.\cite{JiangJW2013diffusion} For temperatures above $T_{z}$, the adsorb is able to move far away from the GNMR surface, and will beat the GNMR frequently. This adsorb migration effect leads to strong reduction in the Q-factor, resulting in the transition of the temperature scaling factor.

We have selectively discussed some energy dissipation mechanisms for the GNMR in the above. We concentrate on the relationship between the flexural mode and the resonant mechanical oscillation of the GNMR, which is the main focus of the present review article. There are many other interesting and important energy dissipation mechanisms for the GNMR (for review, e.g. see Ref.~\onlinecite{ImbodenM2013pr}).

\section{Summary and Future Prospects}
In this review, we have introduced the basic lattice dynamics of the flexural mode in graphene, and summarized its important contribution to four of graphene's thermal and mechanical properties: thermal conduction, thermal expansion, elasticity, and nanomechanical resonance.

Although fruitful progress has been achieved in the study of the flexural mode in graphene, there are still some challenges and opportunities. For example, it is a long-term and important objective to understand and develop the lattice dynamic theory, particularly in combination with the ongoing explosion in computational speed and power. As CPU speed increases, lattice dynamical properties will be able to be calculated based on first-principles simulations. These first-principles calculations can serve as an effective examination and prediction for existing results that were obtained by applying lattice dynamical theory to classical atomistic simulations.

Furthermore, the basic lattice dynamic theory can be used in conjunction to study other practical scientific problems. For example, quasi-particle phonons can be coupled with other particles like electrons and photons,\cite{BoniniN,MarianiE} which will be useful for studying the electronic or thermal behavior of transistor devices, whose properties and performance are governed by such interactions.  

Finally, it will be important to transfer the knowledge gained regarding the important role of the flexural mode in graphene to other two-dimensional layered materials.\cite{NovoselovKS2005pnas} Inspired by the novel physical properties of two-dimensional graphene, there has been increasing interest in studying other similar two-dimensional layered materials, because these other two-dimensional materials may have superior properties to graphene. For instance, MoS$_{2}$ is a semiconductor with a bulk bandgap above 1.2 eV,\cite{KamKK,FengJ2012npho,LuP2012pccp} which has attracted considerable attention in recent years.\cite{RadisavljevicB2011nn,WangQH2012nn,ChhowallaM,ConleyHJ,SangwanVK,Ghorbani-AslM,CheiwchanchamnangijT,HuangW,VarshneyV,BertolazziS,CooperRC2013prb1} It is of practical significance to determine the contribution from the flexural mode to the physical properties of this, and other two-dimensional materials.

\section*{Acknowledgements}
This work was supported by the Recruitment Program of Global Youth Experts of China and the start-up funding from Shanghai University. Jin-Wu Jiang are grateful to Lin-Ping Yang, Hui Tang, and Zhao-Bing Su for critical help during his Ph.D. study on lattice dynamics in ITP-CAS. He also thanks Xiao-Xi Ni, Li-Hong Shi, Jie Chen, Jing-Hua Lan, Nuo Yang, and Baowen Li for valuable collaborations on the thermal transport work at NUS. He also thanks Bo Liu, Kun Zhou, Zenan Qi, Yan-Cheng Zhang, Jun-Hua Zhao, and Timon Rabczuk for helpful discussions on the mechanical work at BUW.


\begin{thebibliography}{250}
\providecommand{\natexlab}[1]{#1}
\providecommand{\url}[1]{\texttt{#1}}
\providecommand{\urlprefix}{URL }
\expandafter\ifx\csname urlstyle\endcsname\relax
  \providecommand{\doi}[1]{doi:\discretionary{}{}{}#1}\else
  \providecommand{\doi}{doi:\discretionary{}{}{}\begingroup
  \urlstyle{rm}\Url}\fi
\providecommand{\eprint}[2][]{\url{#2}}
\providecommand{\BIBand}{and}
\providecommand{\bibinfo}[2]{#2}
\ifx\xfnm\undefined \def\xfnm[#1]{\unskip,\space#1}\fi
\bibitem[{Born and Huang(1954)}]{BornM}
\bibinfo{author}{Born\xfnm[ M.]}, \bibinfo{author}{Huang\xfnm[ K.]}.
\newblock \bibinfo{title}{Dynamical Theory of Crystal Lattices}.
\newblock \bibinfo{publisher}{Oxford University Press, Oxford};
  \bibinfo{year}{1954}.
\bibitem[{Landau and Lifshitz(1995)}]{LandauLD}
\bibinfo{author}{Landau\xfnm[ L.D.]}, \bibinfo{author}{Lifshitz\xfnm[ E.M.]}.
\newblock \bibinfo{title}{Theory of Elasticity}.
\newblock \bibinfo{publisher}{Pergamon,Oxford}; \bibinfo{year}{1995}.
\bibitem[{Kirchhoff(1850)}]{KirchhoffG1850}
\bibinfo{author}{Kirchhoff\xfnm[ G.]}.
\newblock \bibinfo{title}{Uber das gleichqewicht und die bewegung einer
  elastichen scheibe}.
\newblock \bibinfo{journal}{J Reine und Angewandte Mathematik}
  \bibinfo{year}{1850};\bibinfo{volume}{40}:\bibinfo{pages}{51--88}.
\bibitem[{Rayleigh(1885)}]{RayleighL1885}
\bibinfo{author}{Rayleigh\xfnm[ L.]}.
\newblock \bibinfo{title}{On waves propagated along the plane surface of an
  elastic solid}.
\newblock \bibinfo{journal}{Proc London Math Soc}
  \bibinfo{year}{1885};\bibinfo{volume}{s1-17}(\bibinfo{number}{1}):\bibinfo{p%
ages}{4--11}.
\bibitem[{Lamb(1917)}]{LambH1917prsl}
\bibinfo{author}{Lamb\xfnm[ H.]}.
\newblock \bibinfo{title}{On waves in an elastic plate}.
\newblock \bibinfo{journal}{Proc Roy Soc London, Ser A}
  \bibinfo{year}{1917};\bibinfo{volume}{93}:\bibinfo{pages}{114--128}.
\bibitem[{Geim and Novoselov(2007)}]{GeimAK2007nm}
\bibinfo{author}{Geim\xfnm[ A.K.]}, \bibinfo{author}{Novoselov\xfnm[ K.S.]}.
\newblock \bibinfo{title}{The rise of graphene}.
\newblock \bibinfo{journal}{Nature Materials}
  \bibinfo{year}{2007};\bibinfo{volume}{6}(\bibinfo{number}{3}):\bibinfo{pages%
}{183--191}.
\bibitem[{Novoselov et~al.(2005{\natexlab{a}})Novoselov, Geim, Morozov, Jiang,
  Katsnelson, Grigorieva et~al.}]{NovoselovKS2005nat}
\bibinfo{author}{Novoselov\xfnm[ K.S.]}, \bibinfo{author}{Geim\xfnm[ A.K.]},
  \bibinfo{author}{Morozov\xfnm[ S.V.]}, \bibinfo{author}{Jiang\xfnm[ D.]},
  \bibinfo{author}{Katsnelson\xfnm[ M.I.]}, \bibinfo{author}{Grigorieva\xfnm[
  I.V.]}, et~al.
\newblock \bibinfo{title}{Two-dimensional gas of massless dirac fermions in
  graphene}.
\newblock \bibinfo{journal}{Nature}
  \bibinfo{year}{2005}{\natexlab{a}};\bibinfo{volume}{438}(\bibinfo{number}{70%
65}):\bibinfo{pages}{197--200}.
\bibitem[{Balandin et~al.(2008)Balandin, Ghosh, Bao, Calizo, Teweldebrhan, Miao
  et~al.}]{BalandinAA2008}
\bibinfo{author}{Balandin\xfnm[ A.A.]}, \bibinfo{author}{Ghosh\xfnm[ S.]},
  \bibinfo{author}{Bao\xfnm[ W.]}, \bibinfo{author}{Calizo\xfnm[ I.]},
  \bibinfo{author}{Teweldebrhan\xfnm[ D.]}, \bibinfo{author}{Miao\xfnm[ F.]},
  et~al.
\newblock \bibinfo{title}{Superior thermal conductivity of single-layer
  graphene}.
\newblock \bibinfo{journal}{Nano Letters}
  \bibinfo{year}{2008};\bibinfo{volume}{8}(\bibinfo{number}{3}):\bibinfo{pages%
}{902--907}.
\bibitem[{Nika et~al.(2009{\natexlab{a}})Nika, Pokatilov, Askerov and
  Balandin}]{NikaDL2009prb}
\bibinfo{author}{Nika\xfnm[ D.L.]}, \bibinfo{author}{Pokatilov\xfnm[ E.P.]},
  \bibinfo{author}{Askerov\xfnm[ A.S.]}, \bibinfo{author}{Balandin\xfnm[
  A.A.]}.
\newblock \bibinfo{title}{Phonon thermal conduction in graphene: Role of
  umklapp and edge roughness scattering}.
\newblock \bibinfo{journal}{Physical Review B}
  \bibinfo{year}{2009}{\natexlab{a}};\bibinfo{volume}{79}:\bibinfo{pages}{1554%
13}.
\bibitem[{Nika et~al.(2009{\natexlab{b}})Nika, Ghosh, Pokatilov and
  Balandin}]{NikaDL2009apl}
\bibinfo{author}{Nika\xfnm[ D.L.]}, \bibinfo{author}{Ghosh\xfnm[ S.]},
  \bibinfo{author}{Pokatilov\xfnm[ E.P.]}, \bibinfo{author}{Balandin\xfnm[
  A.A.]}.
\newblock \bibinfo{title}{Lattice thermal conductivity of graphene flakes:
  Comparison with bulk graphite}.
\newblock \bibinfo{journal}{Applied Physics Letters}
  \bibinfo{year}{2009}{\natexlab{b}};\bibinfo{volume}{94}(\bibinfo{number}{20}%
):\bibinfo{pages}{203103}.
\bibitem[{Balandin(2011)}]{Balandin2011nm}
\bibinfo{author}{Balandin\xfnm[ A.A.]}.
\newblock \bibinfo{title}{Thermal properties of graphene and nanostructured
  carbon materials}.
\newblock \bibinfo{journal}{Nature Materials}
  \bibinfo{year}{2011};\bibinfo{volume}{10}:\bibinfo{pages}{569--581}.
\bibitem[{Jiang et~al.(2009{\natexlab{a}})Jiang, Wang and
  Li}]{JiangJW2009direction}
\bibinfo{author}{Jiang\xfnm[ J.W.]}, \bibinfo{author}{Wang\xfnm[ J.S.]},
  \bibinfo{author}{Li\xfnm[ B.]}.
\newblock \bibinfo{title}{Thermal conductance of graphene and dimerite}.
\newblock \bibinfo{journal}{Physical Review B}
  \bibinfo{year}{2009}{\natexlab{a}};\bibinfo{volume}{79}(\bibinfo{number}{20}%
):\bibinfo{pages}{205418}.
\bibitem[{Bao et~al.(2009)Bao, Miao, Chen, Zhang, Jang, Dames
  et~al.}]{BaoW2009nn}
\bibinfo{author}{Bao\xfnm[ W.]}, \bibinfo{author}{Miao\xfnm[ F.]},
  \bibinfo{author}{Chen\xfnm[ Z.]}, \bibinfo{author}{Zhang\xfnm[ H.]},
  \bibinfo{author}{Jang\xfnm[ W.]}, \bibinfo{author}{Dames\xfnm[ C.]}, et~al.
\newblock \bibinfo{title}{Controlled ripple texturing of suspended graphene and
  ultrathin graphite membranes}.
\newblock \bibinfo{journal}{Nature Nanotechnology}
  \bibinfo{year}{2009};\bibinfo{volume}{4}:\bibinfo{pages}{562--566}.
\bibitem[{Jensen et~al.(2008)Jensen, Kim and Zettl}]{JensenK}
\bibinfo{author}{Jensen\xfnm[ K.]}, \bibinfo{author}{Kim\xfnm[ K.]},
  \bibinfo{author}{Zettl\xfnm[ A.]}.
\newblock \bibinfo{title}{An atomic-resolution nanomechanical mass sensor}.
\newblock \bibinfo{journal}{Nature Nanotechnology}
  \bibinfo{year}{2008};\bibinfo{volume}{3}:\bibinfo{pages}{533--537}.
\bibitem[{Tang et~al.(2011)Tang, Wang and Su}]{TangH2011book}
\bibinfo{author}{Tang\xfnm[ H.]}, \bibinfo{author}{Wang\xfnm[ B.S.]},
  \bibinfo{author}{Su\xfnm[ Z.B.]}.
\newblock \bibinfo{title}{Graphene Simulation, Chapter 10: Symmetry and Lattice
  Dynamics}.
\newblock \bibinfo{publisher}{InTech, Shanghai China}; \bibinfo{year}{2011}.
\bibitem[{Saito et~al.(1998)Saito, Dresselhaus and Dresselhaus}]{SaitoR}
\bibinfo{author}{Saito\xfnm[ R.]}, \bibinfo{author}{Dresselhaus\xfnm[ G.]},
  \bibinfo{author}{Dresselhaus\xfnm[ M.S.]}.
\newblock \bibinfo{title}{Physical Properties of Carbon Nanotubes}.
\newblock \bibinfo{publisher}{Imperial College, London}; \bibinfo{year}{1998}.
\bibitem[{Rotman(1995)}]{RotmanJJ1995book}
\bibinfo{author}{Rotman\xfnm[ J.J.]}.
\newblock \bibinfo{title}{An Introduction to the Theory of Groups}.
\newblock \bibinfo{publisher}{Springer-Verlag New York, Inc};
  \bibinfo{year}{1995}.
\newblock ISBN \bibinfo{isbn}{0-387-94285-8}.
\bibitem[{Miloevic and Damnjanovic(1993)}]{MiloevicI1993prb}
\bibinfo{author}{Miloevic\xfnm[ I.]}, \bibinfo{author}{Damnjanovic\xfnm[ M.]}.
\newblock \bibinfo{title}{Normal vibrations and {J}ahn-{T}eller effect for
  polymers and quasi-one-dimensional systems}.
\newblock \bibinfo{journal}{Physical Review B}
  \bibinfo{year}{1993};\bibinfo{volume}{47}(\bibinfo{number}{13}):\bibinfo{pag%
es}{7805--7818}.
\bibitem[{Popov et~al.(1999)Popov, Doren and Balkanski}]{PopovVN1999prb}
\bibinfo{author}{Popov\xfnm[ V.N.]}, \bibinfo{author}{Doren\xfnm[ V.E.V.]},
  \bibinfo{author}{Balkanski\xfnm[ M.]}.
\newblock \bibinfo{title}{Lattice dynamics of single-walled carbon nanotubes}.
\newblock \bibinfo{journal}{Physical Review B}
  \bibinfo{year}{1999};\bibinfo{volume}{59}(\bibinfo{number}{13}):\bibinfo{pag%
es}{8355}.
\bibitem[{Dobardzic et~al.(2003)Dobardzic, Milosevic, Nikolic, Vukovi and
  Damnjanovic}]{DobardzicE2003prb}
\bibinfo{author}{Dobardzic\xfnm[ E.]}, \bibinfo{author}{Milosevic\xfnm[ I.]},
  \bibinfo{author}{Nikolic\xfnm[ B.]}, \bibinfo{author}{Vukovi\xfnm[ T.]},
  \bibinfo{author}{Damnjanovic\xfnm[ M.]}.
\newblock \bibinfo{title}{Single-wall carbon nanotubes phonon
  spectra: symmetry-based calculations}.
\newblock \bibinfo{journal}{Physical Review B}
  \bibinfo{year}{2003};\bibinfo{volume}{68}(\bibinfo{number}{4}):\bibinfo{page%
s}{045408}.
\bibitem[{Jiang et~al.(2006)Jiang, Tang, Wang and Su}]{JiangJW2006}
\bibinfo{author}{Jiang\xfnm[ J.W.]}, \bibinfo{author}{Tang\xfnm[ H.]},
  \bibinfo{author}{Wang\xfnm[ B.S.]}, \bibinfo{author}{Su\xfnm[ Z.B.]}.
\newblock \bibinfo{title}{Chiral symmetry analysis and rigid rotational
  invariance for the lattice dynamics of single-wall carbon nanotubes}.
\newblock \bibinfo{journal}{Physical Review B}
  \bibinfo{year}{2006};\bibinfo{volume}{73}(\bibinfo{number}{23}):\bibinfo{pag%
es}{235434}.
\bibitem[{Dakic et~al.(2009)Dakic, Damnjanovic and Miloevic}]{DakicB2009jpamt}
\bibinfo{author}{Dakic\xfnm[ B.]}, \bibinfo{author}{Damnjanovic\xfnm[ M.]},
  \bibinfo{author}{Miloevic\xfnm[ I.]}.
\newblock \bibinfo{title}{Generalized bloch states and potentials of nanotubes
  and other quasi-1d systems ii}.
\newblock \bibinfo{journal}{Journal of Physics A: Mathematical and Theoretical}
  \bibinfo{year}{2009};\bibinfo{volume}{42}(\bibinfo{number}{12}):\bibinfo{pag%
es}{125202}.
\bibitem[{Peelaers et~al.(2009)Peelaers, Partoens and Peeters}]{PeelaersH}
\bibinfo{author}{Peelaers\xfnm[ H.]}, \bibinfo{author}{Partoens\xfnm[ B.]},
  \bibinfo{author}{Peeters\xfnm[ F.M.]}.
\newblock \bibinfo{title}{Phonon band structure of si nanowires: A stability
  analysis}.
\newblock \bibinfo{journal}{Nano Letters}
  \bibinfo{year}{2009};\bibinfo{volume}{9}(\bibinfo{number}{1}):\bibinfo{pages%
}{107--111}.
\bibitem[{Liu et~al.(2007)Liu, Ming and Li}]{LiuF2007prb}
\bibinfo{author}{Liu\xfnm[ F.]}, \bibinfo{author}{Ming\xfnm[ P.]},
  \bibinfo{author}{Li\xfnm[ J.]}.
\newblock \bibinfo{title}{Ab initio calculation of ideal strength and phonon
  instability of graphene under tension}.
\newblock \bibinfo{journal}{Physical Review B}
  \bibinfo{year}{2007};\bibinfo{volume}{76}:\bibinfo{pages}{064120}.
\bibitem[{Jiang(2014{\natexlab{a}})}]{JiangJW2014mos2bandgap}
\bibinfo{author}{Jiang\xfnm[ J.W.]}.
\newblock \bibinfo{title}{Phonon bandgap engineering of strained monolayer
  mos2}.
\newblock \bibinfo{journal}{Nanoscale}
  \bibinfo{year}{2014}{\natexlab{a}};\bibinfo{volume}{6}:\bibinfo{pages}{8326}.
\bibitem[{Kubo et~al.(1992)Kubo, Toda and Hashitsume}]{KuboR}
\bibinfo{author}{Kubo\xfnm[ R.]}, \bibinfo{author}{Toda\xfnm[ M.]},
  \bibinfo{author}{Hashitsume\xfnm[ N.]}.
\newblock \bibinfo{title}{Statistical Physics II}.
\newblock \bibinfo{publisher}{Springer, Berlin}; \bibinfo{year}{1992}.
\bibitem[{Wang and Li(2004)}]{WangJS2004pre}
\bibinfo{author}{Wang\xfnm[ J.S.]}, \bibinfo{author}{Li\xfnm[ B.]}.
\newblock \bibinfo{title}{Mode-coupling theory and molecular dynamics
  simulation for heat conduction in a chain with transverse motions}.
\newblock \bibinfo{journal}{Physical Review E}
  \bibinfo{year}{2004};\bibinfo{volume}{70}(\bibinfo{number}{2}):\bibinfo{page%
s}{021204}.
\bibitem[{Bonini et~al.(2007)Bonini, Lazzeri, Marzari and Mauri}]{BoniniN}
\bibinfo{author}{Bonini\xfnm[ N.]}, \bibinfo{author}{Lazzeri\xfnm[ M.]},
  \bibinfo{author}{Marzari\xfnm[ N.]}, \bibinfo{author}{Mauri\xfnm[ F.]}.
\newblock \bibinfo{title}{Phonon anharmonicities in graphite and graphene}.
\newblock \bibinfo{journal}{Physical Review Letters}
  \bibinfo{year}{2007};\bibinfo{volume}{99}(\bibinfo{number}{17}):\bibinfo{pag%
es}{176802}.
\bibitem[{Mariani and Von~Oppen(2008)}]{MarianiE}
\bibinfo{author}{Mariani\xfnm[ E.]}, \bibinfo{author}{Von~Oppen\xfnm[ F.]}.
\newblock \bibinfo{title}{Flexural phonons in free-standing graphene}.
\newblock \bibinfo{journal}{Physical Review Letters}
  \bibinfo{year}{2008};\bibinfo{volume}{100}(\bibinfo{number}{7}):\bibinfo{pag%
es}{076801}.
\bibitem[{Ziman(1960)}]{ZimanJM}
\bibinfo{author}{Ziman\xfnm[ J.M.]}.
\newblock \bibinfo{title}{Electrons and Phonons}.
\newblock \bibinfo{publisher}{Clarendon Press, Oxford}; \bibinfo{year}{1960}.
\bibitem[{Cochran(1959)}]{CochranW}
\bibinfo{author}{Cochran\xfnm[ W.]}.
\newblock \bibinfo{title}{Theory of the lattice vibrations of germanium}.
\newblock \bibinfo{journal}{Proc R Soc Ser A}
  \bibinfo{year}{1959};\bibinfo{volume}{253}:\bibinfo{pages}{260--276}.
\bibitem[{Martin(1969)}]{MartinRM}
\bibinfo{author}{Martin\xfnm[ R.M.]}.
\newblock \bibinfo{title}{Dielectric screening model for lattice vibrations of
  diamondstructure crystals}.
\newblock \bibinfo{journal}{Physical Review B}
  \bibinfo{year}{1969};\bibinfo{volume}{186}:\bibinfo{pages}{871}.
\bibitem[{Gale(1997)}]{gulp}
\bibinfo{author}{Gale\xfnm[ J.D.]}.
\newblock \bibinfo{title}{Gulp: A computer program for the symmetry-adapted
  simulation of solids}.
\newblock \bibinfo{journal}{J Chem Soc, Faraday Trans}
  \bibinfo{year}{1997};\bibinfo{volume}{93}(\bibinfo{number}{4}):\bibinfo{page%
s}{629--637. Code available from https://projects.ivec.org/gulp/}.
\bibitem[{Soler et~al.(2002)Soler, Artacho, Gale, Garcia, Junquera, Ordejon
  et~al.}]{siesta}
\bibinfo{author}{Soler\xfnm[ J.M.]}, \bibinfo{author}{Artacho\xfnm[ E.]},
  \bibinfo{author}{Gale\xfnm[ J.D.]}, \bibinfo{author}{Garcia\xfnm[ A.]},
  \bibinfo{author}{Junquera\xfnm[ J.]}, \bibinfo{author}{Ordejon\xfnm[ P.]},
  et~al.
\newblock \bibinfo{title}{The siesta method for ab initio order-n materials
  simulation}.
\newblock \bibinfo{journal}{Journal of Physics: Condensed Matter}
  \bibinfo{year}{2002};\bibinfo{volume}{14}(\bibinfo{number}{11}):\bibinfo{pag%
es}{2745. Code available from http://www.icmab.es/dmmis/leem/siesta/}.
\bibitem[{Jishi et~al.(1993)Jishi, Venkataraman, Dresselhaus and
  Dresselhaus}]{JishiRA1993cpl}
\bibinfo{author}{Jishi\xfnm[ R.A.]}, \bibinfo{author}{Venkataraman\xfnm[ L.]},
  \bibinfo{author}{Dresselhaus\xfnm[ M.S.]}, \bibinfo{author}{Dresselhaus\xfnm[
  G.]}.
\newblock \bibinfo{title}{Phonon modes in carbon nanotubules}.
\newblock \bibinfo{journal}{Chemical Physics Letters}
  \bibinfo{year}{1993};\bibinfo{volume}{209}(\bibinfo{number}{1-2}):\bibinfo{p%
ages}{77--82}.
\bibitem[{Aizawa et~al.(1990)Aizawa, Souda, Otani and Ishizawa}]{AizawaT}
\bibinfo{author}{Aizawa\xfnm[ T.]}, \bibinfo{author}{Souda\xfnm[ R.]},
  \bibinfo{author}{Otani\xfnm[ S.]}, \bibinfo{author}{Ishizawa\xfnm[ Y.]}.
\newblock \bibinfo{title}{Bond softening in monolayer graphite formed on
  transition-metal carbide surfaces}.
\newblock \bibinfo{journal}{Physical Review B}
  \bibinfo{year}{1990};\bibinfo{volume}{42}(\bibinfo{number}{18}):\bibinfo{pag%
es}{11469--11478}.
\bibitem[{Yu(2010)}]{YuPY}
\bibinfo{author}{Yu\xfnm[ P.Y.]}.
\newblock \bibinfo{title}{Fundamentals of Semiconductors}.
\newblock \bibinfo{publisher}{Springer, New York}; \bibinfo{year}{2010}.
\newblock ISBN \bibinfo{isbn}{978-3-642-00709-5}.
\bibitem[{Musgrave and Pople(1962)}]{MusgraveMJP}
\bibinfo{author}{Musgrave\xfnm[ M.J.P.]}, \bibinfo{author}{Pople\xfnm[ J.A.]}.
\newblock \bibinfo{title}{A general valence force field for diamond}.
\newblock \bibinfo{journal}{Proc R Soc Ser A}
  \bibinfo{year}{1962};\bibinfo{volume}{268}:\bibinfo{pages}{474--484}.
\bibitem[{Nusimovici and Birman(1967)}]{NusimoviciMA}
\bibinfo{author}{Nusimovici\xfnm[ M.A.]}, \bibinfo{author}{Birman\xfnm[ J.L.]}.
\newblock \bibinfo{title}{Lattice dynamics of wurtzite: Cds}.
\newblock \bibinfo{journal}{Physical Review}
  \bibinfo{year}{1967};\bibinfo{volume}{156}(\bibinfo{number}{3}):\bibinfo{pag%
es}{925--938}.
\bibitem[{Keating(1966)}]{KeatingPN}
\bibinfo{author}{Keating\xfnm[ P.N.]}.
\newblock \bibinfo{title}{Effect of invariance requirements on the elastic
  strain energy of crystals with application to the diamond structure}.
\newblock \bibinfo{journal}{Physical Review}
  \bibinfo{year}{1966};\bibinfo{volume}{145}(\bibinfo{number}{2}):\bibinfo{pag%
es}{637--645}.
\bibitem[{Jiang et~al.(2008)Jiang, Tang, Wang and Su}]{JiangJW2008}
\bibinfo{author}{Jiang\xfnm[ J.W.]}, \bibinfo{author}{Tang\xfnm[ H.]},
  \bibinfo{author}{Wang\xfnm[ B.S.]}, \bibinfo{author}{Su\xfnm[ Z.B.]}.
\newblock \bibinfo{title}{A lattice dynamical treatment for the total potential
  energy of single-walled carbon nanotubes and its applications: relaxed
  equilibrium structure, elastic properties, and vibrational modes of
  ultra-narrow tubes}.
\newblock \bibinfo{journal}{Journal of Physics: Condensed Matter}
  \bibinfo{year}{2008};\bibinfo{volume}{20}:\bibinfo{pages}{045228}.
\bibitem[{Brenner et~al.(2002)Brenner, Shenderova, Harrison, Stuart, Ni and
  Sinnott}]{brennerJPCM2002}
\bibinfo{author}{Brenner\xfnm[ D.W.]}, \bibinfo{author}{Shenderova\xfnm[
  O.A.]}, \bibinfo{author}{Harrison\xfnm[ J.A.]}, \bibinfo{author}{Stuart\xfnm[
  S.J.]}, \bibinfo{author}{Ni\xfnm[ B.]}, \bibinfo{author}{Sinnott\xfnm[
  S.B.]}.
\newblock \bibinfo{title}{A second-generation reactive empirical bond order
  ({REBO}) potential energy expression for hydrocarbons}.
\newblock \bibinfo{journal}{Journal of Physics: Condensed Matter}
  \bibinfo{year}{2002};\bibinfo{volume}{14}:\bibinfo{pages}{783--802}.
\bibitem[{Mahan and Jeon(2004)}]{MahanGD2004prb}
\bibinfo{author}{Mahan\xfnm[ G.D.]}, \bibinfo{author}{Jeon\xfnm[ G.S.]}.
\newblock \bibinfo{title}{Flexure modes in carbon nanotubes}.
\newblock \bibinfo{journal}{Physical Review B}
  \bibinfo{year}{2004};\bibinfo{volume}{70}:\bibinfo{pages}{075405}.
\bibitem[{Jiang and Wang(2010{\natexlab{a}})}]{JiangJWcorssover}
\bibinfo{author}{Jiang\xfnm[ J.W.]}, \bibinfo{author}{Wang\xfnm[ J.S.]}.
\newblock \bibinfo{title}{A universal exponential factor in the dimensional
  crossover from graphene to graphite}.
\newblock \bibinfo{journal}{Journal of Applied Physics}
  \bibinfo{year}{2010}{\natexlab{a}};\bibinfo{volume}{108}:\bibinfo{pages}{124%
311}.
\bibitem[{Dubay and Kresse(2003)}]{DubayO2003prb}
\bibinfo{author}{Dubay\xfnm[ O.]}, \bibinfo{author}{Kresse\xfnm[ G.]}.
\newblock \bibinfo{title}{Accurate density functional calculations for the
  phonon dispersion relations of graphite layer and carbon nanotubes}.
\newblock \bibinfo{journal}{Physical Review B}
  \bibinfo{year}{2003};\bibinfo{volume}{67}:\bibinfo{pages}{035401}.
\bibitem[{Mounet and Marzari(2005)}]{MounetN}
\bibinfo{author}{Mounet\xfnm[ N.]}, \bibinfo{author}{Marzari\xfnm[ N.]}.
\newblock \bibinfo{title}{First-principles determination of the structural,
  vibrational and thermodynamic properties of diamond, graphite, and
  derivatives}.
\newblock \bibinfo{journal}{Physical Review B}
  \bibinfo{year}{2005};\bibinfo{volume}{71}(\bibinfo{number}{20}):\bibinfo{pag%
es}{205214}.
\bibitem[{Gillen et~al.(2009)Gillen, Mohr, Thomsen and
  Maultzsch}]{GillenR2009prb}
\bibinfo{author}{Gillen\xfnm[ R.]}, \bibinfo{author}{Mohr\xfnm[ M.]},
  \bibinfo{author}{Thomsen\xfnm[ C.]}, \bibinfo{author}{Maultzsch\xfnm[ J.]}.
\newblock \bibinfo{title}{Vibrational properties of graphene nanoribbons by
  first-principles calculations}.
\newblock \bibinfo{journal}{Physical Review B}
  \bibinfo{year}{2009};\bibinfo{volume}{80}:\bibinfo{pages}{155418}.
\bibitem[{Metlov(2010)}]{MetlovKL2010prb}
\bibinfo{author}{Metlov\xfnm[ K.L.]}.
\newblock \bibinfo{title}{Model for flexural phonon dispersion in graphite and
  graphene}.
\newblock \bibinfo{journal}{Physical Review B}
  \bibinfo{year}{2010};\bibinfo{volume}{82}:\bibinfo{pages}{033404}.
\bibitem[{Al-Jishi and Dresselhaus(1982)}]{JishiR1982prb}
\bibinfo{author}{Al-Jishi\xfnm[ R.]}, \bibinfo{author}{Dresselhaus\xfnm[ G.]}.
\newblock \bibinfo{title}{Lattice-dynamical model for graphite}.
\newblock \bibinfo{journal}{Physical Review B}
  \bibinfo{year}{1982};\bibinfo{volume}{26}(\bibinfo{number}{8}):\bibinfo{page%
s}{4514}.
\bibitem[{Mohr et~al.(2007)Mohr, Maultzsch, Dobardzic, Reich, Milosevic,
  Damnjanovic et~al.}]{MohrM2007prb}
\bibinfo{author}{Mohr\xfnm[ M.]}, \bibinfo{author}{Maultzsch\xfnm[ J.]},
  \bibinfo{author}{Dobardzic\xfnm[ E.]}, \bibinfo{author}{Reich\xfnm[ S.]},
  \bibinfo{author}{Milosevic\xfnm[ I.]}, \bibinfo{author}{Damnjanovic\xfnm[
  M.]}, et~al.
\newblock \bibinfo{title}{Phonon dispersion of graphite by inelastic x-ray
  scattering}.
\newblock \bibinfo{journal}{Physical Review B}
  \bibinfo{year}{2007};\bibinfo{volume}{76}:\bibinfo{pages}{035439}.
\bibitem[{Stillinger and Weber(1985)}]{StillingerFH}
\bibinfo{author}{Stillinger\xfnm[ F.H.]}, \bibinfo{author}{Weber\xfnm[ T.A.]}.
\newblock \bibinfo{title}{Computer simulation of local order in condensed
  phases of silicon}.
\newblock \bibinfo{journal}{Physical Review B}
  \bibinfo{year}{1985};\bibinfo{volume}{31}(\bibinfo{number}{8}):\bibinfo{page%
s}{5262}.
\bibitem[{Yi$\breve{g}$en et~al.(2013)Yi$\breve{g}$en, Tayari, Island, Porter
  and Champagne}]{YigenS}
\bibinfo{author}{Yi$\breve{g}$en\xfnm[ S.]}, \bibinfo{author}{Tayari\xfnm[
  V.]}, \bibinfo{author}{Island\xfnm[ J.O.]}, \bibinfo{author}{Porter\xfnm[
  J.M.]}, \bibinfo{author}{Champagne\xfnm[ A.R.]}.
\newblock \bibinfo{title}{Electronic thermal conductivity measurements in
  intrinsic graphene}.
\newblock \bibinfo{journal}{Physical Review B}
  \bibinfo{year}{2013};\bibinfo{volume}{87}(\bibinfo{number}{24}):\bibinfo{pag%
es}{241411}.
\bibitem[{Saito et~al.(2007)Saito, Nakamura and Natori}]{SaitoK}
\bibinfo{author}{Saito\xfnm[ K.]}, \bibinfo{author}{Nakamura\xfnm[ J.]},
  \bibinfo{author}{Natori\xfnm[ A.]}.
\newblock \bibinfo{title}{Ballistic thermal conductance of a graphene sheet}.
\newblock \bibinfo{journal}{Physical Review B}
  \bibinfo{year}{2007};\bibinfo{volume}{76}(\bibinfo{number}{11}):\bibinfo{pag%
es}{115409}.
\bibitem[{Schwab et~al.(2000)Schwab, Henriksen, Worlock and Roukes}]{SchwabK}
\bibinfo{author}{Schwab\xfnm[ K.]}, \bibinfo{author}{Henriksen\xfnm[ E.A.]},
  \bibinfo{author}{Worlock\xfnm[ J.M.]}, \bibinfo{author}{Roukes\xfnm[ M.L.]}.
\newblock \bibinfo{title}{Measurement of the quantum of thermal conductance}.
\newblock \bibinfo{journal}{Nature}
  \bibinfo{year}{2000};\bibinfo{volume}{404}:\bibinfo{pages}{974}.
\bibitem[{Ghosh et~al.(2008)Ghosh, Calizo, Teweldebrhan, Pokatilov, Nika,
  Balandin et~al.}]{GhoshS2008apl}
\bibinfo{author}{Ghosh\xfnm[ S.]}, \bibinfo{author}{Calizo\xfnm[ I.]},
  \bibinfo{author}{Teweldebrhan\xfnm[ D.]}, \bibinfo{author}{Pokatilov\xfnm[
  E.P.]}, \bibinfo{author}{Nika\xfnm[ D.L.]}, \bibinfo{author}{Balandin\xfnm[
  A.A.]}, et~al.
\newblock \bibinfo{title}{Extremely high thermal conductivity of graphene:
  Prospects for thermal management applications in nanoelectronic circuits}.
\newblock \bibinfo{journal}{Applied Physics Letters}
  \bibinfo{year}{2008};\bibinfo{volume}{92}(\bibinfo{number}{15}):\bibinfo{pag%
es}{151911}.
\bibitem[{Xu et~al.(2010)Xu, Wang, Zhang, Zhao, Bae, Heinrich
  et~al.}]{XuXF2010}
\bibinfo{author}{Xu\xfnm[ X.]}, \bibinfo{author}{Wang\xfnm[ Y.]},
  \bibinfo{author}{Zhang\xfnm[ K.]}, \bibinfo{author}{Zhao\xfnm[ X.]},
  \bibinfo{author}{Bae\xfnm[ S.]}, \bibinfo{author}{Heinrich\xfnm[ M.]}, et~al.
\newblock \bibinfo{title}{Phonon transport in suspended single layer graphene}.
\newblock \bibinfo{journal}{Arxivorg} \bibinfo{year}{2010};.
\bibitem[{Mingo and Broido(2005)}]{MingoN2005prl}
\bibinfo{author}{Mingo\xfnm[ N.]}, \bibinfo{author}{Broido\xfnm[ D.A.]}.
\newblock \bibinfo{title}{Carbon nanotube ballistic thermal conductance and its
  limits}.
\newblock \bibinfo{journal}{Physical Review Letters}
  \bibinfo{year}{2005};\bibinfo{volume}{95}(\bibinfo{number}{9}):\bibinfo{page%
s}{096105}.
\bibitem[{Xu et~al.(2009)Xu, Chen, Gu and Duan}]{XuY2009apl}
\bibinfo{author}{Xu\xfnm[ Y.]}, \bibinfo{author}{Chen\xfnm[ X.]},
  \bibinfo{author}{Gu\xfnm[ B.L.]}, \bibinfo{author}{Duan\xfnm[ W.]}.
\newblock \bibinfo{title}{Intrinsic anisotropy of thermal conductance in
  graphene nanoribbons}.
\newblock \bibinfo{journal}{Applied Physics Letters}
  \bibinfo{year}{2009};\bibinfo{volume}{95}(\bibinfo{number}{23}):\bibinfo{pag%
es}{233116}.
\bibitem[{Mun$\tilde{n}$oz et~al.(2010)Mun$\tilde{n}$oz, Lu and
  Yakobson}]{MunozE}
\bibinfo{author}{Mun$\tilde{n}$oz\xfnm[ E.]}, \bibinfo{author}{Lu\xfnm[ J.]},
  \bibinfo{author}{Yakobson\xfnm[ B.I.]}.
\newblock \bibinfo{title}{Ballistic thermal conductance of graphene ribbons}.
\newblock \bibinfo{journal}{Nano Letters}
  \bibinfo{year}{2010};\bibinfo{volume}{10}:\bibinfo{pages}{1652--1656}.
\bibitem[{Wang et~al.(2012{\natexlab{a}})Wang, Wang, Chen and
  Wang}]{WangJ2012jpcm}
\bibinfo{author}{Wang\xfnm[ J.]}, \bibinfo{author}{Wang\xfnm[ X.M.]},
  \bibinfo{author}{Chen\xfnm[ Y.F.]}, \bibinfo{author}{Wang\xfnm[ J.S.]}.
\newblock \bibinfo{title}{Dimensional crossover of thermal conductance in
  graphene nanoribbons: a first-principles approach}.
\newblock \bibinfo{journal}{Journal of Physics: Condensed Matter}
  \bibinfo{year}{2012}{\natexlab{a}};\bibinfo{volume}{24}:\bibinfo{pages}{2954%
03}.
\bibitem[{Nika et~al.(2012)Nika, Askerov and Balandin}]{NikaDL2012nl}
\bibinfo{author}{Nika\xfnm[ D.L.]}, \bibinfo{author}{Askerov\xfnm[ A.S.]},
  \bibinfo{author}{Balandin\xfnm[ A.A.]}.
\newblock \bibinfo{title}{Anomalous size dependence of the thermal conductivity
  of graphene ribbons}.
\newblock \bibinfo{journal}{Nano Letters}
  \bibinfo{year}{2012};\bibinfo{volume}{12}(\bibinfo{number}{6}):\bibinfo{page%
s}{3238--3244}.
\bibitem[{Xu et~al.(2014)Xu, Pereira, Wang, Wu, Zhang, Zhao et~al.}]{XuX2014nc}
\bibinfo{author}{Xu\xfnm[ X.]}, \bibinfo{author}{Pereira\xfnm[ L.F.]},
  \bibinfo{author}{Wang\xfnm[ Y.]}, \bibinfo{author}{Wu\xfnm[ J.]},
  \bibinfo{author}{Zhang\xfnm[ K.]}, \bibinfo{author}{Zhao\xfnm[ X.]}, et~al.
\newblock \bibinfo{title}{Length-dependent thermal conductivity in suspended
  single-layer graphene}.
\newblock \bibinfo{journal}{Nature Communications}
  \bibinfo{year}{2014};\bibinfo{volume}{5}:\bibinfo{pages}{3689}.
\bibitem[{Xiong et~al.(2010{\natexlab{a}})Xiong, Wang, Zhang and Zhao}]{XiongD}
\bibinfo{author}{Xiong\xfnm[ D.]}, \bibinfo{author}{Wang\xfnm[ J.]},
  \bibinfo{author}{Zhang\xfnm[ Y.]}, \bibinfo{author}{Zhao\xfnm[ H.]}.
\newblock \bibinfo{title}{Heat conduction in two-dimensional disk models}.
\newblock \bibinfo{journal}{Physical Review E}
  \bibinfo{year}{2010}{\natexlab{a}};\bibinfo{volume}{82}:\bibinfo{pages}{0301%
01(R)}.
\bibitem[{Jiang et~al.(2010{\natexlab{a}})Jiang, Lan, Wang and
  Li}]{JiangJW2010isotopic}
\bibinfo{author}{Jiang\xfnm[ J.W.]}, \bibinfo{author}{Lan\xfnm[ J.]},
  \bibinfo{author}{Wang\xfnm[ J.S.]}, \bibinfo{author}{Li\xfnm[ B.]}.
\newblock \bibinfo{title}{Isotopic effects on the thermal conductivity of
  graphene nanoribbons: Localization mechanism}.
\newblock \bibinfo{journal}{Journal of Applied Physics}
  \bibinfo{year}{2010}{\natexlab{a}};\bibinfo{volume}{107}(\bibinfo{number}{5}%
):\bibinfo{pages}{054314}.
\bibitem[{Chen et~al.(2012{\natexlab{a}})Chen, Wu, Mishra, Kang, Zhang, Cho
  et~al.}]{ChenSS}
\bibinfo{author}{Chen\xfnm[ S.]}, \bibinfo{author}{Wu\xfnm[ Q.]},
  \bibinfo{author}{Mishra\xfnm[ C.]}, \bibinfo{author}{Kang\xfnm[ J.]},
  \bibinfo{author}{Zhang\xfnm[ H.]}, \bibinfo{author}{Cho\xfnm[ K.]}, et~al.
\newblock \bibinfo{title}{Thermal conductivity of isotopically modified
  graphene}.
\newblock \bibinfo{journal}{Nature Materials}
  \bibinfo{year}{2012}{\natexlab{a}};\bibinfo{volume}{11}:\bibinfo{pages}{203-%
-207}.
\bibitem[{Yang et~al.(2009)Yang, Zhang and Li}]{YangN2009apl}
\bibinfo{author}{Yang\xfnm[ N.]}, \bibinfo{author}{Zhang\xfnm[ G.]},
  \bibinfo{author}{Li\xfnm[ B.]}.
\newblock \bibinfo{title}{Thermal rectification in asymmetric graphene
  ribbons}.
\newblock \bibinfo{journal}{Applied Physics Letters}
  \bibinfo{year}{2009};\bibinfo{volume}{95}(\bibinfo{number}{3}):\bibinfo{page%
s}{033107}.
\bibitem[{Hu et~al.(2009)Hu, Ruan and Chen}]{HuJ}
\bibinfo{author}{Hu\xfnm[ J.]}, \bibinfo{author}{Ruan\xfnm[ X.]},
  \bibinfo{author}{Chen\xfnm[ Y.P.]}.
\newblock \bibinfo{title}{Thermal conductivity and thermal rectification in
  graphene nanoribbons: A molecular dynamics study}.
\newblock \bibinfo{journal}{Nano Letters}
  \bibinfo{year}{2009};\bibinfo{volume}{9}(\bibinfo{number}{7}):\bibinfo{pages%
}{2730--2735}.
\bibitem[{Jiang et~al.(2010{\natexlab{b}})Jiang, Wang and Li}]{JiangJW2010epl}
\bibinfo{author}{Jiang\xfnm[ J.W.]}, \bibinfo{author}{Wang\xfnm[ J.S.]},
  \bibinfo{author}{Li\xfnm[ B.]}.
\newblock \bibinfo{title}{Topology-induced thermal rectification in carbon
  nanodevice}.
\newblock \bibinfo{journal}{Europhysics Letters}
  \bibinfo{year}{2010}{\natexlab{b}};\bibinfo{volume}{89}:\bibinfo{pages}{4600%
5}.
\bibitem[{Jiang et~al.(2010{\natexlab{c}})Jiang, Wang and
  Li}]{JiangJW2010topology}
\bibinfo{author}{Jiang\xfnm[ J.W.]}, \bibinfo{author}{Wang\xfnm[ J.S.]},
  \bibinfo{author}{Li\xfnm[ B.]}.
\newblock \bibinfo{title}{Topological effect on thermal conductivity in
  graphene}.
\newblock \bibinfo{journal}{Journal of Applied Physics}
  \bibinfo{year}{2010}{\natexlab{c}};\bibinfo{volume}{108}:\bibinfo{pages}{064%
307}.
\bibitem[{Hu et~al.(2011)Hu, Wang, Vallabhaneni, Ruan and Chen}]{HuJ2011apl}
\bibinfo{author}{Hu\xfnm[ J.]}, \bibinfo{author}{Wang\xfnm[ Y.]},
  \bibinfo{author}{Vallabhaneni\xfnm[ A.]}, \bibinfo{author}{Ruan\xfnm[ X.]},
  \bibinfo{author}{Chen\xfnm[ Y.P.]}.
\newblock \bibinfo{title}{Nonlinear thermal transport and negative differential
  thermal conductance in graphene nanoribbons}.
\newblock \bibinfo{journal}{Applied Physics Letters}
  \bibinfo{year}{2011};\bibinfo{volume}{99}(\bibinfo{number}{11}):\bibinfo{pag%
es}{113101}.
\bibitem[{Cheh and Zhao(2012)}]{ChehJ2012}
\bibinfo{author}{Cheh\xfnm[ J.]}, \bibinfo{author}{Zhao\xfnm[ H.]}.
\newblock \bibinfo{title}{Thermal rectification in asymmetric u-shaped graphene
  flakes}.
\newblock \bibinfo{journal}{Journal of Statistical Mechanics: Theory and
  Experiment}
  \bibinfo{year}{2012};\bibinfo{volume}{2012}:\bibinfo{pages}{06011}.
\bibitem[{Ghosh et~al.(2010)Ghosh, Bao, Nika, Subrina, Pokatilov, Lau
  et~al.}]{GhoshS}
\bibinfo{author}{Ghosh\xfnm[ S.]}, \bibinfo{author}{Bao\xfnm[ W.]},
  \bibinfo{author}{Nika\xfnm[ D.L.]}, \bibinfo{author}{Subrina\xfnm[ S.]},
  \bibinfo{author}{Pokatilov\xfnm[ E.P.]}, \bibinfo{author}{Lau\xfnm[ C.N.]},
  et~al.
\newblock \bibinfo{title}{Dimensional crossover of thermal transport in
  few-layer graphene}.
\newblock \bibinfo{journal}{Nature Materials}
  \bibinfo{year}{2010};\bibinfo{volume}{9}:\bibinfo{pages}{555--558}.
\bibitem[{Singh et~al.(2011)Singh, Murthy and Fisher}]{SinghD}
\bibinfo{author}{Singh\xfnm[ D.]}, \bibinfo{author}{Murthy\xfnm[ J.Y.]},
  \bibinfo{author}{Fisher\xfnm[ T.S.]}.
\newblock \bibinfo{title}{Mechanism of thermal conductivity reduction in
  few-layer graphene}.
\newblock \bibinfo{journal}{Journal of Applied Physics}
  \bibinfo{year}{2011};\bibinfo{volume}{110}(\bibinfo{number}{4}):\bibinfo{pag%
es}{044317}.
\bibitem[{Lindsay et~al.(2011)Lindsay, Broido and Mingo}]{LindsayL}
\bibinfo{author}{Lindsay\xfnm[ L.]}, \bibinfo{author}{Broido\xfnm[ D.A.]},
  \bibinfo{author}{Mingo\xfnm[ N.]}.
\newblock \bibinfo{title}{Flexural phonons and thermal transport in multilayer
  graphene and graphite}.
\newblock \bibinfo{journal}{Physical Review B}
  \bibinfo{year}{2011};\bibinfo{volume}{83}:\bibinfo{pages}{235428}.
\bibitem[{Kong et~al.(2009)Kong, Paul, Nardelli and Kim}]{KongBD}
\bibinfo{author}{Kong\xfnm[ B.D.]}, \bibinfo{author}{Paul\xfnm[ S.]},
  \bibinfo{author}{Nardelli\xfnm[ M.B.]}, \bibinfo{author}{Kim\xfnm[ K.W.]}.
\newblock \bibinfo{title}{First-principles analysis of lattice thermal
  conductivity in monolayer and bilayer graphene}.
\newblock \bibinfo{journal}{Physical Review B}
  \bibinfo{year}{2009};\bibinfo{volume}{80}:\bibinfo{pages}{033406}.
\bibitem[{Zhang and Zhang(2011)}]{ZhangG2011nns}
\bibinfo{author}{Zhang\xfnm[ G.]}, \bibinfo{author}{Zhang\xfnm[ H.]}.
\newblock \bibinfo{title}{Thermal conduction and rectification in few-layer
  graphene y junctions}.
\newblock \bibinfo{journal}{Nanoscale}
  \bibinfo{year}{2011};\bibinfo{volume}{3}:\bibinfo{pages}{4604}.
\bibitem[{Zhong et~al.(2011{\natexlab{a}})Zhong, Ai and Zheng}]{ZhongWR20111}
\bibinfo{author}{Zhong\xfnm[ W.R.]}, \bibinfo{author}{Ai\xfnm[ M.P.Z.B.Q.]},
  \bibinfo{author}{Zheng\xfnm[ D.Q.]}.
\newblock \bibinfo{title}{Chirality and thickness-dependent thermal
  conductivity of few-layer graphene: A molecular dynamics study}.
\newblock \bibinfo{journal}{Applied Physics Letters}
  \bibinfo{year}{2011}{\natexlab{a}};\bibinfo{volume}{98}(\bibinfo{number}{11}%
):\bibinfo{pages}{113107}.
\bibitem[{Zhong et~al.(2011{\natexlab{b}})Zhong, Huang, Deng and
  Ai}]{ZhongWR20112}
\bibinfo{author}{Zhong\xfnm[ W.R.]}, \bibinfo{author}{Huang\xfnm[ W.H.]},
  \bibinfo{author}{Deng\xfnm[ X.R.]}, \bibinfo{author}{Ai\xfnm[ B.Q.]}.
\newblock \bibinfo{title}{Thermal rectification in thickness-asymmetric
  graphene nanoribbons}.
\newblock \bibinfo{journal}{Applied Physics Letters}
  \bibinfo{year}{2011}{\natexlab{b}};\bibinfo{volume}{99}(\bibinfo{number}{19}%
):\bibinfo{pages}{193104}.
\bibitem[{Rajabpour and Allaei(2012)}]{RajabpourA}
\bibinfo{author}{Rajabpour\xfnm[ A.]}, \bibinfo{author}{Allaei\xfnm[ S.M.V.]}.
\newblock \bibinfo{title}{Tuning thermal conductivity of bilayer graphene by
  inter-layer sp3 bonding: A molecular dynamics study}.
\newblock \bibinfo{journal}{Applied Physics Letters}
  \bibinfo{year}{2012};\bibinfo{volume}{101}(\bibinfo{number}{5}):\bibinfo{pag%
es}{053115}.
\bibitem[{Cao et~al.(2012)Cao, Guo, Xiang and Gong}]{CaoHY}
\bibinfo{author}{Cao\xfnm[ H.Y.]}, \bibinfo{author}{Guo\xfnm[ Z.X.]},
  \bibinfo{author}{Xiang\xfnm[ H.]}, \bibinfo{author}{Gong\xfnm[ X.G.]}.
\newblock \bibinfo{title}{Layer and size dependence of thermal conductivity in
  multilayer graphene nanoribbons}.
\newblock \bibinfo{journal}{Physics Letters A}
  \bibinfo{year}{2012};\bibinfo{volume}{376}(\bibinfo{number}{9}):\bibinfo{pag%
es}{525--528}.
\bibitem[{Sun et~al.(2013)Sun, Wang and Kang}]{SunT}
\bibinfo{author}{Sun\xfnm[ T.]}, \bibinfo{author}{Wang\xfnm[ J.]},
  \bibinfo{author}{Kang\xfnm[ W.]}.
\newblock \bibinfo{title}{Van der waals interaction-tuned heat transfer in
  nanostructures}.
\newblock \bibinfo{journal}{Nanoscale}
  \bibinfo{year}{2013};\bibinfo{volume}{5}:\bibinfo{pages}{128}.
\bibitem[{Jiang(2014{\natexlab{b}})}]{JiangJW2014flgthermal}
\bibinfo{author}{Jiang\xfnm[ J.W.]}.
\newblock \bibinfo{title}{Registry effect on the thermal conductivity of
  few-layer graphene}.
\newblock \bibinfo{journal}{Journal of Applied Physics}
  \bibinfo{year}{2014}{\natexlab{b}};\bibinfo{volume}{116}:\bibinfo{pages}{164%
313}.
\bibitem[{Yan et~al.(2012)Yan, Liu, Khan and Balandin}]{YangZ}
\bibinfo{author}{Yan\xfnm[ Z.]}, \bibinfo{author}{Liu\xfnm[ G.]},
  \bibinfo{author}{Khan\xfnm[ J.M.]}, \bibinfo{author}{Balandin\xfnm[ A.A.]}.
\newblock \bibinfo{title}{Graphene quilts for thermal management of high-power
  gan transistors}.
\newblock \bibinfo{journal}{Nature Communications}
  \bibinfo{year}{2012};\bibinfo{volume}{3}:\bibinfo{pages}{827}.
\bibitem[{Goyal and Balandin(2012)}]{GoyalV}
\bibinfo{author}{Goyal\xfnm[ V.]}, \bibinfo{author}{Balandin\xfnm[ A.A.]}.
\newblock \bibinfo{title}{Thermal properties of the hybrid graphene-metal
  nano-micro-composites: Applications in thermal interface materials}.
\newblock \bibinfo{journal}{Applied Physics Letters}
  \bibinfo{year}{2012};\bibinfo{volume}{100}(\bibinfo{number}{7}):\bibinfo{pag%
es}{073113}.
\bibitem[{Shahil and Balandin(2012)}]{ShahilKMF}
\bibinfo{author}{Shahil\xfnm[ K.M.F.]}, \bibinfo{author}{Balandin\xfnm[ A.A.]}.
\newblock \bibinfo{title}{Graphene−multilayer graphene nanocomposites as
  highly efficient thermal interface materials}.
\newblock \bibinfo{journal}{Nano Letters}
  \bibinfo{year}{2012};\bibinfo{volume}{12}:\bibinfo{pages}{861--867}.
\bibitem[{Li et~al.(2010)Li, Maute, Dunn and Yang}]{LiX}
\bibinfo{author}{Li\xfnm[ X.]}, \bibinfo{author}{Maute\xfnm[ K.]},
  \bibinfo{author}{Dunn\xfnm[ M.L.]}, \bibinfo{author}{Yang\xfnm[ R.]}.
\newblock \bibinfo{title}{Strain effects on the thermal conductivity of
  nanostructures}.
\newblock \bibinfo{journal}{Physical Review B}
  \bibinfo{year}{2010};\bibinfo{volume}{81}(\bibinfo{number}{24}):\bibinfo{pag%
es}{245318}.
\bibitem[{Jiang et~al.(2011{\natexlab{a}})Jiang, Wang and Li}]{JiangJW2011negf}
\bibinfo{author}{Jiang\xfnm[ J.W.]}, \bibinfo{author}{Wang\xfnm[ J.S.]},
  \bibinfo{author}{Li\xfnm[ B.]}.
\newblock \bibinfo{title}{A nonequilibrium green’s function study of
  thermoelectric properties in single-walled carbon nanotubes}.
\newblock \bibinfo{journal}{Journal of Applied Physics}
  \bibinfo{year}{2011}{\natexlab{a}};\bibinfo{volume}{109}:\bibinfo{pages}{014%
326}.
\bibitem[{Wei et~al.(2011)Wei, Xu, Wang and Zheng}]{WeiN}
\bibinfo{author}{Wei\xfnm[ N.]}, \bibinfo{author}{Xu\xfnm[ L.]},
  \bibinfo{author}{Wang\xfnm[ H.Q.]}, \bibinfo{author}{Zheng\xfnm[ J.C.]}.
\newblock \bibinfo{title}{Strain engineering of thermal conductivity in
  graphene sheets and nanoribbons: a demonstration of magic flexibility}.
\newblock \bibinfo{journal}{Nanotechnology}
  \bibinfo{year}{2011};\bibinfo{volume}{22}(\bibinfo{number}{10}):\bibinfo{pag%
es}{105705}.
\bibitem[{Gunawardana et~al.(2012)Gunawardana, Mullen, Hu, Chen and
  Ruan}]{GunawardanaKGSH}
\bibinfo{author}{Gunawardana\xfnm[ K.G.S.H.]}, \bibinfo{author}{Mullen\xfnm[
  K.]}, \bibinfo{author}{Hu\xfnm[ J.]}, \bibinfo{author}{Chen\xfnm[ Y.P.]},
  \bibinfo{author}{Ruan\xfnm[ X.]}.
\newblock \bibinfo{title}{Tunable thermal transport and thermal rectification
  in strained graphene nanoribbons}.
\newblock \bibinfo{journal}{Physical Review B}
  \bibinfo{year}{2012};\bibinfo{volume}{85}(\bibinfo{number}{24}):\bibinfo{pag%
es}{245417}.
\bibitem[{Zhang and Wang(2013)}]{ZhangJ}
\bibinfo{author}{Zhang\xfnm[ J.]}, \bibinfo{author}{Wang\xfnm[ X.]}.
\newblock \bibinfo{title}{Thermal transport in bent graphene nanoribbons}.
\newblock \bibinfo{journal}{Nanoscale}
  \bibinfo{year}{2013};\bibinfo{volume}{5}(\bibinfo{number}{2}):\bibinfo{pages%
}{734--743}.
\bibitem[{Jiang et~al.(2009{\natexlab{b}})Jiang, Chen, Wang and
  Li}]{JiangJW2009edge}
\bibinfo{author}{Jiang\xfnm[ J.W.]}, \bibinfo{author}{Chen\xfnm[ J.]},
  \bibinfo{author}{Wang\xfnm[ J.S.]}, \bibinfo{author}{Li\xfnm[ B.]}.
\newblock \bibinfo{title}{Edge states induce boundary temperature jump in
  molecular dynamics simulation of heat conduction}.
\newblock \bibinfo{journal}{Physical Review B}
  \bibinfo{year}{2009}{\natexlab{b}};\bibinfo{volume}{80}(\bibinfo{number}{5})%
:\bibinfo{pages}{052301}.
\bibitem[{Lan et~al.(2009)Lan, Wang, Gan and Chin}]{LanJH2009}
\bibinfo{author}{Lan\xfnm[ J.]}, \bibinfo{author}{Wang\xfnm[ J.S.]},
  \bibinfo{author}{Gan\xfnm[ C.K.]}, \bibinfo{author}{Chin\xfnm[ S.K.]}.
\newblock \bibinfo{title}{Edge effects on quantum thermal transport in graphene
  nanoribbons: Tight-binding calculations}.
\newblock \bibinfo{journal}{Physical Review B}
  \bibinfo{year}{2009};\bibinfo{volume}{79}(\bibinfo{number}{11}):\bibinfo{pag%
es}{115401}.
\bibitem[{Jiang and Wang(2010{\natexlab{b}})}]{JiangJW2010edgemode}
\bibinfo{author}{Jiang\xfnm[ J.W.]}, \bibinfo{author}{Wang\xfnm[ J.S.]}.
\newblock \bibinfo{title}{Conditions for the existence of phonon localized
  edge-modes}.
\newblock \bibinfo{journal}{Physical Review B}
  \bibinfo{year}{2010}{\natexlab{b}};\bibinfo{volume}{81}(\bibinfo{number}{17}%
):\bibinfo{pages}{174117}.
\bibitem[{Savin et~al.(2010)Savin, Kivshar and Hu}]{SavinAV}
\bibinfo{author}{Savin\xfnm[ A.V.]}, \bibinfo{author}{Kivshar\xfnm[ Y.S.]},
  \bibinfo{author}{Hu\xfnm[ B.]}.
\newblock \bibinfo{title}{Suppression of thermal conductivity in graphene
  nanoribbons with rough edges}.
\newblock \bibinfo{journal}{Physical Review B}
  \bibinfo{year}{2010};\bibinfo{volume}{82}(\bibinfo{number}{19}):\bibinfo{pag%
es}{195422}.
\bibitem[{Tan et~al.(2010)Tan, Wang and Gan}]{TanZW}
\bibinfo{author}{Tan\xfnm[ Z.W.]}, \bibinfo{author}{Wang\xfnm[ J.S.]},
  \bibinfo{author}{Gan\xfnm[ C.K.]}.
\newblock \bibinfo{title}{First-principles study of heat transport properties
  of graphene nanoribbons}.
\newblock \bibinfo{journal}{Nano Letters}
  \bibinfo{year}{2010};\bibinfo{volume}{11}(\bibinfo{number}{1}):\bibinfo{page%
s}{214--219}.
\bibitem[{Hu et~al.(2010)Hu, Schiffli, Vallabhaneni, Ruan and
  Chen}]{HuJ2010apl}
\bibinfo{author}{Hu\xfnm[ J.]}, \bibinfo{author}{Schiffli\xfnm[ S.]},
  \bibinfo{author}{Vallabhaneni\xfnm[ A.]}, \bibinfo{author}{Ruan\xfnm[ X.]},
  \bibinfo{author}{Chen\xfnm[ Y.P.]}.
\newblock \bibinfo{title}{Tuning the thermal conductivity of graphene
  nanoribbons by edge passivation and isotope engineering: A molecular dynamics
  study}.
\newblock \bibinfo{journal}{Applied Physics Letters}
  \bibinfo{year}{2010};\bibinfo{volume}{97}:\bibinfo{pages}{133107}.
\bibitem[{Jiang et~al.(2011{\natexlab{b}})Jiang, Wang and
  Wang}]{JiangJW2011minimum}
\bibinfo{author}{Jiang\xfnm[ J.W.]}, \bibinfo{author}{Wang\xfnm[ J.S.]},
  \bibinfo{author}{Wang\xfnm[ B.S.]}.
\newblock \bibinfo{title}{Minimum thermal conductance in graphene and boron
  nitride superlattice}.
\newblock \bibinfo{journal}{Applied Physics Letters}
  \bibinfo{year}{2011}{\natexlab{b}};\bibinfo{volume}{99}:\bibinfo{pages}{0431%
09}.
\bibitem[{Jiang and Wang(2011{\natexlab{a}})}]{JiangJW2011bngra}
\bibinfo{author}{Jiang\xfnm[ J.W.]}, \bibinfo{author}{Wang\xfnm[ J.S.]}.
\newblock \bibinfo{title}{Manipulation of heat current by the interface between
  graphene and white graphene}.
\newblock \bibinfo{journal}{Europhysics Letters}
  \bibinfo{year}{2011}{\natexlab{a}};\bibinfo{volume}{96}:\bibinfo{pages}{1600%
3}.
\bibitem[{Cheh and Zhao(2011)}]{ChehJ}
\bibinfo{author}{Cheh\xfnm[ J.]}, \bibinfo{author}{Zhao\xfnm[ H.]}.
\newblock \bibinfo{title}{Heat conduction in graphene flakes with inhomogeneous
  mass interface}.
\newblock \bibinfo{journal}{Journal of Statistical Mechanics: Theory and
  Experiment}
  \bibinfo{year}{2011};\bibinfo{volume}{2011}:\bibinfo{pages}{10031}.
\bibitem[{Jiang et~al.(2011{\natexlab{c}})Jiang, Wang and
  Wang}]{JiangJW2011defect}
\bibinfo{author}{Jiang\xfnm[ J.W.]}, \bibinfo{author}{Wang\xfnm[ B.S.]},
  \bibinfo{author}{Wang\xfnm[ J.S.]}.
\newblock \bibinfo{title}{First principle study of the thermal conductance in
  graphene nanoribbon with vacancy and substitutional silicon defects}.
\newblock \bibinfo{journal}{Applied Physics Letters}
  \bibinfo{year}{2011}{\natexlab{c}};\bibinfo{volume}{98}:\bibinfo{pages}{1131%
14}.
\bibitem[{Hao et~al.(2011)Hao, Fang and Xu}]{HaoF}
\bibinfo{author}{Hao\xfnm[ F.]}, \bibinfo{author}{Fang\xfnm[ D.]},
  \bibinfo{author}{Xu\xfnm[ Z.]}.
\newblock \bibinfo{title}{Mechanical and thermal transport properties of
  graphene with defects}.
\newblock \bibinfo{journal}{Applied Physics Letters}
  \bibinfo{year}{2011};\bibinfo{volume}{99}:\bibinfo{pages}{041901}.
\bibitem[{Zhang et~al.(2011)Zhang, Lee and Cho}]{ZhangH}
\bibinfo{author}{Zhang\xfnm[ H.]}, \bibinfo{author}{Lee\xfnm[ G.]},
  \bibinfo{author}{Cho\xfnm[ K.]}.
\newblock \bibinfo{title}{Thermal transport in graphene and effects of vacancy
  defects}.
\newblock \bibinfo{journal}{Physical Review B}
  \bibinfo{year}{2011};\bibinfo{volume}{84}:\bibinfo{pages}{115460}.
\bibitem[{Adamyan and Zavalniuk(2012)}]{AdamyanV}
\bibinfo{author}{Adamyan\xfnm[ V.]}, \bibinfo{author}{Zavalniuk\xfnm[ V.]}.
\newblock \bibinfo{title}{Lattice thermal conductivity of graphene with
  conventionally isotopic defects}.
\newblock \bibinfo{journal}{Journal of Physics: Condensed Matter}
  \bibinfo{year}{2012};\bibinfo{volume}{24}:\bibinfo{pages}{415401}.
\bibitem[{Serov et~al.(2013)Serov, Ong and Pop}]{SerovAY}
\bibinfo{author}{Serov\xfnm[ A.Y.]}, \bibinfo{author}{Ong\xfnm[ Z.Y.]},
  \bibinfo{author}{Pop\xfnm[ E.]}.
\newblock \bibinfo{title}{Effect of grain boundaries on thermal transport in
  graphene}.
\newblock \bibinfo{journal}{Applied Physics Letters}
  \bibinfo{year}{2013};\bibinfo{volume}{102}:\bibinfo{pages}{033104}.
\bibitem[{Chen et~al.(2012{\natexlab{b}})Chen, Li, Zhang, Qu, Ji, Ruoff
  et~al.}]{ChenSS2012nano}
\bibinfo{author}{Chen\xfnm[ S.]}, \bibinfo{author}{Li\xfnm[ Q.]},
  \bibinfo{author}{Zhang\xfnm[ Q.]}, \bibinfo{author}{Qu\xfnm[ Y.]},
  \bibinfo{author}{Ji\xfnm[ H.]}, \bibinfo{author}{Ruoff\xfnm[ R.S.]}, et~al.
\newblock \bibinfo{title}{Thermal conductivity measurements of suspended
  graphene with and without wrinkles by micro-raman mapping}.
\newblock \bibinfo{journal}{Nanotechnology}
  \bibinfo{year}{2012}{\natexlab{b}};\bibinfo{volume}{23}:\bibinfo{pages}{3657%
01}.
\bibitem[{Cai et~al.(2010)Cai, Moore, Zhu, Li, Chen, Shi et~al.}]{CaiWW2010nl}
\bibinfo{author}{Cai\xfnm[ W.]}, \bibinfo{author}{Moore\xfnm[ A.L.]},
  \bibinfo{author}{Zhu\xfnm[ Y.]}, \bibinfo{author}{Li\xfnm[ X.]},
  \bibinfo{author}{Chen\xfnm[ S.]}, \bibinfo{author}{Shi\xfnm[ L.]}, et~al.
\newblock \bibinfo{title}{Thermal transport in suspended and supported
  monolayer graphene grown by chemical vapor deposition}.
\newblock \bibinfo{journal}{Nano Letters}
  \bibinfo{year}{2010};\bibinfo{volume}{10}:\bibinfo{pages}{1645--1651}.
\bibitem[{Chen et~al.(2011)Chen, Moore, Cai, Suk, An, Mishra
  et~al.}]{ChenSS2011acsn}
\bibinfo{author}{Chen\xfnm[ S.]}, \bibinfo{author}{Moore\xfnm[ A.L.]},
  \bibinfo{author}{Cai\xfnm[ W.]}, \bibinfo{author}{Suk\xfnm[ J.W.]},
  \bibinfo{author}{An\xfnm[ J.]}, \bibinfo{author}{Mishra\xfnm[ C.]}, et~al.
\newblock \bibinfo{title}{Raman measurements of thermal transport in suspended
  monolayer graphene of variable sizes in vacuum and gaseous environments}.
\newblock \bibinfo{journal}{ACS Nano}
  \bibinfo{year}{2011};\bibinfo{volume}{5}(\bibinfo{number}{1}):\bibinfo{pages%
}{321--328}.
\bibitem[{Lee et~al.(2011)Lee, Yoon, Kim, Lee and Cheong}]{LeeJU}
\bibinfo{author}{Lee\xfnm[ J.U.]}, \bibinfo{author}{Yoon\xfnm[ D.]},
  \bibinfo{author}{Kim\xfnm[ H.]}, \bibinfo{author}{Lee\xfnm[ S.W.]},
  \bibinfo{author}{Cheong\xfnm[ H.]}.
\newblock \bibinfo{title}{Thermal conductivity of suspended pristine graphene
  measured by raman spectroscopy}.
\newblock \bibinfo{journal}{Physical Review B}
  \bibinfo{year}{2011};\bibinfo{volume}{83}:\bibinfo{pages}{081419(R)}.
\bibitem[{Guo et~al.(2011)Guo, Zhang and Gong}]{GuoZX2011}
\bibinfo{author}{Guo\xfnm[ Z.X.]}, \bibinfo{author}{Zhang\xfnm[ D.]},
  \bibinfo{author}{Gong\xfnm[ X.G.]}.
\newblock \bibinfo{title}{Manipulating thermal conductivity through substrate
  coupling}.
\newblock \bibinfo{journal}{Physical Review B}
  \bibinfo{year}{2011};\bibinfo{volume}{84}:\bibinfo{pages}{075470}.
\bibitem[{Ong and Pop(2011)}]{OngZY2011}
\bibinfo{author}{Ong\xfnm[ Z.Y.]}, \bibinfo{author}{Pop\xfnm[ E.]}.
\newblock \bibinfo{title}{Effect of substrate modes on thermal transport in
  supported graphene}.
\newblock \bibinfo{journal}{Physical Review B}
  \bibinfo{year}{2011};\bibinfo{volume}{84}(\bibinfo{number}{7}):\bibinfo{page%
s}{075471}.
\bibitem[{Huang et~al.(2011)Huang, Wang, Kwo and Chien}]{HuangSY}
\bibinfo{author}{Huang\xfnm[ S.Y.]}, \bibinfo{author}{Wang\xfnm[ W.G.]},
  \bibinfo{author}{Kwo\xfnm[ S.F.L.J.]}, \bibinfo{author}{Chien\xfnm[ C.L.]}.
\newblock \bibinfo{title}{Intrinsic spin-dependent thermal transport}.
\newblock \bibinfo{journal}{Physical Review Letters}
  \bibinfo{year}{2011};\bibinfo{volume}{107}:\bibinfo{pages}{216604}.
\bibitem[{Qiu and Ruan(2012)}]{QiuB}
\bibinfo{author}{Qiu\xfnm[ B.]}, \bibinfo{author}{Ruan\xfnm[ X.]}.
\newblock \bibinfo{title}{Reduction of spectral phonon relaxation times from
  suspended to supported graphene}.
\newblock \bibinfo{journal}{Applied Physics Letters}
  \bibinfo{year}{2012};\bibinfo{volume}{100}(\bibinfo{number}{19}):\bibinfo{pa%
ges}{193101}.
\bibitem[{Guo et~al.(2012)Guo, Ding and Gong}]{GuoZX}
\bibinfo{author}{Guo\xfnm[ Z.X.]}, \bibinfo{author}{Ding\xfnm[ J.W.]},
  \bibinfo{author}{Gong\xfnm[ X.G.]}.
\newblock \bibinfo{title}{Substrate effects on the thermal conductivity of
  epitaxial graphene nanoribbons}.
\newblock \bibinfo{journal}{Physical Review B}
  \bibinfo{year}{2012};\bibinfo{volume}{85}(\bibinfo{number}{23}):\bibinfo{pag%
es}{235429}.
\bibitem[{Chen et~al.(2012{\natexlab{c}})Chen, Zhang, Wang and
  Zhao}]{ChenS2012arxiv}
\bibinfo{author}{Chen\xfnm[ S.]}, \bibinfo{author}{Zhang\xfnm[ Y.]},
  \bibinfo{author}{Wang\xfnm[ J.]}, \bibinfo{author}{Zhao\xfnm[ H.]}.
\newblock \bibinfo{title}{Breakdown of the power-law decay prediction of the
  heat current correlation in one-dimensional momentum conserving lattices}.
\newblock \bibinfo{journal}{arXiv:12045933}
  \bibinfo{year}{2012}{\natexlab{c}};.
\bibitem[{Chen et~al.(2013{\natexlab{a}})Chen, Zhang, Wang and
  Zhao}]{ChenS2013arxiv}
\bibinfo{author}{Chen\xfnm[ S.]}, \bibinfo{author}{Zhang\xfnm[ Y.]},
  \bibinfo{author}{Wang\xfnm[ J.]}, \bibinfo{author}{Zhao\xfnm[ H.]}.
\newblock \bibinfo{title}{Why asymmetric interparticle interaction can result
  in convergent heat conductivity}.
\newblock \bibinfo{journal}{arXiv:13097146}
  \bibinfo{year}{2013}{\natexlab{a}};.
\bibitem[{Wang et~al.(2008)Wang, Wang and L$\ddot{u}$}]{WangJSnegf}
\bibinfo{author}{Wang\xfnm[ J.S.]}, \bibinfo{author}{Wang\xfnm[ J.]},
  \bibinfo{author}{L$\ddot{u}$\xfnm[ J.T.]}.
\newblock \bibinfo{title}{Quantum thermal transport in nanostructures}.
\newblock \bibinfo{journal}{Eur Phys J B}
  \bibinfo{year}{2008};\bibinfo{volume}{62}(\bibinfo{number}{4}):\bibinfo{page%
s}{381--404}.
\bibitem[{Wang et~al.(2013)Wang, Agarwalla, Li and Thingna}]{WangJS2013fp}
\bibinfo{author}{Wang\xfnm[ J.S.]}, \bibinfo{author}{Agarwalla\xfnm[ B.K.]},
  \bibinfo{author}{Li\xfnm[ H.]}, \bibinfo{author}{Thingna\xfnm[ J.]}.
\newblock \bibinfo{title}{Nonequilibrium green’s function method for quantum
  thermal transport}.
\newblock \bibinfo{journal}{Front Phys}
  \bibinfo{year}{2013};:\bibinfo{pages}{DOI:10.1007/s11467--013--0340--x}.
\bibitem[{Nika and Balandin(2012)}]{NikaDL2012jpcm}
\bibinfo{author}{Nika\xfnm[ D.L.]}, \bibinfo{author}{Balandin\xfnm[ A.A.]}.
\newblock \bibinfo{title}{Two-dimensional phonon transport in graphene}.
\newblock \bibinfo{journal}{Journal of Physics: Condensed Matter}
  \bibinfo{year}{2012};\bibinfo{volume}{24}(\bibinfo{number}{23}):\bibinfo{pag%
es}{233203}.
\bibitem[{Cahill et~al.(2003)Cahill, Ford, Goodson, Mahan, Majumdar, Maris
  et~al.}]{CahillDG}
\bibinfo{author}{Cahill\xfnm[ D.G.]}, \bibinfo{author}{Ford\xfnm[ W.K.]},
  \bibinfo{author}{Goodson\xfnm[ K.E.]}, \bibinfo{author}{Mahan\xfnm[ G.D.]},
  \bibinfo{author}{Majumdar\xfnm[ A.]}, \bibinfo{author}{Maris\xfnm[ H.J.]},
  et~al.
\newblock \bibinfo{title}{Nanoscale thermal transport}.
\newblock \bibinfo{journal}{Journal of Applied Physics}
  \bibinfo{year}{2003};\bibinfo{volume}{93}(\bibinfo{number}{2}):\bibinfo{page%
s}{793}.
\bibitem[{Dhar(2008)}]{DharA2008}
\bibinfo{author}{Dhar\xfnm[ A.]}.
\newblock \bibinfo{title}{Heat transport in low-dimensional systems}.
\newblock \bibinfo{journal}{Advances in Physics}
  \bibinfo{year}{2008};\bibinfo{volume}{57}(\bibinfo{number}{5}):\bibinfo{page%
s}{457--537}.
\bibitem[{Liu et~al.(2012)Liu, Xu, Xie, Zhang and Li}]{LiuS}
\bibinfo{author}{Liu\xfnm[ S.]}, \bibinfo{author}{Xu\xfnm[ X.]},
  \bibinfo{author}{Xie\xfnm[ R.]}, \bibinfo{author}{Zhang\xfnm[ G.]},
  \bibinfo{author}{Li\xfnm[ B.]}.
\newblock \bibinfo{title}{Anomalous heat conduction and anomalous diffusion in
  low dimensional nanoscale systems}.
\newblock \bibinfo{journal}{Eur Phys J B}
  \bibinfo{year}{2012};\bibinfo{volume}{85}:\bibinfo{pages}{337}.
\bibitem[{Li et~al.(2012)Li, Ren, Wang, Zhang, H$\ddot{a}$nggi and
  Li}]{LiNB2012rmp}
\bibinfo{author}{Li\xfnm[ N.]}, \bibinfo{author}{Ren\xfnm[ J.]},
  \bibinfo{author}{Wang\xfnm[ L.]}, \bibinfo{author}{Zhang\xfnm[ G.]},
  \bibinfo{author}{H$\ddot{a}$nggi\xfnm[ P.]}, \bibinfo{author}{Li\xfnm[ B.]}.
\newblock \bibinfo{title}{Colloquium: Phononics: Manipulating heat flow with
  electronic analogs and beyond}.
\newblock \bibinfo{journal}{Rev Mod Phys}
  \bibinfo{year}{2012};\bibinfo{volume}{84}:\bibinfo{pages}{1045--1066}.
\bibitem[{Yang et~al.(2012{\natexlab{a}})Yang, Xu, Zhang and
  Li}]{YangN2012aipa}
\bibinfo{author}{Yang\xfnm[ N.]}, \bibinfo{author}{Xu\xfnm[ X.]},
  \bibinfo{author}{Zhang\xfnm[ G.]}, \bibinfo{author}{Li\xfnm[ B.]}.
\newblock \bibinfo{title}{Thermal transport in nanostructures}.
\newblock \bibinfo{journal}{AIP Advances}
  \bibinfo{year}{2012}{\natexlab{a}};\bibinfo{volume}{2}(\bibinfo{number}{4}):%
\bibinfo{pages}{041410}.
\bibitem[{Luo and Chen(2013)}]{LuoTF}
\bibinfo{author}{Luo\xfnm[ T.]}, \bibinfo{author}{Chen\xfnm[ G.]}.
\newblock \bibinfo{title}{anoscale heat transfer - from computation to
  experiment}.
\newblock \bibinfo{journal}{Phys Chem Chem Phys}
  \bibinfo{year}{2013};\bibinfo{volume}{15}(\bibinfo{number}{10}):\bibinfo{pag%
es}{3389--3412}.
\bibitem[{Marconnet et~al.(2013)Marconnet, Panzer and Goodson}]{MarconnetAM}
\bibinfo{author}{Marconnet\xfnm[ A.M.]}, \bibinfo{author}{Panzer\xfnm[ M.A.]},
  \bibinfo{author}{Goodson\xfnm[ K.E.]}.
\newblock \bibinfo{title}{Thermal conduction phenomena in carbon nanotubes and
  related nanostructured materials}.
\newblock \bibinfo{journal}{Rev Mod Phys}
  \bibinfo{year}{2013};\bibinfo{volume}{85}:\bibinfo{pages}{1295--1326}.
\bibitem[{Dubi and Ventra(2011)}]{DubiY}
\bibinfo{author}{Dubi\xfnm[ Y.]}, \bibinfo{author}{Ventra\xfnm[ M.D.]}.
\newblock \bibinfo{title}{Colloquium: Heat flow and thermoelectricity in atomic
  and molecular junctions}.
\newblock \bibinfo{journal}{Rev Mod Phys}
  \bibinfo{year}{2011};\bibinfo{volume}{83}:\bibinfo{pages}{131}.
\bibitem[{Zhang and Li(2010)}]{ZhangG2010nns}
\bibinfo{author}{Zhang\xfnm[ G.]}, \bibinfo{author}{Li\xfnm[ B.]}.
\newblock \bibinfo{title}{Impacts of doping on thermal and thermoelectric
  properties of nanomaterials}.
\newblock \bibinfo{journal}{Nanoscale}
  \bibinfo{year}{2010};\bibinfo{volume}{2}(\bibinfo{number}{7}):\bibinfo{pages%
}{1058--1068}.
\bibitem[{Pop(2010)}]{PopE2010}
\bibinfo{author}{Pop\xfnm[ E.]}.
\newblock \bibinfo{title}{Energy dissipation and transport in nanoscale
  devices}.
\newblock \bibinfo{journal}{Nano Research}
  \bibinfo{year}{2010};\bibinfo{volume}{3}(\bibinfo{number}{3}):\bibinfo{pages%
}{147--169}.
\bibitem[{Ozpineci and Ciraci(2001)}]{OzpineciA}
\bibinfo{author}{Ozpineci\xfnm[ A.]}, \bibinfo{author}{Ciraci\xfnm[ S.]}.
\newblock \bibinfo{title}{Quantum effects of thermal conductance through atomic
  chains}.
\newblock \bibinfo{journal}{Physical Review B}
  \bibinfo{year}{2001};\bibinfo{volume}{63}(\bibinfo{number}{12}):\bibinfo{pag%
es}{125415}.
\bibitem[{Mingo and Yang(2003)}]{MingoN2003prb}
\bibinfo{author}{Mingo\xfnm[ N.]}, \bibinfo{author}{Yang\xfnm[ L.]}.
\newblock \bibinfo{title}{Phonon transport in nanowires coated with an
  amorphous material: An atomistic green’s function approach}.
\newblock \bibinfo{journal}{Physical Review B}
  \bibinfo{year}{2003};\bibinfo{volume}{68}(\bibinfo{number}{24}):\bibinfo{pag%
es}{245406}.
\bibitem[{Yamamoto and Watanabe(2006)}]{YamamotoT}
\bibinfo{author}{Yamamoto\xfnm[ T.]}, \bibinfo{author}{Watanabe\xfnm[ K.]}.
\newblock \bibinfo{title}{Nonequilibrium green’s function approach to phonon
  transport in defective carbon nanotubes}.
\newblock \bibinfo{journal}{Physical Review Letters}
  \bibinfo{year}{2006};\bibinfo{volume}{96}(\bibinfo{number}{25}):\bibinfo{pag%
es}{255503}.
\bibitem[{Mingo(2006)}]{MingoN2006prb}
\bibinfo{author}{Mingo\xfnm[ N.]}.
\newblock \bibinfo{title}{Anharmonic phonon flow through molecular-sized
  junctions}.
\newblock \bibinfo{journal}{Physical Review B}
  \bibinfo{year}{2006};\bibinfo{volume}{74}(\bibinfo{number}{12}):\bibinfo{pag%
es}{125402}.
\bibitem[{Spohn(2006)}]{SpohnH}
\bibinfo{author}{Spohn\xfnm[ H.]}.
\newblock \bibinfo{title}{The phonon boltzmann equation, properties and link to
  weakly anharmonic lattice dynamics}.
\newblock \bibinfo{journal}{Journal of Statistical Physics}
  \bibinfo{year}{2006};\bibinfo{volume}{124}:\bibinfo{pages}{1041--1104}.
\bibitem[{Lepri et~al.(2003)Lepri, Livi and Politi}]{LepriS}
\bibinfo{author}{Lepri\xfnm[ S.]}, \bibinfo{author}{Livi\xfnm[ R.]},
  \bibinfo{author}{Politi\xfnm[ A.]}.
\newblock \bibinfo{title}{Thermal conduction in classical low-dimensional
  lattices}.
\newblock \bibinfo{journal}{Physical Review}
  \bibinfo{year}{2003};\bibinfo{volume}{337}(\bibinfo{number}{1}):\bibinfo{pag%
es}{1--80}.
\bibitem[{Wang(2007)}]{WangJS2007prl}
\bibinfo{author}{Wang\xfnm[ J.S.]}.
\newblock \bibinfo{title}{Quantum thermal transport from classical molecular
  dynamics}.
\newblock \bibinfo{journal}{Physical Review Letters}
  \bibinfo{year}{2007};\bibinfo{volume}{99}:\bibinfo{pages}{160601}.
\bibitem[{Wang et~al.(2009)Wang, Ni and Jiang}]{WangJS2009prb}
\bibinfo{author}{Wang\xfnm[ J.S.]}, \bibinfo{author}{Ni\xfnm[ X.]},
  \bibinfo{author}{Jiang\xfnm[ J.W.]}.
\newblock \bibinfo{title}{Molecular dynamics with quantum heat baths:
  Application to nanoribbons and nanotubes}.
\newblock \bibinfo{journal}{Physical Review B}
  \bibinfo{year}{2009};\bibinfo{volume}{80}(\bibinfo{number}{22}):\bibinfo{pag%
es}{224302}.
\bibitem[{Ceriotti et~al.(2009{\natexlab{a}})Ceriotti, Bussi and
  Parrinello}]{CeriottiM}
\bibinfo{author}{Ceriotti\xfnm[ M.]}, \bibinfo{author}{Bussi\xfnm[ G.]},
  \bibinfo{author}{Parrinello\xfnm[ M.]}.
\newblock \bibinfo{title}{Langevin equation with colored noise for
  constant-temperature molecular dynamics simulations}.
\newblock \bibinfo{journal}{Physical Review Letters}
  \bibinfo{year}{2009}{\natexlab{a}};\bibinfo{volume}{102}:\bibinfo{pages}{020%
601}.
\bibitem[{Ceriotti et~al.(2009{\natexlab{b}})Ceriotti, Bussi and
  Parrinello}]{CeriottiM2}
\bibinfo{author}{Ceriotti\xfnm[ M.]}, \bibinfo{author}{Bussi\xfnm[ G.]},
  \bibinfo{author}{Parrinello\xfnm[ M.]}.
\newblock \bibinfo{title}{Nuclear quantum effects in solids using a
  colored-noise thermostat}.
\newblock \bibinfo{journal}{Physical Review Letters}
  \bibinfo{year}{2009}{\natexlab{b}};\bibinfo{volume}{103}:\bibinfo{pages}{030%
603}.
\bibitem[{Dammak et~al.(2009)Dammak, Chalopin, Laroche, Hayoun and
  Greffet}]{DammakH}
\bibinfo{author}{Dammak\xfnm[ H.]}, \bibinfo{author}{Chalopin\xfnm[ Y.]},
  \bibinfo{author}{Laroche\xfnm[ M.]}, \bibinfo{author}{Hayoun\xfnm[ M.]},
  \bibinfo{author}{Greffet\xfnm[ J.J.]}.
\newblock \bibinfo{title}{Quantum thermal bath for molecular dynamics
  simulation}.
\newblock \bibinfo{journal}{Physical Review Letters}
  \bibinfo{year}{2009};\bibinfo{volume}{103}:\bibinfo{pages}{190601}.
\bibitem[{Lammps(2012)}]{lammps}
\bibinfo{author}{Lammps\xfnm[]}.
\newblock \bibinfo{journal}{http://wwwcssandiagov/$\sim$sjplimp/lammpshtml}
  \bibinfo{year}{2012};.
\bibitem[{Chang et~al.(2008)Chang, Okawa, Garcia, Majumdar and
  Zettl}]{ChangCW2008prl}
\bibinfo{author}{Chang\xfnm[ C.W.]}, \bibinfo{author}{Okawa\xfnm[ D.]},
  \bibinfo{author}{Garcia\xfnm[ H.]}, \bibinfo{author}{Majumdar\xfnm[ A.]},
  \bibinfo{author}{Zettl\xfnm[ A.]}.
\newblock \bibinfo{title}{Breakdown of fourier’s law in nanotube thermal
  conductors}.
\newblock \bibinfo{journal}{Physical Review Letters}
  \bibinfo{year}{2008};\bibinfo{volume}{101}:\bibinfo{pages}{075903}.
\bibitem[{Dubi and Ventra(2009)}]{DubiY2009pre}
\bibinfo{author}{Dubi\xfnm[ Y.]}, \bibinfo{author}{Ventra\xfnm[ M.D.]}.
\newblock \bibinfo{title}{Fourier’s law: Insight from a simple derivation}.
\newblock \bibinfo{journal}{Physical Review E}
  \bibinfo{year}{2009};\bibinfo{volume}{79}:\bibinfo{pages}{042101}.
\bibitem[{Yang et~al.(2010)Yang, Zhang and Li}]{YangN2010nt}
\bibinfo{author}{Yang\xfnm[ N.]}, \bibinfo{author}{Zhang\xfnm[ G.]},
  \bibinfo{author}{Li\xfnm[ B.]}.
\newblock \bibinfo{title}{Violation of fourier's law and anomalous heat
  diffusion in silicon nanowires}.
\newblock \bibinfo{journal}{Nanotoday}
  \bibinfo{year}{2010};\bibinfo{volume}{5}(\bibinfo{number}{2}):\bibinfo{pages%
}{85--90}.
\bibitem[{Xiong et~al.(2010{\natexlab{b}})Xiong, Wang, Zhang and
  Zhao}]{XiongDX2010pre}
\bibinfo{author}{Xiong\xfnm[ D.]}, \bibinfo{author}{Wang\xfnm[ J.]},
  \bibinfo{author}{Zhang\xfnm[ Y.]}, \bibinfo{author}{Zhao\xfnm[ H.]}.
\newblock \bibinfo{title}{Heat conduction in two-dimensional disk models}.
\newblock \bibinfo{journal}{Physical Review E}
  \bibinfo{year}{2010}{\natexlab{b}};\bibinfo{volume}{82}(\bibinfo{number}{3})%
:\bibinfo{pages}{030101}.
\bibitem[{Chen et~al.(2013{\natexlab{b}})Chen, Zhang, Wang and
  Zhao}]{ChenSD2013pre}
\bibinfo{author}{Chen\xfnm[ S.]}, \bibinfo{author}{Zhang\xfnm[ Y.]},
  \bibinfo{author}{Wang\xfnm[ J.]}, \bibinfo{author}{Zhao\xfnm[ H.]}.
\newblock \bibinfo{title}{Diffusion of heat, energy, momentum, and mass in
  one-dimensional systems}.
\newblock \bibinfo{journal}{Physical Review E}
  \bibinfo{year}{2013}{\natexlab{b}};\bibinfo{volume}{87}(\bibinfo{number}{3})%
:\bibinfo{pages}{032153}.
\bibitem[{Nose(1984)}]{Nose}
\bibinfo{author}{Nose\xfnm[ S.]}.
\newblock \bibinfo{title}{A unified formulation of the constant temperature
  molecular dynamics methods}.
\newblock \bibinfo{journal}{Journal of Chemical Physics}
  \bibinfo{year}{1984};\bibinfo{volume}{81}(\bibinfo{number}{1}):\bibinfo{page%
s}{511}.
\bibitem[{Hoover(1985)}]{Hoover}
\bibinfo{author}{Hoover\xfnm[ W.G.]}.
\newblock \bibinfo{title}{Canonical dynamics: Equilibrium phase-space
  distributions}.
\newblock \bibinfo{journal}{Physical Review A}
  \bibinfo{year}{1985};\bibinfo{volume}{31}(\bibinfo{number}{3}):\bibinfo{page%
s}{1695}.
\bibitem[{Poetzsch and B$\ddot{o}$ttger(1994)}]{PoetzschRHH}
\bibinfo{author}{Poetzsch\xfnm[ R.H.H.]},
  \bibinfo{author}{B$\ddot{o}$ttger\xfnm[ H.]}.
\newblock \bibinfo{title}{Interplay of disorder and anharmonicity in heat
  conduction: Molecular-dynamics study}.
\newblock \bibinfo{journal}{Physical Review B}
  \bibinfo{year}{1994};\bibinfo{volume}{50}(\bibinfo{number}{21}):\bibinfo{pag%
es}{15757--15763}.
\bibitem[{Jiang and Wang(2011{\natexlab{b}})}]{JiangJW2011bntube}
\bibinfo{author}{Jiang\xfnm[ J.W.]}, \bibinfo{author}{Wang\xfnm[ J.S.]}.
\newblock \bibinfo{title}{Theoretical study of thermal conductivity in
  single-walled boron nitride nanotubes}.
\newblock \bibinfo{journal}{Physical Review B}
  \bibinfo{year}{2011}{\natexlab{b}};\bibinfo{volume}{84}(\bibinfo{number}{8})%
:\bibinfo{pages}{085439}.
\bibitem[{Hone et~al.(1999)Hone, Whitney and Zettl}]{HoneJ}
\bibinfo{author}{Hone\xfnm[ J.]}, \bibinfo{author}{Whitney\xfnm[ M.]},
  \bibinfo{author}{Zettl\xfnm[ A.]}.
\newblock \bibinfo{title}{Thermal conductivity of single-walled carbon
  nanotubes}.
\newblock \bibinfo{journal}{Synthetic Metals}
  \bibinfo{year}{1999};\bibinfo{volume}{103}:\bibinfo{pages}{2498--2499}.
\bibitem[{Gu and Chen(2007)}]{GuY}
\bibinfo{author}{Gu\xfnm[ Y.]}, \bibinfo{author}{Chen\xfnm[ Y.]}.
\newblock \bibinfo{title}{Thermal conductivities of single-walled carbon
  nanotubes calculated from the complete phonon dispersion relations}.
\newblock \bibinfo{journal}{Physical Review B}
  \bibinfo{year}{2007};\bibinfo{volume}{76}(\bibinfo{number}{13}):\bibinfo{pag%
es}{134110}.
\bibitem[{Aksamija and Knezevic(2011)}]{AksamijaZ}
\bibinfo{author}{Aksamija\xfnm[ Z.]}, \bibinfo{author}{Knezevic\xfnm[ I.]}.
\newblock \bibinfo{title}{Lattice thermal conductivity of graphene nanoribbons:
  Anisotropy and edge roughness scattering}.
\newblock \bibinfo{journal}{Applied Physics Letters}
  \bibinfo{year}{2011};\bibinfo{volume}{98}:\bibinfo{pages}{141919}.
\bibitem[{Chen and Kumar(2012)}]{ChenL2012jap}
\bibinfo{author}{Chen\xfnm[ L.]}, \bibinfo{author}{Kumar\xfnm[ S.]}.
\newblock \bibinfo{title}{Thermal transport in graphene supported on copper}.
\newblock \bibinfo{journal}{Journal of Applied Physics}
  \bibinfo{year}{2012};\bibinfo{volume}{112}:\bibinfo{pages}{043502}.
\bibitem[{Ouyang et~al.(2011)Ouyang, Chen, Xie, Stocks and
  Zhong}]{OuyangT2011apl}
\bibinfo{author}{Ouyang\xfnm[ T.]}, \bibinfo{author}{Chen\xfnm[ Y.]},
  \bibinfo{author}{Xie\xfnm[ Y.]}, \bibinfo{author}{Stocks\xfnm[ G.M.]},
  \bibinfo{author}{Zhong\xfnm[ J.]}.
\newblock \bibinfo{title}{Thermal conductance modulator based on folded
  graphene nanoribbons}.
\newblock \bibinfo{journal}{Applied Physics Letters}
  \bibinfo{year}{2011};\bibinfo{volume}{99}:\bibinfo{pages}{233101}.
\bibitem[{Yang et~al.(2012{\natexlab{b}})Yang, Ni, Jiang and Li}]{YangNfold}
\bibinfo{author}{Yang\xfnm[ N.]}, \bibinfo{author}{Ni\xfnm[ X.]},
  \bibinfo{author}{Jiang\xfnm[ J.W.]}, \bibinfo{author}{Li\xfnm[ B.]}.
\newblock \bibinfo{title}{How does folding modulate thermal conductivity of
  graphene?}
\newblock \bibinfo{journal}{Applied Physics Letters}
  \bibinfo{year}{2012}{\natexlab{b}};\bibinfo{volume}{100}(\bibinfo{number}{9}%
):\bibinfo{pages}{093107}.
\bibitem[{Amorim and Guinea(2013)}]{AmorimB2013prb}
\bibinfo{author}{Amorim\xfnm[ B.]}, \bibinfo{author}{Guinea\xfnm[ F.]}.
\newblock \bibinfo{title}{Flexural mode of graphene on a substrate}.
\newblock \bibinfo{journal}{Physical Review B}
  \bibinfo{year}{2013};\bibinfo{volume}{88}(\bibinfo{number}{11}):\bibinfo{pag%
es}{115418}.
\bibitem[{Xu et~al.(1991)Xu, Wang, Chan and Ho}]{XuCH1991prb}
\bibinfo{author}{Xu\xfnm[ C.H.]}, \bibinfo{author}{Wang\xfnm[ C.Z.]},
  \bibinfo{author}{Chan\xfnm[ C.T.]}, \bibinfo{author}{Ho\xfnm[ K.M.]}.
\newblock \bibinfo{title}{Theory of the thermal expansion of {S}i and diamond}.
\newblock \bibinfo{journal}{Physical Review B}
  \bibinfo{year}{1991};\bibinfo{volume}{43}(\bibinfo{number}{6}):\bibinfo{page%
s}{5024--5027}.
\bibitem[{Wei et~al.(1994)Wei, Li and Chou}]{WeiS}
\bibinfo{author}{Wei\xfnm[ S.]}, \bibinfo{author}{Li\xfnm[ C.]},
  \bibinfo{author}{Chou\xfnm[ M.Y.]}.
\newblock \bibinfo{title}{Ab initio calculation of thermodynamic properties of
  silicon}.
\newblock \bibinfo{journal}{Physical Review B}
  \bibinfo{year}{1994};\bibinfo{volume}{50}(\bibinfo{number}{19}):\bibinfo{pag%
es}{14587--14590}.
\bibitem[{Barrera et~al.(2005)Barrera, Bruno, Barron and Allan}]{BarreraGD}
\bibinfo{author}{Barrera\xfnm[ G.D.]}, \bibinfo{author}{Bruno\xfnm[ J.A.O.]},
  \bibinfo{author}{Barron\xfnm[ T.H.K.]}, \bibinfo{author}{Allan\xfnm[ N.L.]}.
\newblock \bibinfo{title}{Negative thermal expansion}.
\newblock \bibinfo{journal}{Journal of Physics: Condensed Matter}
  \bibinfo{year}{2005};\bibinfo{volume}{17}:\bibinfo{pages}{R217--R252}.
\bibitem[{Yoon et~al.(2011)Yoon, Son and Cheong}]{YoonD}
\bibinfo{author}{Yoon\xfnm[ D.]}, \bibinfo{author}{Son\xfnm[ Y.W.]},
  \bibinfo{author}{Cheong\xfnm[ H.]}.
\newblock \bibinfo{title}{Negative thermal expansion coefficient of graphene
  measured by raman spectroscopy}.
\newblock \bibinfo{journal}{Nano Letters}
  \bibinfo{year}{2011};\bibinfo{volume}{11}(\bibinfo{number}{8}):\bibinfo{page%
s}{3227--3231}.
\bibitem[{Singh et~al.(2010)Singh, Sengupta, Solanki, Dhall, Allain, Dhara
  et~al.}]{SinghV}
\bibinfo{author}{Singh\xfnm[ V.]}, \bibinfo{author}{Sengupta\xfnm[ S.]},
  \bibinfo{author}{Solanki\xfnm[ H.S.]}, \bibinfo{author}{Dhall\xfnm[ R.]},
  \bibinfo{author}{Allain\xfnm[ A.]}, \bibinfo{author}{Dhara\xfnm[ S.]}, et~al.
\newblock \bibinfo{title}{Probing thermal expansion of graphene and modal
  dispersion at low-temperature using graphene nanoelectromechanical systems
  resonators}.
\newblock \bibinfo{journal}{Nanotechnology}
  \bibinfo{year}{2010};\bibinfo{volume}{21}(\bibinfo{number}{16}):\bibinfo{pag%
es}{165204}.
\bibitem[{Zakharchenko et~al.(2009)Zakharchenko, Katsnelson and
  Fasolino}]{ZakharchenkoKV}
\bibinfo{author}{Zakharchenko\xfnm[ K.V.]}, \bibinfo{author}{Katsnelson\xfnm[
  M.I.]}, \bibinfo{author}{Fasolino\xfnm[ A.]}.
\newblock \bibinfo{title}{Finite temperature lattice properties of graphene
  beyond the quasiharmonic approximation}.
\newblock \bibinfo{journal}{Physical Review Letters}
  \bibinfo{year}{2009};\bibinfo{volume}{102}(\bibinfo{number}{4}):\bibinfo{pag%
es}{046808}.
\bibitem[{Pozzo et~al.(2011)Pozzo, Alfe, Lacovig, Hofmann, Lizzit and
  Baraldi}]{PozzoM}
\bibinfo{author}{Pozzo\xfnm[ M.]}, \bibinfo{author}{Alfe\xfnm[ D.]},
  \bibinfo{author}{Lacovig\xfnm[ P.]}, \bibinfo{author}{Hofmann\xfnm[ P.]},
  \bibinfo{author}{Lizzit\xfnm[ S.]}, \bibinfo{author}{Baraldi\xfnm[ A.]}.
\newblock \bibinfo{title}{Thermal expansion of supported and freestanding
  graphene: Lattice constant versus interatomic distance}.
\newblock \bibinfo{journal}{Physical Review Letters}
  \bibinfo{year}{2011};\bibinfo{volume}{106}:\bibinfo{pages}{135501}.
\bibitem[{Jiang et~al.(2009{\natexlab{c}})Jiang, Wang and
  Li}]{JiangJW2009expansion}
\bibinfo{author}{Jiang\xfnm[ J.W.]}, \bibinfo{author}{Wang\xfnm[ J.S.]},
  \bibinfo{author}{Li\xfnm[ B.]}.
\newblock \bibinfo{title}{Thermal expansion in single-walled carbon nanotubes
  and graphene: Nonequilibrium green’s function approach}.
\newblock \bibinfo{journal}{Physical Review B}
  \bibinfo{year}{2009}{\natexlab{c}};\bibinfo{volume}{80}(\bibinfo{number}{20}%
):\bibinfo{pages}{205429}.
\bibitem[{Gruneisen(1926)}]{GruneisenE1926}
\bibinfo{author}{Gruneisen\xfnm[ E.]}.
\newblock \bibinfo{title}{Handbuch der Physik}; vol.~\bibinfo{volume}{10}.
\newblock \bibinfo{publisher}{pp 1-52; eds Geiger H (Springer, Berlin)};
  \bibinfo{year}{1926}.
\bibitem[{Wang and Jiang(2014)}]{WangJS2014thermalexpansion}
\bibinfo{author}{Wang\xfnm[ J.S.]}, \bibinfo{author}{Jiang\xfnm[ J.W.]}.
\newblock \bibinfo{title}{{NEGF} approach and {G}runeisen method for thermal
  expansion}.
\newblock \bibinfo{journal}{in prepare} \bibinfo{year}{2014};.
\bibitem[{Li et~al.(2009)Li, Xu, Srivastava and Banerjee}]{LiH2009ieee}
\bibinfo{author}{Li\xfnm[ H.]}, \bibinfo{author}{Xu\xfnm[ C.]},
  \bibinfo{author}{Srivastava\xfnm[ N.]}, \bibinfo{author}{Banerjee\xfnm[ K.]}.
\newblock \bibinfo{title}{Carbon nanomaterials for next-generation
  interconnects and passives: Physics, status, and prospects}.
\newblock \bibinfo{journal}{IEEE Transactions on Electron Devices}
  \bibinfo{year}{2009};\bibinfo{volume}{56}(\bibinfo{number}{9}):\bibinfo{page%
s}{1799--1821}.
\bibitem[{Terrones et~al.(2010)Terrones, Botello-Méndez, Campos-Delgado,
  López-Urías, Vega-Cantú, Rodríguez-Macías et~al.}]{TerronesM2010nt}
\bibinfo{author}{Terrones\xfnm[ M.]}, \bibinfo{author}{Botello-Méndez\xfnm[
  A.R.]}, \bibinfo{author}{Campos-Delgado\xfnm[ J.]},
  \bibinfo{author}{López-Urías\xfnm[ F.]}, \bibinfo{author}{Vega-Cantú\xfnm[
  Y.I.]}, \bibinfo{author}{Rodríguez-Macías\xfnm[ F.J.]}, et~al.
\newblock \bibinfo{title}{Graphene and graphite nanoribbons: Morphology,
  properties, synthesis, defects and applications}.
\newblock \bibinfo{journal}{Nanotoday}
  \bibinfo{year}{2010};\bibinfo{volume}{5}(\bibinfo{number}{4}):\bibinfo{pages%
}{351--372}.
\bibitem[{Yang et~al.(2012{\natexlab{c}})Yang, Gao, Hu, Chai, Cheng, Zhang
  et~al.}]{YangZ2012nml}
\bibinfo{author}{Yang\xfnm[ Z.]}, \bibinfo{author}{Gao\xfnm[ R.]},
  \bibinfo{author}{Hu\xfnm[ N.]}, \bibinfo{author}{Chai\xfnm[ J.]},
  \bibinfo{author}{Cheng\xfnm[ Y.]}, \bibinfo{author}{Zhang\xfnm[ L.]}, et~al.
\newblock \bibinfo{title}{The prospective two-dimensional graphene nanosheets:
  Preparation, functionalization, and applications}.
\newblock \bibinfo{journal}{Nano-Micro Letters}
  \bibinfo{year}{2012}{\natexlab{c}};\bibinfo{volume}{4}(\bibinfo{number}{1}):%
\bibinfo{pages}{1--9}.
\bibitem[{Lau et~al.(2012)Lau, Bao and Velasco~Jr.}]{LauCN2012mt}
\bibinfo{author}{Lau\xfnm[ C.N.]}, \bibinfo{author}{Bao\xfnm[ W.]},
  \bibinfo{author}{Velasco~Jr.\xfnm[ J.]}.
\newblock \bibinfo{title}{Properties of suspended graphene membranes}.
\newblock \bibinfo{journal}{Materials Today}
  \bibinfo{year}{2012};\bibinfo{volume}{15}(\bibinfo{number}{6}):\bibinfo{page%
s}{238--245}.
\bibitem[{Tu and Ou-Yang(2008)}]{TuZC2008jctn}
\bibinfo{author}{Tu\xfnm[ Z.C.]}, \bibinfo{author}{Ou-Yang\xfnm[ Z.C.]}.
\newblock \bibinfo{title}{Elastic theory of low-dimensional continua and its
  applications in bio- and nano-structures}.
\newblock \bibinfo{journal}{Journal of Computational and Theoretical
  Nanoscience}
  \bibinfo{year}{2008};\bibinfo{volume}{5}(\bibinfo{number}{4}):\bibinfo{pages%
}{422--448}.
\bibitem[{Lu(1997)}]{luPRL1997}
\bibinfo{author}{Lu\xfnm[ J.P.]}.
\newblock \bibinfo{title}{Elastic properties of carbon nanotubes and
  nanoropes}.
\newblock \bibinfo{journal}{Physical Review Letters}
  \bibinfo{year}{1997};\bibinfo{volume}{79}(\bibinfo{number}{7}):\bibinfo{page%
s}{1297--1300}.
\bibitem[{Lier et~al.(2000)Lier, Alsenoy, Doren and Geerlings}]{LierGV2000cpl}
\bibinfo{author}{Lier\xfnm[ G.V.]}, \bibinfo{author}{Alsenoy\xfnm[ C.V.]},
  \bibinfo{author}{Doren\xfnm[ V.V.]}, \bibinfo{author}{Geerlings\xfnm[ P.]}.
\newblock \bibinfo{title}{Ab initio study of the elastic properties of
  single-walled carbon nanotubes and graphene}.
\newblock \bibinfo{journal}{Chemical Physics Letters}
  \bibinfo{year}{2000};\bibinfo{volume}{326}:\bibinfo{pages}{181--185}.
\bibitem[{Kudin et~al.(2001)Kudin, Scuseria and Yakobson}]{KudinKN2001prb}
\bibinfo{author}{Kudin\xfnm[ K.N.]}, \bibinfo{author}{Scuseria\xfnm[ G.E.]},
  \bibinfo{author}{Yakobson\xfnm[ B.I.]}.
\newblock \bibinfo{title}{C2f, bn, and c nanoshell elasticity from ab initio
  computations}.
\newblock \bibinfo{journal}{Physical Review B}
  \bibinfo{year}{2001};\bibinfo{volume}{64}(\bibinfo{number}{23}):\bibinfo{pag%
es}{235406}.
\bibitem[{Konstantinova et~al.(2006)Konstantinova, Dantas and
  Barone}]{KonstantinovaE2006prb}
\bibinfo{author}{Konstantinova\xfnm[ E.]}, \bibinfo{author}{Dantas\xfnm[
  S.O.]}, \bibinfo{author}{Barone\xfnm[ P.M.V.B.]}.
\newblock \bibinfo{title}{lectronic and elastic properties of two-dimensional
  carbon planes}.
\newblock \bibinfo{journal}{Physical Review B}
  \bibinfo{year}{2006};\bibinfo{volume}{74}(\bibinfo{number}{3}):\bibinfo{page%
s}{035417}.
\bibitem[{Reddy et~al.(2006)Reddy, Rajendran and Liew}]{ReddyCD2006nano}
\bibinfo{author}{Reddy\xfnm[ C.D.]}, \bibinfo{author}{Rajendran\xfnm[ S.]},
  \bibinfo{author}{Liew\xfnm[ K.M.]}.
\newblock \bibinfo{title}{Equilibrium configuration and continuum elastic
  properties of finite sized graphene}.
\newblock \bibinfo{journal}{Nanotechnology}
  \bibinfo{year}{2006};\bibinfo{volume}{17}(\bibinfo{number}{3}):\bibinfo{page%
s}{864}.
\bibitem[{Huang et~al.(2006)Huang, Wu and Hwang}]{huangPRB2006}
\bibinfo{author}{Huang\xfnm[ Y.]}, \bibinfo{author}{Wu\xfnm[ J.]},
  \bibinfo{author}{Hwang\xfnm[ K.C.]}.
\newblock \bibinfo{title}{Thickness of graphene and single-wall carbon
  nanotubes}.
\newblock \bibinfo{journal}{Physical Review B}
  \bibinfo{year}{2006};\bibinfo{volume}{74}:\bibinfo{pages}{245413}.
\bibitem[{Khare et~al.(2007)Khare, Mielke, Paci, Zhang, Ballarini, Schatz
  et~al.}]{KhareR2007prb}
\bibinfo{author}{Khare\xfnm[ R.]}, \bibinfo{author}{Mielke\xfnm[ S.L.]},
  \bibinfo{author}{Paci\xfnm[ J.T.]}, \bibinfo{author}{Zhang\xfnm[ S.]},
  \bibinfo{author}{Ballarini\xfnm[ R.]}, \bibinfo{author}{Schatz\xfnm[ G.C.]},
  et~al.
\newblock \bibinfo{title}{Coupled quantum mechanical/molecular mechanical
  modeling of the fracture of defective carbon nanotubes and graphene sheets}.
\newblock \bibinfo{journal}{Physical Review B}
  \bibinfo{year}{2007};\bibinfo{volume}{75}(\bibinfo{number}{7}):\bibinfo{page%
s}{075412}.
\bibitem[{Wei et~al.(2009)Wei, Fragneaud, Marianetti and Kysar}]{WeiXD2009prb}
\bibinfo{author}{Wei\xfnm[ X.]}, \bibinfo{author}{Fragneaud\xfnm[ B.]},
  \bibinfo{author}{Marianetti\xfnm[ C.A.]}, \bibinfo{author}{Kysar\xfnm[
  J.W.]}.
\newblock \bibinfo{title}{Nonlinear elastic behavior of graphene: Ab initio
  calculations to continuum description}.
\newblock \bibinfo{journal}{Physical Review B}
  \bibinfo{year}{2009};\bibinfo{volume}{80}:\bibinfo{pages}{205407}.
\bibitem[{Jiang et~al.(2010{\natexlab{d}})Jiang, Wang and
  Li}]{JiangJW2010young}
\bibinfo{author}{Jiang\xfnm[ J.W.]}, \bibinfo{author}{Wang\xfnm[ J.S.]},
  \bibinfo{author}{Li\xfnm[ B.]}.
\newblock \bibinfo{title}{Elastic and nonlinear stiffness of graphene: a simple
  approach}.
\newblock \bibinfo{journal}{Physical Review B}
  \bibinfo{year}{2010}{\natexlab{d}};\bibinfo{volume}{81}(\bibinfo{number}{7})%
:\bibinfo{pages}{073405}.
\bibitem[{Yi and Chang(2012)}]{YiLJ2012scpma}
\bibinfo{author}{Yi\xfnm[ L.]}, \bibinfo{author}{Chang\xfnm[ T.]}.
\newblock \bibinfo{title}{Loading direction dependent mechanical behavior of
  graphene under shear strain}.
\newblock \bibinfo{journal}{Science China Physics, Mechanics and Astronomy}
  \bibinfo{year}{2012};\bibinfo{volume}{55}(\bibinfo{number}{6}):\bibinfo{page%
s}{1083--1087}.
\bibitem[{Zhou et~al.(2013)Zhou, Wang and Cao}]{ZhouL2013jpcm}
\bibinfo{author}{Zhou\xfnm[ L.]}, \bibinfo{author}{Wang\xfnm[ Y.]},
  \bibinfo{author}{Cao\xfnm[ G.]}.
\newblock \bibinfo{title}{Elastic properties of monolayer graphene with
  different chiralities}.
\newblock \bibinfo{journal}{Journal of Physics: Condensed Matter}
  \bibinfo{year}{2013};\bibinfo{volume}{25}:\bibinfo{pages}{125302}.
\bibitem[{Shenoy et~al.(2008)Shenoy, Reddy, Ramasubramaniam and
  Zhang}]{ShenoyVB}
\bibinfo{author}{Shenoy\xfnm[ V.B.]}, \bibinfo{author}{Reddy\xfnm[ C.D.]},
  \bibinfo{author}{Ramasubramaniam\xfnm[ A.]}, \bibinfo{author}{Zhang\xfnm[
  Y.W.]}.
\newblock \bibinfo{title}{Edge-stress-induced warping of graphene sheets and
  nanoribbons}.
\newblock \bibinfo{journal}{Physical Review Letters}
  \bibinfo{year}{2008};\bibinfo{volume}{101}(\bibinfo{number}{24}):\bibinfo{pa%
ges}{245501}.
\bibitem[{Duan and Wang(2009)}]{DuanWH2009nano}
\bibinfo{author}{Duan\xfnm[ W.H.]}, \bibinfo{author}{Wang\xfnm[ C.M.]}.
\newblock \bibinfo{title}{Nonlinear bending and stretching of a circular
  graphene sheet under a central point load}.
\newblock \bibinfo{journal}{Nanotechnology}
  \bibinfo{year}{2009};\bibinfo{volume}{20}:\bibinfo{pages}{075702}.
\bibitem[{Treacy et~al.(1996)Treacy, Ebbesen and Gibson}]{TreacyMMJ1996nat}
\bibinfo{author}{Treacy\xfnm[ M.M.J.]}, \bibinfo{author}{Ebbesen\xfnm[ T.W.]},
  \bibinfo{author}{Gibson\xfnm[ J.M.]}.
\newblock \bibinfo{title}{Exceptionally high young's modulus observed for
  individual carbon nanotubes}.
\newblock \bibinfo{journal}{Nature}
  \bibinfo{year}{1996};\bibinfo{volume}{381}:\bibinfo{pages}{678 -- 680}.
\bibitem[{Krishnan et~al.(1998)Krishnan, Dujardin, Ebbesen, Yianilos and
  Treacy}]{KrishnanA}
\bibinfo{author}{Krishnan\xfnm[ A.]}, \bibinfo{author}{Dujardin\xfnm[ E.]},
  \bibinfo{author}{Ebbesen\xfnm[ T.W.]}, \bibinfo{author}{Yianilos\xfnm[
  P.N.]}, \bibinfo{author}{Treacy\xfnm[ M.M.J.]}.
\newblock \bibinfo{title}{Young$’$s modulus of single-walled nanotubes}.
\newblock \bibinfo{journal}{Physical Review B}
  \bibinfo{year}{1998};\bibinfo{volume}{58}(\bibinfo{number}{20}):\bibinfo{pag%
es}{14013--14019}.
\bibitem[{Jiang et~al.(2009{\natexlab{d}})Jiang, Wang and
  Li}]{JiangJW2009young}
\bibinfo{author}{Jiang\xfnm[ J.W.]}, \bibinfo{author}{Wang\xfnm[ J.S.]},
  \bibinfo{author}{Li\xfnm[ B.]}.
\newblock \bibinfo{title}{Young’s modulus of graphene: a molecular dynamics
  study}.
\newblock \bibinfo{journal}{Physical Review B}
  \bibinfo{year}{2009}{\natexlab{d}};\bibinfo{volume}{80}(\bibinfo{number}{11}%
):\bibinfo{pages}{113405}.
\bibitem[{Polyanin(2002)}]{PolyaninAD}
\bibinfo{author}{Polyanin\xfnm[ A.D.]}.
\newblock \bibinfo{title}{Handbook of Linear Partial Differential Equations for
  Engineers and Scientists}.
\newblock \bibinfo{publisher}{CRC Press/C$\&$H}; \bibinfo{year}{2002}.
\bibitem[{Blakslee et~al.(1970)Blakslee, Proctor, Seldin, Spence,  and
  Weng}]{BlaksleeOL1970jap}
\bibinfo{author}{Blakslee\xfnm[ O.L.]}, \bibinfo{author}{Proctor\xfnm[ D.G.]},
  \bibinfo{author}{Seldin\xfnm[ E.J.]}, \bibinfo{author}{Spence\xfnm[ G.B.]},
  \bibinfo{author}{\xfnm[]}, \bibinfo{author}{Weng\xfnm[ T.]}.
\newblock \bibinfo{title}{Elastic constants of compression‐annealed pyrolytic
  graphite}.
\newblock \bibinfo{journal}{Journal of Applied Physics}
  \bibinfo{year}{1970};\bibinfo{volume}{41}(\bibinfo{number}{8}):\bibinfo{page%
s}{3373}.
\bibitem[{Portal et~al.(1999)Portal, Artacho, Soler, Rubio and
  Ordej\'{o}n}]{PortalD1999prb}
\bibinfo{author}{Portal\xfnm[ D.]}, \bibinfo{author}{Artacho\xfnm[ E.]},
  \bibinfo{author}{Soler\xfnm[ J.M.]}, \bibinfo{author}{Rubio\xfnm[ A.]},
  \bibinfo{author}{Ordej\'{o}n\xfnm[ P.]}.
\newblock \bibinfo{title}{Ab initio structural, elastic, and vibrational
  properties of carbon nanotubes}.
\newblock \bibinfo{journal}{Physical Review B}
  \bibinfo{year}{1999};\bibinfo{volume}{59}(\bibinfo{number}{19}):\bibinfo{pag%
es}{12678–12688}.
\bibitem[{Reddy et~al.(2009)Reddy, Ramasubramaniam, Shenoy and Zhang}]{ReddyCD}
\bibinfo{author}{Reddy\xfnm[ C.D.]}, \bibinfo{author}{Ramasubramaniam\xfnm[
  A.]}, \bibinfo{author}{Shenoy\xfnm[ V.B.]}, \bibinfo{author}{Zhang\xfnm[
  Y..]}.
\newblock \bibinfo{title}{Edge elastic properties of defect-free single-layer
  graphene sheets}.
\newblock \bibinfo{journal}{Applied Physics Letters}
  \bibinfo{year}{2009};\bibinfo{volume}{94}(\bibinfo{number}{10}):\bibinfo{pag%
es}{101904}.
\bibitem[{Cadelano et~al.(2009)Cadelano, Palla, Giordano and
  Colombo}]{CadelanoE2009prl}
\bibinfo{author}{Cadelano\xfnm[ E.]}, \bibinfo{author}{Palla\xfnm[ P.L.]},
  \bibinfo{author}{Giordano\xfnm[ S.]}, \bibinfo{author}{Colombo\xfnm[ L.]}.
\newblock \bibinfo{title}{Nonlinear elasticity of monolayer graphene}.
\newblock \bibinfo{journal}{Physical Review Letters}
  \bibinfo{year}{2009};\bibinfo{volume}{102}(\bibinfo{number}{23}):\bibinfo{pa%
ges}{235502}.
\bibitem[{Zhao et~al.(2009)Zhao, Min and Aluru}]{ZhaoH2009nl}
\bibinfo{author}{Zhao\xfnm[ H.]}, \bibinfo{author}{Min\xfnm[ K.]},
  \bibinfo{author}{Aluru\xfnm[ N.R.]}.
\newblock \bibinfo{title}{Size and chirality dependent elastic properties of
  graphene nanoribbons under uniaxial tension}.
\newblock \bibinfo{journal}{Nano Letters}
  \bibinfo{year}{2009};\bibinfo{volume}{9}(\bibinfo{number}{8}):\bibinfo{pages%
}{3012--3015}.
\bibitem[{Lee et~al.(2008)Lee, Wei, Kysar and Hone}]{LeeC2008sci}
\bibinfo{author}{Lee\xfnm[ C.]}, \bibinfo{author}{Wei\xfnm[ X.]},
  \bibinfo{author}{Kysar\xfnm[ J.W.]}, \bibinfo{author}{Hone\xfnm[ J.]}.
\newblock \bibinfo{title}{Measurement of the elastic properties and intrinsic
  strength of monolayer graphene}.
\newblock \bibinfo{journal}{Science}
  \bibinfo{year}{2008};\bibinfo{volume}{321}:\bibinfo{pages}{385}.
\bibitem[{Zhao and Xue(2013)}]{ZhaoS2013jpdap}
\bibinfo{author}{Zhao\xfnm[ S.]}, \bibinfo{author}{Xue\xfnm[ J.]}.
\newblock \bibinfo{title}{Mechanical properties of hybrid graphene and
  hexagonal boron nitride sheets as revealed by molecular dynamic simulations}.
\newblock \bibinfo{journal}{Journal of Physics D: Applied Physics}
  \bibinfo{year}{2013};\bibinfo{volume}{46}:\bibinfo{pages}{135303}.
\bibitem[{Jiang and Park(2014)}]{JiangJW2014gmgyoung}
\bibinfo{author}{Jiang\xfnm[ J.W.]}, \bibinfo{author}{Park\xfnm[ H.S.]}.
\newblock \bibinfo{title}{Mechanical properties of mos$_{2}$/graphene
  heterostructures}.
\newblock \bibinfo{journal}{Applied Physics Letters}
  \bibinfo{year}{2014};\bibinfo{volume}{105}:\bibinfo{pages}{033108}.
\bibitem[{O$'$Connell et~al.(2010)O$'$Connell, Hofheinz, Ansmann, Bialczak,
  Lenander, Lucero et~al.}]{ConnellADO}
\bibinfo{author}{O$'$Connell\xfnm[ A.D.]}, \bibinfo{author}{Hofheinz\xfnm[
  M.]}, \bibinfo{author}{Ansmann\xfnm[ M.]}, \bibinfo{author}{Bialczak\xfnm[
  R.C.]}, \bibinfo{author}{Lenander\xfnm[ M.]}, \bibinfo{author}{Lucero\xfnm[
  E.]}, et~al.
\newblock \bibinfo{title}{Quantum ground state and single-phonon control of a
  mechanical resonator}.
\newblock \bibinfo{journal}{Nature}
  \bibinfo{year}{2010};\bibinfo{volume}{464}:\bibinfo{pages}{697}.
\bibitem[{Palomaki et~al.(2013)Palomaki, Harlow, Teufel, Simmonds and
  Lehnert}]{PalomakiTA}
\bibinfo{author}{Palomaki\xfnm[ T.A.]}, \bibinfo{author}{Harlow\xfnm[ J.W.]},
  \bibinfo{author}{Teufel\xfnm[ J.D.]}, \bibinfo{author}{Simmonds\xfnm[ R.W.]},
  \bibinfo{author}{Lehnert\xfnm[ K.W.]}.
\newblock \bibinfo{title}{Coherent state transfer between itinerant microwave
  fields and a mechanical oscillator}.
\newblock \bibinfo{journal}{Nature}
  \bibinfo{year}{2013};\bibinfo{volume}{495}:\bibinfo{pages}{210}.
\bibitem[{Low et~al.(2012)Low, Jiang, Katsnelson and Guinea}]{LowT}
\bibinfo{author}{Low\xfnm[ T.]}, \bibinfo{author}{Jiang\xfnm[ Y.]},
  \bibinfo{author}{Katsnelson\xfnm[ M.]}, \bibinfo{author}{Guinea\xfnm[ F.]}.
\newblock \bibinfo{title}{Electron pumping in graphene mechanical resonators}.
\newblock \bibinfo{journal}{Nano Letters}
  \bibinfo{year}{2012};\bibinfo{volume}{12}(\bibinfo{number}{2}):\bibinfo{page%
s}{850--854}.
\bibitem[{Sakhaee-Pour et~al.(2008)Sakhaee-Pour, Ahmadian and
  Vafai}]{Sakhaee-PourA}
\bibinfo{author}{Sakhaee-Pour\xfnm[ A.]}, \bibinfo{author}{Ahmadian\xfnm[ M.]},
  \bibinfo{author}{Vafai\xfnm[ A.]}.
\newblock \bibinfo{title}{Applications of single-layered graphene sheets as
  mass sensors and atomistic dust detectors}.
\newblock \bibinfo{journal}{Solid State Communications}
  \bibinfo{year}{2008};\bibinfo{volume}{145}:\bibinfo{pages}{168--172}.
\bibitem[{Rumyantsev et~al.(2012)Rumyantsev, Liu, Shur, Potyrailo and
  Balandin}]{RumyantsevS}
\bibinfo{author}{Rumyantsev\xfnm[ S.]}, \bibinfo{author}{Liu\xfnm[ G.]},
  \bibinfo{author}{Shur\xfnm[ M.S.]}, \bibinfo{author}{Potyrailo\xfnm[ R.A.]},
  \bibinfo{author}{Balandin\xfnm[ A.A.]}.
\newblock \bibinfo{title}{Selective gas sensing with a single pristine graphene
  transistor}.
\newblock \bibinfo{journal}{Nano Letters}
  \bibinfo{year}{2012};\bibinfo{volume}{12}(\bibinfo{number}{5}):\bibinfo{page%
s}{2294--2298}.
\bibitem[{Avdoshenko et~al.(2012)Avdoshenko, da~Rocha and
  Cuniberti}]{AvdoshenkoSM}
\bibinfo{author}{Avdoshenko\xfnm[ S.M.]}, \bibinfo{author}{da~Rocha\xfnm[
  C.G.]}, \bibinfo{author}{Cuniberti\xfnm[ G.]}.
\newblock \bibinfo{title}{Nanoscale ear drum: Graphene based nanoscale
  sensors}.
\newblock \bibinfo{journal}{Nanoscale}
  \bibinfo{year}{2012};\bibinfo{volume}{4}:\bibinfo{pages}{3168--3174}.
\bibitem[{Midtvedt et~al.(2014)Midtvedt, Croy, Isacsson, Qi and
  Park}]{MidtvedtD2014prl}
\bibinfo{author}{Midtvedt\xfnm[ D.]}, \bibinfo{author}{Croy\xfnm[ A.]},
  \bibinfo{author}{Isacsson\xfnm[ A.]}, \bibinfo{author}{Qi\xfnm[ Z.]},
  \bibinfo{author}{Park\xfnm[ H.S.]}.
\newblock \bibinfo{title}{Fermi-pasta-ulam physics with nanomechanical graphene
  resonators: Intrinsic relaxation and thermalization from flexural mode
  coupling}.
\newblock \bibinfo{journal}{Physical Review Letters}
  \bibinfo{year}{2014};\bibinfo{volume}{112}:\bibinfo{pages}{145503}.
\bibitem[{Bunch et~al.(2007)Bunch, van~der Zande, Verbridge, Frank, Tanenbaum,
  Parpia et~al.}]{BunchJS2007sci}
\bibinfo{author}{Bunch\xfnm[ J.S.]}, \bibinfo{author}{van~der Zande\xfnm[
  A.M.]}, \bibinfo{author}{Verbridge\xfnm[ S.S.]}, \bibinfo{author}{Frank\xfnm[
  I.W.]}, \bibinfo{author}{Tanenbaum\xfnm[ D.M.]},
  \bibinfo{author}{Parpia\xfnm[ J.M.]}, et~al.
\newblock \bibinfo{title}{Electromechanical resonators from graphene sheets}.
\newblock \bibinfo{journal}{Science}
  \bibinfo{year}{2007};\bibinfo{volume}{315}:\bibinfo{pages}{490}.
\bibitem[{Robinson et~al.(2008)Robinson, Zalalutdinov, Baldwin, Snow, Wei,
  Sheehan et~al.}]{RobinsonJT}
\bibinfo{author}{Robinson\xfnm[ J.T.]}, \bibinfo{author}{Zalalutdinov\xfnm[
  M.]}, \bibinfo{author}{Baldwin\xfnm[ J.W.]}, \bibinfo{author}{Snow\xfnm[
  E.S.]}, \bibinfo{author}{Wei\xfnm[ Z.]}, \bibinfo{author}{Sheehan\xfnm[ P.]},
  et~al.
\newblock \bibinfo{title}{Wafer-scale reduced graphene oxide films for
  nanomechanical devices}.
\newblock \bibinfo{journal}{Nano Letters}
  \bibinfo{year}{2008};\bibinfo{volume}{8}(\bibinfo{number}{10}):\bibinfo{page%
s}{3441--3445}.
\bibitem[{van~der Zande et~al.(2010)van~der Zande, Barton, Alden, Ruiz-Vargas,
  Whitney, Pham et~al.}]{ZandeAMVD}
\bibinfo{author}{van~der Zande\xfnm[ A.M.]}, \bibinfo{author}{Barton\xfnm[
  R.A.]}, \bibinfo{author}{Alden\xfnm[ J.S.]},
  \bibinfo{author}{Ruiz-Vargas\xfnm[ C.S.]}, \bibinfo{author}{Whitney\xfnm[
  W.S.]}, \bibinfo{author}{Pham\xfnm[ P.H.Q.]}, et~al.
\newblock \bibinfo{title}{Large-scale arrays of single-layer graphene
  resonators}.
\newblock \bibinfo{journal}{Nano Letters}
  \bibinfo{year}{2010};\bibinfo{volume}{10}(\bibinfo{number}{12}):\bibinfo{pag%
es}{4869--4873}.
\bibitem[{Reserbat-Plantey et~al.(2012)Reserbat-Plantey, Marty, Arcizet,
  Bendiab and Bouchiat}]{Reserbat-PlanteyA}
\bibinfo{author}{Reserbat-Plantey\xfnm[ A.]}, \bibinfo{author}{Marty\xfnm[
  L.]}, \bibinfo{author}{Arcizet\xfnm[ O.]}, \bibinfo{author}{Bendiab\xfnm[
  N.]}, \bibinfo{author}{Bouchiat\xfnm[ V.]}.
\newblock \bibinfo{title}{A local optical probe for measuring motion and stress
  in a nanoelectromechanical system}.
\newblock \bibinfo{journal}{Nature Nanotechnology}
  \bibinfo{year}{2012};\bibinfo{volume}{7}:\bibinfo{pages}{151--155}.
\bibitem[{Ruiz-Vargas et~al.(2011)Ruiz-Vargas, Zhuang, Huang, van~der Zande,
  Garg, McEuen et~al.}]{Ruiz-VargasCS}
\bibinfo{author}{Ruiz-Vargas\xfnm[ C.S.]}, \bibinfo{author}{Zhuang\xfnm[
  H.L.]}, \bibinfo{author}{Huang\xfnm[ P.Y.]}, \bibinfo{author}{van~der
  Zande\xfnm[ A.M.]}, \bibinfo{author}{Garg\xfnm[ S.]},
  \bibinfo{author}{McEuen\xfnm[ P.L.]}, et~al.
\newblock \bibinfo{title}{Softened elastic response and unzipping in chemical
  vapor deposition graphene membranes}.
\newblock \bibinfo{journal}{Nano Letters}
  \bibinfo{year}{2011};\bibinfo{volume}{11}:\bibinfo{pages}{2259--2263}.
\bibitem[{Garcia-Sanchez et~al.(2008)Garcia-Sanchez, van~der Zande, Paulo,
  Lassagne, McEuen and Bachtold}]{SanchezDG}
\bibinfo{author}{Garcia-Sanchez\xfnm[ D.]}, \bibinfo{author}{van~der
  Zande\xfnm[ A.M.]}, \bibinfo{author}{Paulo\xfnm[ A.S.]},
  \bibinfo{author}{Lassagne\xfnm[ B.]}, \bibinfo{author}{McEuen\xfnm[ P.L.]},
  \bibinfo{author}{Bachtold\xfnm[ A.]}.
\newblock \bibinfo{title}{Imaging mechanical vibrations in suspended graphene
  sheets}.
\newblock \bibinfo{journal}{Nano Letters}
  \bibinfo{year}{2008};\bibinfo{volume}{8}(\bibinfo{number}{5}):\bibinfo{pages%
}{1399}.
\bibitem[{$\ddot{O}$. Girit et~al.(2009)$\ddot{O}$. Girit, Meyer, Erni,
  Rossell, Kisielowski, Yang et~al.}]{GiritCO}
\bibinfo{author}{$\ddot{O}$. Girit\xfnm[ C.]}, \bibinfo{author}{Meyer\xfnm[
  J.C.]}, \bibinfo{author}{Erni\xfnm[ R.]}, \bibinfo{author}{Rossell\xfnm[
  M.D.]}, \bibinfo{author}{Kisielowski\xfnm[ C.]}, \bibinfo{author}{Yang\xfnm[
  L.]}, et~al.
\newblock \bibinfo{title}{Graphene at the edge: Stability and dynamics}.
\newblock \bibinfo{journal}{Science}
  \bibinfo{year}{2009};\bibinfo{volume}{323}:\bibinfo{pages}{1705}.
\bibitem[{Chen et~al.(2009)Chen, Rosenblatt, Bolotin, Kalb, Kim, Kymissis
  et~al.}]{ChenC2009nn}
\bibinfo{author}{Chen\xfnm[ C.]}, \bibinfo{author}{Rosenblatt\xfnm[ S.]},
  \bibinfo{author}{Bolotin\xfnm[ K.I.]}, \bibinfo{author}{Kalb\xfnm[ W.]},
  \bibinfo{author}{Kim\xfnm[ P.]}, \bibinfo{author}{Kymissis\xfnm[ I.]}, et~al.
\newblock \bibinfo{title}{Performance of monolayer graphene nanomechanical
  resonators with electrical readout}.
\newblock \bibinfo{journal}{Nature Nanotechnology}
  \bibinfo{year}{2009};\bibinfo{volume}{4}:\bibinfo{pages}{861}.
\bibitem[{Barton et~al.(2011)Barton, Ilic, van~der Zande, Whitney, Parpia and
  Craighead}]{BartonRA}
\bibinfo{author}{Barton\xfnm[ R.A.]}, \bibinfo{author}{Ilic\xfnm[ B.]},
  \bibinfo{author}{van~der Zande\xfnm[ A.M.]}, \bibinfo{author}{Whitney\xfnm[
  W.S.]}, \bibinfo{author}{Parpia\xfnm[ P.L.M.J.M.]},
  \bibinfo{author}{Craighead\xfnm[ H.G.]}.
\newblock \bibinfo{title}{High, size-dependent quality factor in an array of
  graphene mechanical resonators}.
\newblock \bibinfo{journal}{Nano Letters}
  \bibinfo{year}{2011};\bibinfo{volume}{11}:\bibinfo{pages}{1232--1236}.
\bibitem[{Croy et~al.(2012)Croy, Midtvedt, Isacsson and Kinaret}]{CroyA}
\bibinfo{author}{Croy\xfnm[ A.]}, \bibinfo{author}{Midtvedt\xfnm[ D.]},
  \bibinfo{author}{Isacsson\xfnm[ A.]}, \bibinfo{author}{Kinaret\xfnm[ J.M.]}.
\newblock \bibinfo{title}{Nonlinear damping in graphene resonators}.
\newblock \bibinfo{journal}{Physical Review B}
  \bibinfo{year}{2012};\bibinfo{volume}{86}(\bibinfo{number}{23}):\bibinfo{pag%
es}{235435}.
\bibitem[{Atalaya et~al.(2008)Atalaya, Isacsson and Kinaret}]{AtalayaJ}
\bibinfo{author}{Atalaya\xfnm[ J.]}, \bibinfo{author}{Isacsson\xfnm[ A.]},
  \bibinfo{author}{Kinaret\xfnm[ J.M.]}.
\newblock \bibinfo{title}{Continuum elastic modeling of graphene resonators}.
\newblock \bibinfo{journal}{Nano Letters}
  \bibinfo{year}{2008};\bibinfo{volume}{8}(\bibinfo{number}{12}):\bibinfo{page%
s}{4196--4200}.
\bibitem[{Atalaya et~al.(2010)Atalaya, Kinaret and Isacsson}]{AtalayaJ2010epl}
\bibinfo{author}{Atalaya\xfnm[ J.]}, \bibinfo{author}{Kinaret\xfnm[ J.M.]},
  \bibinfo{author}{Isacsson\xfnm[ A.]}.
\newblock \bibinfo{title}{Nanomechanical mass measurement using nonlinear
  response of a graphene membrane}.
\newblock \bibinfo{journal}{Europhysics Letters}
  \bibinfo{year}{2010};\bibinfo{volume}{91}:\bibinfo{pages}{48001}.
\bibitem[{Kim and Park(2010)}]{KimSYnanotechnology}
\bibinfo{author}{Kim\xfnm[ S.Y.]}, \bibinfo{author}{Park\xfnm[ H.S.]}.
\newblock \bibinfo{title}{On the utility of vacancies and tensile
  strain-induced quality factor enhancement for mass sensing using graphene
  monolayers}.
\newblock \bibinfo{journal}{Nanotechnology}
  \bibinfo{year}{2010};\bibinfo{volume}{21}:\bibinfo{pages}{105710}.
\bibitem[{Qi and Park(2012)}]{QiZ2012nns}
\bibinfo{author}{Qi\xfnm[ Z.]}, \bibinfo{author}{Park\xfnm[ H.S.]}.
\newblock \bibinfo{title}{Intrinsic energy dissipation in cvd-grown graphene
  nanoresonators}.
\newblock \bibinfo{journal}{Nanoscale}
  \bibinfo{year}{2012};\bibinfo{volume}{4}:\bibinfo{pages}{3460--3465}.
\bibitem[{Kwona et~al.(2012)Kwona, Leeb, Hwangc and Kang}]{KwonaOK}
\bibinfo{author}{Kwona\xfnm[ O.K.]}, \bibinfo{author}{Leeb\xfnm[ G.Y.]},
  \bibinfo{author}{Hwangc\xfnm[ H.J.]}, \bibinfo{author}{Kang\xfnm[ J.W.]}.
\newblock \bibinfo{title}{Molecular dynamics modeling and simulations to
  understand gate-tunable graphene-nanoribbon-resonator}.
\newblock \bibinfo{journal}{Physica E: Low-dimensional Systems and
  Nanostructures}
  \bibinfo{year}{2012};\bibinfo{volume}{45}:\bibinfo{pages}{194--200}.
\bibitem[{He et~al.(2005)He, Kitipornchai and Liew}]{HeXQ2005nano}
\bibinfo{author}{He\xfnm[ X.Q.]}, \bibinfo{author}{Kitipornchai\xfnm[ S.]},
  \bibinfo{author}{Liew\xfnm[ K.M.]}.
\newblock \bibinfo{title}{Resonance analysis of multi-layered graphene sheets
  used as nanoscale resonators}.
\newblock \bibinfo{journal}{Nanotechnology}
  \bibinfo{year}{2005};\bibinfo{volume}{16}:\bibinfo{pages}{2086--2091}.
\bibitem[{Kim and Park(2009{\natexlab{a}})}]{KimSY2009apl}
\bibinfo{author}{Kim\xfnm[ S.Y.]}, \bibinfo{author}{Park\xfnm[ H.S.]}.
\newblock \bibinfo{title}{Multilayer friction and attachment effects on energy
  dissipation in graphene nanoresonators}.
\newblock \bibinfo{journal}{Applied Physics Letters}
  \bibinfo{year}{2009}{\natexlab{a}};\bibinfo{volume}{94}:\bibinfo{pages}{1019%
18}.
\bibitem[{Seo$\acute{a}$nez et~al.(2007)Seo$\acute{a}$nez, Guinea and
  Neto}]{SeoanezC}
\bibinfo{author}{Seo$\acute{a}$nez\xfnm[ C.]}, \bibinfo{author}{Guinea\xfnm[
  F.]}, \bibinfo{author}{Neto\xfnm[ A.H.C.]}.
\newblock \bibinfo{title}{Dissipation in graphene and nanotube resonators}.
\newblock \bibinfo{journal}{Physical Review B}
  \bibinfo{year}{2007};\bibinfo{volume}{76}:\bibinfo{pages}{125427}.
\bibitem[{Kim and Park(2009{\natexlab{b}})}]{KimSY2009nl}
\bibinfo{author}{Kim\xfnm[ S.Y.]}, \bibinfo{author}{Park\xfnm[ H.S.]}.
\newblock \bibinfo{title}{The importance of edge effects on the intrinsic loss
  mechanisms of graphene nanoresonators}.
\newblock \bibinfo{journal}{Nano Letters}
  \bibinfo{year}{2009}{\natexlab{b}};\bibinfo{volume}{9}(\bibinfo{number}{3}):%
\bibinfo{pages}{969--974}.
\bibitem[{Jiang and Wang(2012)}]{JiangJW2012jap}
\bibinfo{author}{Jiang\xfnm[ J.W.]}, \bibinfo{author}{Wang\xfnm[ J.S.]}.
\newblock \bibinfo{title}{Why edge effects are important on the intrinsic loss
  mechanisms of graphene nanoresonators}.
\newblock \bibinfo{journal}{Journal of Applied Physics}
  \bibinfo{year}{2012};\bibinfo{volume}{111}(\bibinfo{number}{5}):\bibinfo{pag%
es}{054314}.
\bibitem[{Jiang et~al.(2014)Jiang, Wang, Park and
  Rabczuk}]{JiangJW2013diffusion}
\bibinfo{author}{Jiang\xfnm[ J.W.]}, \bibinfo{author}{Wang\xfnm[ B.S.]},
  \bibinfo{author}{Park\xfnm[ H.S.]}, \bibinfo{author}{Rabczuk\xfnm[ T.]}.
\newblock \bibinfo{title}{Adsorbate migration effects on continuous and
  discontinuous temperature-dependent transitions in the quality factors of
  graphene nanoresonators}.
\newblock \bibinfo{journal}{Nanotechnology}
  \bibinfo{year}{2014};\bibinfo{volume}{25}(\bibinfo{number}{2}):\bibinfo{page%
s}{025501}.
\bibitem[{Ekinci and Roukes(2005)}]{EkinciKL}
\bibinfo{author}{Ekinci\xfnm[ K.L.]}, \bibinfo{author}{Roukes\xfnm[ M.L.]}.
\newblock \bibinfo{title}{Nanoelectromechanical systems}.
\newblock \bibinfo{journal}{Rev Sci Instrum}
  \bibinfo{year}{2005};\bibinfo{volume}{76}:\bibinfo{pages}{061101}.
\bibitem[{Varghese et~al.(2010)Varghese, Nair, Nair, Hanajiri, Maekawa, Yoshida
  et~al.}]{VargheseSH}
\bibinfo{author}{Varghese\xfnm[ S.H.]}, \bibinfo{author}{Nair\xfnm[ R.]},
  \bibinfo{author}{Nair\xfnm[ B.G.]}, \bibinfo{author}{Hanajiri\xfnm[ T.]},
  \bibinfo{author}{Maekawa\xfnm[ T.]}, \bibinfo{author}{Yoshida\xfnm[ Y.]},
  et~al.
\newblock \bibinfo{title}{Sensors based on carbon nanotubes and their
  applications: A review}.
\newblock \bibinfo{journal}{Current Nanoscience}
  \bibinfo{year}{2010};\bibinfo{volume}{6}:\bibinfo{pages}{331--346}.
\bibitem[{Eom et~al.(2011)Eom, Park, Yoon and Kwon}]{EomK}
\bibinfo{author}{Eom\xfnm[ K.]}, \bibinfo{author}{Park\xfnm[ H.S.]},
  \bibinfo{author}{Yoon\xfnm[ D.S.]}, \bibinfo{author}{Kwon\xfnm[ T.]}.
\newblock \bibinfo{title}{Nanomechanical resonators and their applications in
  biological/chemical detection: Nanomechanics principles}.
\newblock \bibinfo{journal}{Physical Review}
  \bibinfo{year}{2011};\bibinfo{volume}{503}:\bibinfo{pages}{115--163}.
\bibitem[{Arlett et~al.(2011)Arlett, Myers and Roukes}]{ArlettJL}
\bibinfo{author}{Arlett\xfnm[ J.]}, \bibinfo{author}{Myers\xfnm[ E.]},
  \bibinfo{author}{Roukes\xfnm[ M.]}.
\newblock \bibinfo{title}{Comparative advantages of mechanical biosensors}.
\newblock \bibinfo{journal}{Nature Nanotechnology}
  \bibinfo{year}{2011};\bibinfo{volume}{6}:\bibinfo{pages}{203}.
\bibitem[{Imboden and Mohanty(2013)}]{ImbodenM2013pr}
\bibinfo{author}{Imboden\xfnm[ M.]}, \bibinfo{author}{Mohanty\xfnm[ P.]}.
\newblock \bibinfo{title}{Dissipation in nanoelectromechanical systems}.
\newblock \bibinfo{journal}{Physics Reports}
  \bibinfo{year}{2013};\bibinfo{volume}{534}(\bibinfo{number}{3}):\bibinfo{pag%
es}{89--146}.
\bibitem[{Paulo et~al.(2008)Paulo, Black, Garcia-Sanchez, Esplandiu, Aguasca,
  Bokor et~al.}]{PauloAS2008jpcs}
\bibinfo{author}{Paulo\xfnm[ A.S.]}, \bibinfo{author}{Black\xfnm[ J.]},
  \bibinfo{author}{Garcia-Sanchez\xfnm[ D.]}, \bibinfo{author}{Esplandiu\xfnm[
  M.J.]}, \bibinfo{author}{Aguasca\xfnm[ A.]}, \bibinfo{author}{Bokor\xfnm[
  J.]}, et~al.
\newblock \bibinfo{title}{Mechanical detection and mode shape imaging of
  vibrational modes of micro and nanomechanical resonators by dynamic force
  microscopy}.
\newblock \bibinfo{journal}{Journal of Physics: Conference Series}
  \bibinfo{year}{2008};\bibinfo{volume}{100}:\bibinfo{pages}{052009}.
\bibitem[{Unterreithmeier et~al.(2010)Unterreithmeier, Faust and
  Kotthaus}]{UnterreithmeierQP2010prl}
\bibinfo{author}{Unterreithmeier\xfnm[ Q.P.]}, \bibinfo{author}{Faust\xfnm[
  T.]}, \bibinfo{author}{Kotthaus\xfnm[ J.P.]}.
\newblock \bibinfo{title}{Damping of nanomechanical resonators}.
\newblock \bibinfo{journal}{Physical Review Letters}
  \bibinfo{year}{2010};\bibinfo{volume}{105}(\bibinfo{number}{2}):\bibinfo{pag%
es}{027205}.
\bibitem[{Jiang et~al.(2004)Jiang, Yu, Liu and Huang}]{JiangH2004prl}
\bibinfo{author}{Jiang\xfnm[ H.]}, \bibinfo{author}{Yu\xfnm[ M.F.]},
  \bibinfo{author}{Liu\xfnm[ B.]}, \bibinfo{author}{Huang\xfnm[ Y.]}.
\newblock \bibinfo{title}{Intrinsic energy loss mechanisms in a cantilevered
  carbon nanotube beam oscillator}.
\newblock \bibinfo{journal}{Physical Review Letters}
  \bibinfo{year}{2004};\bibinfo{volume}{93}(\bibinfo{number}{18}):\bibinfo{pag%
es}{185501}.
\bibitem[{Jiang and Wang(2011{\natexlab{c}})}]{JiangJWtorsion}
\bibinfo{author}{Jiang\xfnm[ J.W.]}, \bibinfo{author}{Wang\xfnm[ J.S.]}.
\newblock \bibinfo{title}{Graphene-based tortional resonator from molecular
  dynamics simulation}.
\newblock \bibinfo{journal}{Europhysics Letters}
  \bibinfo{year}{2011}{\natexlab{c}};\bibinfo{volume}{96}(\bibinfo{number}{6})%
:\bibinfo{pages}{66007}.
\bibitem[{Matheny et~al.(2013)Matheny, Villanueva, Karabalin, Sader and
  Roukes}]{MathenyMH}
\bibinfo{author}{Matheny\xfnm[ M.H.]}, \bibinfo{author}{Villanueva\xfnm[
  L.G.]}, \bibinfo{author}{Karabalin\xfnm[ R.B.]}, \bibinfo{author}{Sader\xfnm[
  J.E.]}, \bibinfo{author}{Roukes\xfnm[ M.L.]}.
\newblock \bibinfo{title}{Nonlinear mode-coupling in nanomechanical systems}.
\newblock \bibinfo{journal}{Nano Letters}
  \bibinfo{year}{2013};:\bibinfo{pages}{DOI: 10.1021/nl400070e}.
\bibitem[{Jiang et~al.(2012)Jiang, Park and
  Rabczuk}]{JiangJW2012nanotechnology}
\bibinfo{author}{Jiang\xfnm[ J.W.]}, \bibinfo{author}{Park\xfnm[ H.S.]},
  \bibinfo{author}{Rabczuk\xfnm[ T.]}.
\newblock \bibinfo{title}{Enhancing the mass sensitivity of graphene
  nanoresonators via nonlinear oscillations: the effective strain mechanism}.
\newblock \bibinfo{journal}{Nanotechnology}
  \bibinfo{year}{2012};\bibinfo{volume}{23}:\bibinfo{pages}{475501}.
\bibitem[{Novoselov et~al.(2005{\natexlab{b}})Novoselov, Jiang, Schedin, Booth,
  Khotkevich, Morozov et~al.}]{NovoselovKS2005pnas}
\bibinfo{author}{Novoselov\xfnm[ K.S.]}, \bibinfo{author}{Jiang\xfnm[ D.]},
  \bibinfo{author}{Schedin\xfnm[ F.]}, \bibinfo{author}{Booth\xfnm[ T.J.]},
  \bibinfo{author}{Khotkevich\xfnm[ V.V.]}, \bibinfo{author}{Morozov\xfnm[
  S.V.]}, et~al.
\newblock \bibinfo{title}{Two-dimensional atomic crystals}.
\newblock \bibinfo{journal}{Proceedings of the National Academy of Science}
  \bibinfo{year}{2005}{\natexlab{b}};\bibinfo{volume}{102}(\bibinfo{number}{30%
}):\bibinfo{pages}{10451--10453}.
\bibitem[{Kam and Parkinson(1982)}]{KamKK}
\bibinfo{author}{Kam\xfnm[ K.K.]}, \bibinfo{author}{Parkinson\xfnm[ B.A.]}.
\newblock \bibinfo{title}{Detailed photocurrent spectroscopy of the
  semiconducting group vi transition metal dichalcogenides}.
\newblock \bibinfo{journal}{Journal of Physical Chemistry}
  \bibinfo{year}{1982};\bibinfo{volume}{86}(\bibinfo{number}{4}):\bibinfo{page%
s}{463--467}.
\bibitem[{Feng et~al.(2012)Feng, Qian, Huang and Li}]{FengJ2012npho}
\bibinfo{author}{Feng\xfnm[ J.]}, \bibinfo{author}{Qian\xfnm[ X.]},
  \bibinfo{author}{Huang\xfnm[ C..]}, \bibinfo{author}{Li\xfnm[ J.]}.
\newblock \bibinfo{title}{Strain-engineered artificial atom as a broad-spectrum
  solar energy funnel}.
\newblock \bibinfo{journal}{Nature Photonics}
  \bibinfo{year}{2012};\bibinfo{volume}{6}(\bibinfo{number}{12}):\bibinfo{page%
s}{866--872}.
\bibitem[{Lu et~al.(2012)Lu, Wu, Guo and Zeng}]{LuP2012pccp}
\bibinfo{author}{Lu\xfnm[ P.]}, \bibinfo{author}{Wu\xfnm[ X.]},
  \bibinfo{author}{Guo\xfnm[ W.]}, \bibinfo{author}{Zeng\xfnm[ X.C.]}.
\newblock \bibinfo{title}{Strain-dependent electronic and magnetic properties
  of mos$_{2}$ monolayer, bilayer, nanoribbons and nanotubes}.
\newblock \bibinfo{journal}{Phys Chem Chem Phys}
  \bibinfo{year}{2012};\bibinfo{volume}{14}(\bibinfo{number}{37}):\bibinfo{pag%
es}{13035--13040}.
\bibitem[{Radisavljevic et~al.(2011)Radisavljevic, Radenovic, Brivio,
  Giacometti and Kis}]{RadisavljevicB2011nn}
\bibinfo{author}{Radisavljevic\xfnm[ B.]}, \bibinfo{author}{Radenovic\xfnm[
  A.]}, \bibinfo{author}{Brivio\xfnm[ J.]}, \bibinfo{author}{Giacometti\xfnm[
  V.]}, \bibinfo{author}{Kis\xfnm[ A.]}.
\newblock \bibinfo{title}{Single-layer mo{S}$_{2}$ transistors}.
\newblock \bibinfo{journal}{Nature Nanotechnology}
  \bibinfo{year}{2011};\bibinfo{volume}{6}:\bibinfo{pages}{147}.
\bibitem[{Wang et~al.(2012{\natexlab{b}})Wang, Kalantar-Zadeh, Kis, Coleman and
  Strano}]{WangQH2012nn}
\bibinfo{author}{Wang\xfnm[ Q.H.]}, \bibinfo{author}{Kalantar-Zadeh\xfnm[ K.]},
  \bibinfo{author}{Kis\xfnm[ A.]}, \bibinfo{author}{Coleman\xfnm[ J.N.]},
  \bibinfo{author}{Strano\xfnm[ M.S.]}.
\newblock \bibinfo{title}{Electronics and optoelectronics of two-dimensional
  transition metal dichalcogenides}.
\newblock \bibinfo{journal}{Nature Nanotechnology}
  \bibinfo{year}{2012}{\natexlab{b}};\bibinfo{volume}{7}(\bibinfo{number}{11})%
:\bibinfo{pages}{699--712}.
\bibitem[{Chhowalla et~al.(2013)Chhowalla, Shin, Eda, Li, Loh and
  Zhang}]{ChhowallaM}
\bibinfo{author}{Chhowalla\xfnm[ M.]}, \bibinfo{author}{Shin\xfnm[ H.S.]},
  \bibinfo{author}{Eda\xfnm[ G.]}, \bibinfo{author}{Li\xfnm[ L..]},
  \bibinfo{author}{Loh\xfnm[ K.P.]}, \bibinfo{author}{Zhang\xfnm[ H.]}.
\newblock \bibinfo{title}{The chemistry of two-dimensional layered transition
  metal dichalcogenide nanosheets}.
\newblock \bibinfo{journal}{Nature Chemistry}
  \bibinfo{year}{2013};\bibinfo{volume}{5}(\bibinfo{number}{4}):\bibinfo{pages%
}{263--275}.
\bibitem[{Conley et~al.(2013)Conley, Wang, Ziegler, Haglund, Pantelides and
  Bolotin}]{ConleyHJ}
\bibinfo{author}{Conley\xfnm[ H.J.]}, \bibinfo{author}{Wang\xfnm[ B.]},
  \bibinfo{author}{Ziegler\xfnm[ J.I.]}, \bibinfo{author}{Haglund\xfnm[ R.F.]},
  \bibinfo{author}{Pantelides\xfnm[ S.T.]}, \bibinfo{author}{Bolotin\xfnm[
  K.I.]}.
\newblock \bibinfo{title}{Bandgap engineering of strained monolayer and bilayer
  mos$_{2}$}.
\newblock \bibinfo{journal}{Nano Letters}
  \bibinfo{year}{2013};\bibinfo{volume}{13}(\bibinfo{number}{8}):\bibinfo{page%
s}{3626--3630}.
\bibitem[{Sangwan et~al.(2013)Sangwan, Arnold, Jariwala, Marks, Lauhon and
  Hersam}]{SangwanVK}
\bibinfo{author}{Sangwan\xfnm[ V.K.]}, \bibinfo{author}{Arnold\xfnm[ H.N.]},
  \bibinfo{author}{Jariwala\xfnm[ D.]}, \bibinfo{author}{Marks\xfnm[ T.J.]},
  \bibinfo{author}{Lauhon\xfnm[ L.J.]}, \bibinfo{author}{Hersam\xfnm[ M.C.]}.
\newblock \bibinfo{title}{Low-frequency electronic noise in single-layer
  mos$_{2}$ transistors}.
\newblock \bibinfo{journal}{Nano Letters}
  \bibinfo{year}{2013};\bibinfo{volume}{13}(\bibinfo{number}{9}):\bibinfo{page%
s}{4351--4355}.
\bibitem[{Ghorbani-Asl et~al.(2013)Ghorbani-Asl, Zibouche, Wahiduzzaman,
  Oliveira, Kuc and Heine}]{Ghorbani-AslM}
\bibinfo{author}{Ghorbani-Asl\xfnm[ M.]}, \bibinfo{author}{Zibouche\xfnm[ N.]},
  \bibinfo{author}{Wahiduzzaman\xfnm[ M.]}, \bibinfo{author}{Oliveira\xfnm[
  A.F.]}, \bibinfo{author}{Kuc\xfnm[ A.]}, \bibinfo{author}{Heine\xfnm[ T.]}.
\newblock \bibinfo{title}{Electromechanics in mos$_{2}$ and ws$_{2}$: Nanotubes
  vs. monolayers}.
\newblock \bibinfo{journal}{Scientific Reports}
  \bibinfo{year}{2013};\bibinfo{volume}{3}:\bibinfo{pages}{2961}.
\bibitem[{Cheiwchanchamnangij et~al.(2013)Cheiwchanchamnangij, Lambrecht, Song
  and Dery}]{CheiwchanchamnangijT}
\bibinfo{author}{Cheiwchanchamnangij\xfnm[ T.]},
  \bibinfo{author}{Lambrecht\xfnm[ W.R.L.]}, \bibinfo{author}{Song\xfnm[ Y.]},
  \bibinfo{author}{Dery\xfnm[ H.]}.
\newblock \bibinfo{title}{Strain effects on the spin-orbit induced band
  structure splittings in monolayer mos$_{2}$ and graphene}.
\newblock \bibinfo{journal}{Physical Review B}
  \bibinfo{year}{2013};\bibinfo{volume}{88}:\bibinfo{pages}{155404}.
\bibitem[{Huang et~al.(2013)Huang, Da and Liang}]{HuangW}
\bibinfo{author}{Huang\xfnm[ W.]}, \bibinfo{author}{Da\xfnm[ H.]},
  \bibinfo{author}{Liang\xfnm[ G.]}.
\newblock \bibinfo{title}{Thermoelectric performance of mx$_{2}$ (m=mo, w; x=s,
  se) monolayers}.
\newblock \bibinfo{journal}{Journal of Applied Physics}
  \bibinfo{year}{2013};\bibinfo{volume}{113}:\bibinfo{pages}{104304}.
\bibitem[{Varshney et~al.(2010)Varshney, Patnaik, Muratore, Roy, Voevodin and
  Farmer}]{VarshneyV}
\bibinfo{author}{Varshney\xfnm[ V.]}, \bibinfo{author}{Patnaik\xfnm[ S.S.]},
  \bibinfo{author}{Muratore\xfnm[ C.]}, \bibinfo{author}{Roy\xfnm[ A.K.]},
  \bibinfo{author}{Voevodin\xfnm[ A.A.]}, \bibinfo{author}{Farmer\xfnm[ B.L.]}.
\newblock \bibinfo{title}{Md simulations of molybdenum disulphide (mos$_{2}$):
  Force-field parameterization and thermal transport behavior}.
\newblock \bibinfo{journal}{Computational Materials Science}
  \bibinfo{year}{2010};\bibinfo{volume}{48}(\bibinfo{number}{1}):\bibinfo{page%
s}{101--108}.
\bibitem[{Bertolazzi et~al.(2011)Bertolazzi, Brivio and Kis}]{BertolazziS}
\bibinfo{author}{Bertolazzi\xfnm[ S.]}, \bibinfo{author}{Brivio\xfnm[ J.]},
  \bibinfo{author}{Kis\xfnm[ A.]}.
\newblock \bibinfo{title}{Stretching and breaking of ultrathin mos$_{2}$}.
\newblock \bibinfo{journal}{ACS Nano}
  \bibinfo{year}{2011};\bibinfo{volume}{5}(\bibinfo{number}{12}):\bibinfo{page%
s}{9703--9709}.
\bibitem[{Cooper et~al.(2013{\natexlab{a}})Cooper, Lee, Marianetti, Wei, Hone
  and Kysar}]{CooperRC2013prb1}
\bibinfo{author}{Cooper\xfnm[ R.C.]}, \bibinfo{author}{Lee\xfnm[ C.]},
  \bibinfo{author}{Marianetti\xfnm[ C.A.]}, \bibinfo{author}{Wei\xfnm[ X.]},
  \bibinfo{author}{Hone\xfnm[ J.]}, \bibinfo{author}{Kysar\xfnm[ J.W.]}.
\newblock \bibinfo{title}{Nonlinear elastic behavior of two-dimensional
  molybdenum disulfide}.
\newblock \bibinfo{journal}{Physical Review B}
  \bibinfo{year}{2013}{\natexlab{a}};\bibinfo{volume}{87}:\bibinfo{pages}{0354%
23}.


\end{thebibliography}

\end{document}